\documentclass[acmsmall]{acmart}

\AtBeginDocument{%
  \providecommand\BibTeX{{%
    \normalfont B\kern-0.5em{\scshape i\kern-0.25em b}\kern-0.8em\TeX}}}

\setcopyright{acmlicensed}
\acmJournal{PACMMOD}
\acmYear{2023} \acmVolume{1} \acmNumber{1} \acmArticle{15} \acmMonth{5} \acmPrice{15.00}\acmDOI{10.1145/3588695}

\acmPrice{15.00}
\acmISBN{978-1-4503-XXXX-X/18/06}

\usepackage{amsthm}
\usepackage{graphicx}
\usepackage{subfigure}

\usepackage[linesnumbered,ruled,vlined]{algorithm2e}
\usepackage{algpseudocode}  
\usepackage{amsmath}  
\usepackage{enumerate}  
\newtheorem{myDef}{Definition}[section]
\newtheorem{Problem}{Problem Statement}
\newtheorem{example}{Example}[section]
\usepackage{multirow}
\usepackage{balance}
\usepackage{booktabs}

\begin{document}

\title{Fast Continuous Subgraph Matching over Streaming Graphs via Backtracking Reduction}

\author{Rongjian Yang}
\affiliation{%
  \institution{School of Data Science, Fudan University}
  \city{Shanghai}
  \country{China}}
\email{rjyang20@fudan.edu.cn}

\author{Zhijie Zhang}
\affiliation{%
  \institution{School of Data Science, Fudan University}
  \city{Shanghai}
  \country{China}}
\email{zhijiezhang18@fudan.edu.cn}

\author{Weiguo Zheng}
\affiliation{%
  \institution{School of Data Science, Fudan University}
  \city{Shanghai}
  \country{China}}
\email{zhengweiguo@fudan.edu.cn}

\author{Jeffrey Xu Yu}
\affiliation{%
  \institution{The Department of Systems Engineering and Engineering Management, The Chinese University of Hong Kong}
  \city{Hong Kong}
  \country{China}}
\email{yu@se.cuhk.edu.hk}

\begin{abstract}
Streaming graphs are drawing increasing attention in both academic and industrial communities as many graphs in real applications evolve over time. Continuous subgraph matching (shorted as CSM) aims to report the incremental matches of a query graph in such streaming graphs. It involves two major steps, i.e., candidate maintenance and incremental match generation, to answer CSM. Throughout the course of continuous subgraph matching, incremental match generation backtracking over the search space dominates the total cost. However, most previous approaches focus on developing techniques for efficient candidate maintenance, while incremental match generation receives less attention despite its importance in CSM. Aiming to minimize the overall cost, we propose two techniques to reduce backtrackings in this paper. We present a cost-effective index CaLiG that yields tighter candidate maintenance, shrinking the search space of backtracking. In addition, 
 we develop a novel incremental matching paradigm KSS that decomposes the query vertices into conditional kernel vertices and shell vertices. With the matches of kernel vertices, the incremental matches can be produced immediately by joining the candidates of shell vertices without any backtrackings. Benefiting from reduced backtrackings, the elapsed time of CSM decreases significantly. Extensive experiments over real graphs show that our method runs faster than the state-of-the-art algorithm orders of magnitude.
\end{abstract}

\ccsdesc[500]{Information systems~Stream management}

\keywords{Subgraph Matching, Streaming Graph, Backtracking Reduction}

\maketitle


\section{Introduction}
Graphs have been widely used to represent complex relations among a variety of objects, spanning from social networks, knowledge graphs, and road networks to electrical networks. In most real-world applications, the graph evolves over time by adding vertices/edges or deleting vertices/edges, called a streaming graph. 
For example, in a social network, new user registration and closing an account can be regarded as a vertex addition and deletion in the graph, respectively; creating and cancelling the interaction (e.g., following) between users correspond to edge addition and deletion, respectively. Such a social network is a typical streaming graph. How to efficiently manage and query streaming graphs is drawing increasing attention~\cite{8509462, graphone, 10.1145/2627692.2627694, beams, 40999f14773f48b7a3e910d0e899be67, risgraph, iturbograph, auxo}.
 
\begin{figure}[t]
    \subfigure[Query Graph]{
         \centering
         \includegraphics[width=0.18\linewidth]{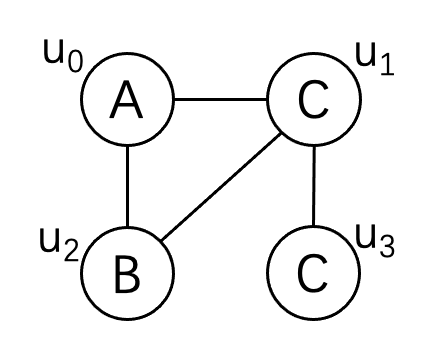}
         \label{fig:q}
         }
    \subfigure[Streaming Graph]{
         \centering
         \includegraphics[width=0.25\linewidth]{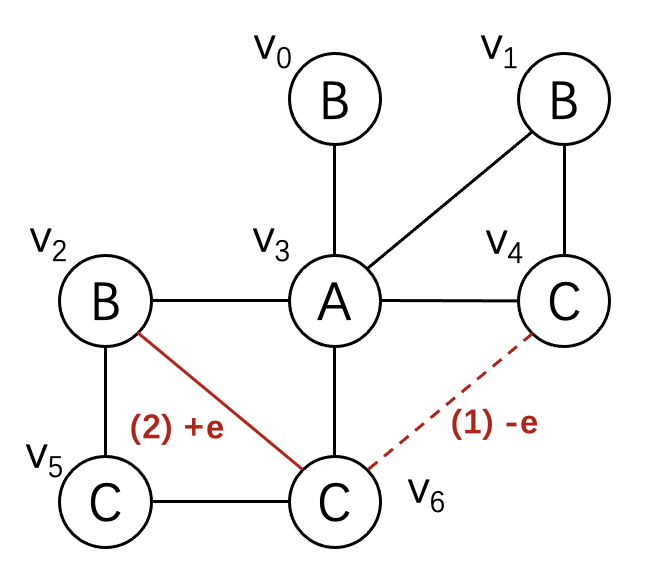}
         \label{fig:g}
         }
         \vspace{-0.05cm}
    \subfigure[Incremental Matches]{
         \centering
         \includegraphics[width=0.3\linewidth]{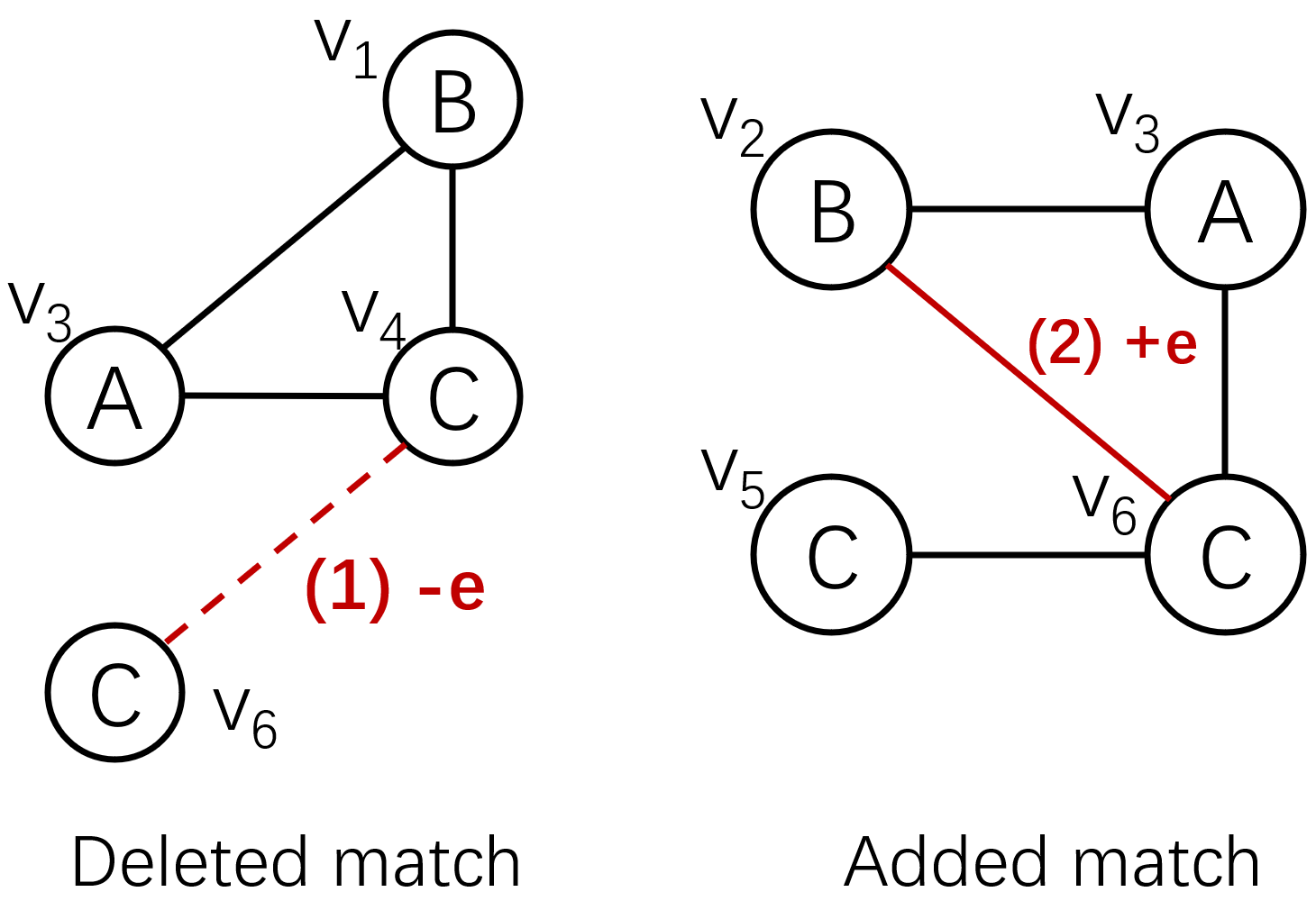}
         \label{fig:um}
         }
    \caption{A running example of CSM.}
    \label{fig:sm}
\end{figure}

As an important task, monitoring a pattern $Q$ of interest over a streaming graph $G$ is to perform continuous subgraph matching (shorted as CSM) to report the incremental matches, i.e., the increased or decreased subgraph matches of $Q$ owing to the graph updates. 
As shown in the running example in Figure~\ref{fig:sm}, one match is deleted after deleting edge $(v_4,v_6)$ and another match is added after adding edge $(v_2,v_6)$.

CSM is useful in a wide range of applications, such as recommendation systems~\cite{10.14778/2733004.2733010,10.1145/2076732.2076746,5741690}, fraud detection~\cite{10.14778/2735496.2735504,10.14778/3229863.3229874}, and cyber security~\cite{osti_1183625, streamworks}, etc.

For example, a cycle pattern can be served as a strong indication of a fake transaction in e-commerce platforms~\cite{10.14778/3229863.3229874}, where the accounts of users (buyers or sellers) are represented as vertices and online transactions, e.g., payment activities, are denoted as dynamic edges. With CSM, suspicious transactions would be detected to generate real-time alerts and trigger prompt actions.

\subsection{The Existing Methods}
With well-developed subgraph matching algotithms~\cite{cfl,ceci,dpiso},
one naive approach for CSM is to enumerate the matches before
and after graph updating respectively.
However, applying subgraph matching directly suffers from intractable cost~\cite{closuretree, 10.1145/543613.543620},
as the characteristics of dynamic graph updates are not fully exploited. 
Considering the fact that the update edge is contained in the desired matches if any, the index-free algorithms, 
e.g., IncIsoMat~\cite{10.1145/2489791}
and Graphflow~\cite{graphflow},  
find the changed matches by expanding the added or deleted edge without generating unnecessary matches.
Such algorithms may waste too much time in exploring the data graph even if there are not any matches since they do not employ any information before updating.

To catch the incremental matches in real-time, more efforts have been devoted to developing index techniques to maintain intermediate results recently~\cite{osti_1183625,turboflux,symbi}.
SJ-Tree~\cite{osti_1183625}
defines a left-deep decomposition tree, in which each node maintains a set of partial matches.
Since the number of matches could be exponential, SJ-tree suffers from an
intractable storage overhead.
Instead of maintaining partial matches, 
TurboFlux~\cite{turboflux} and SymBi~\cite{symbi} 
elaborate on the candidate generation and maintenance by exploiting auxiliary data structures and carefully designed filter rules.
TurboFlux~\cite{turboflux} generates a spanning tree $T_Q$ of $Q$
and introduces an auxiliary
data structure called data-centric graph (DCG), to maintain whether an edge in $G$ matches an edge in $T_Q$ or not.
It uses the bottom-up method to conduct filtration 
with only the edges in $T_Q$ taken into account.
For graph updates, TurboFlux renews the states of edges in DCG to determine 
whether every two edges could be matched and whether each edge in $G$ could be contained in a match of $Q$.  
SymBi~\cite{symbi} adopts a directed acyclic graph (DAG) built from $Q$ and checks forward or backward \emph{neighbors}, covering the non-tree edges overlooked by TurboFlux.
SymBi employs top-down and bottom-up dynamic programming on the DAG. Only those vertices (that) pass the top-down filer need to perform the bottom-up filter
and only those (that) have passed both filters
are the valid candidates.
Both TurboFlux and SymBi emphasize the succinctness and update efficiency of the index, even if the time cost of maintaining candidates usually accounts for a very small part of the overall cost (as discussed in Section~\ref{bottleneck}).
In the search stage, TurboFlux adopts a 
backtracking algorithm 
with a fixed matching order that depends on the estimated size of query paths.
SymBi selects a dynamic match order,
which could be adaptively
changed during the backtracking, 
according to the estimated size of extendable
candidates of query vertices. 
They just focus on the matching order
rather than improving the backtracking framework.

{
\begin{table}[tbp]
  \centering
  \caption{Proportion of candidate maintenance (\%)}
    \begin{tabular}{cccccccccc}
    \toprule
    \textbf{Method} &  Deezer  & Email &  Github  &  Lastfm  &  Skitter  &  Twitch \\
    \midrule
TurboFlux &
 0.13 &  0.093 &	0.072 & 	0.19 & 	0.091 &	0.091
 \\
    SymBi &
0.306 &	0.380 & 0.072 &	0.971 &	0.476 &	0.003 \\
    CaLiG &
1.402 &	2.652 &	0.913 &	1.623 &	1.112 &	0.235 \\
    \bottomrule
    \end{tabular}
  \label{tab:backtracking rate}%
\end{table}
}

{
\subsection{Bottleneck of CSM}
\label{bottleneck}
Among all the aforementioned approaches, the index-based methods often perform better than the others. 
To further accelerate continuous subgraph matching, a crucial question needs to be answered: what is the bottleneck of improving the efficiency performance? As discussed above, the existing methods mainly focus on 
how to maintain candidate vertices fast by using succinct indexes.
}
\textit{Surprisingly, incremental match generation, a more important and crucial issue, has not received much attention up to now,  despite its dominating role in answering CSM.} Specifically, finding incremental
matches involves two major steps, namely candidate maintenance (including candidate generation and index update)
and incremental match generation. Once obtaining  new candidate vertices caused by graph updates, 
{incremental match generation}
is conducted to find the increased or decreased matches of $Q$. Due to the NP-hardness of subgraph isomorphism, incremental matching generation dominates the overall cost. Table~\ref{tab:backtracking rate} reports the proportion of candidate maintenance over the total time cost of different algorithms on 6 real graphs from SNAP~\cite{snapnets}, where CaLiG is our proposed method.
The results are averaged on 50 randomly generated queries for each data graph. 
It is observed that the time cost of
candidate maintenance {in TurboFlux and SymBi} is 
less than
0.5\% on most graphs.

Actually, the target of an index for CSM is to locate the candidate vertices, further reducing the cost of incremental match generation. A better index should not just focus on fast candidate generation at the cost of delivering more false-positive candidates, degrading the incremental match generation.  Hence,  there is a trade-off between update efficiency and candidate accuracy for index designing. 
{
To measure the accuracy of an index, we use the ratio of false positives, i.e., the ratio of vertices in the candidate set but not in any delivered matches.
The fewer false positives, the tighter the candidates generated.
Following the principle of cost-effectiveness, we find the indexes of TurboFlux and SymBi are not tight enough.
It is worth spending a little more time
to 
obtain tighter candidates, achieving
 considerable speedups in the subsequent incremental match generation.
}

\begin{figure}[t]
    \centering
    \includegraphics[width=0.55\linewidth]{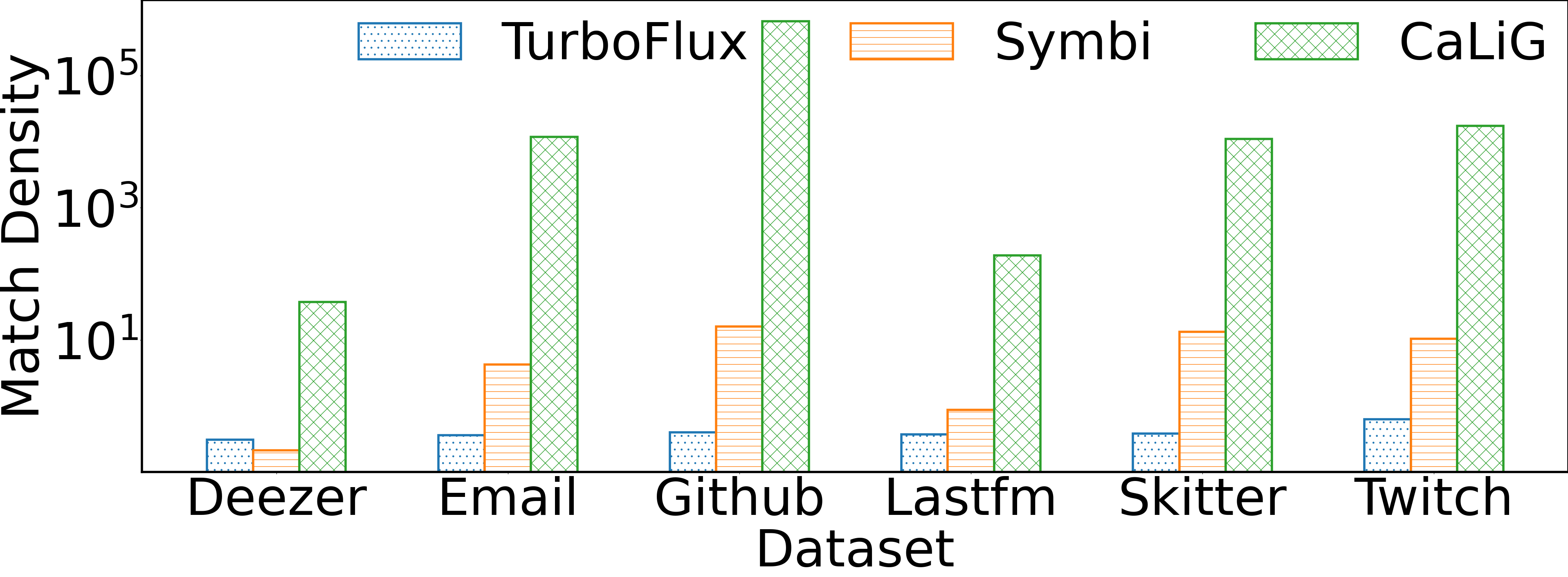}
    \caption{{Averaged match density of different algorithms on 50 random queries for each graph.}}
    \label{fig:match_density}
    \vspace{-0.15cm}
\end{figure}

A common approach to incremental match generation is backtracking
over search space, that is,
it extends the partial match by one vertex each time
and backtracks if failing or finally succeeding.
To reduce backtrackings, it is expected that more matches can be returned by fewer backtrackings, i.e., one backtracking is supposed to generate more matches. Thus, we define a metric \textit{match density} as the average number of matches generated by one backtracking.
Larger match density indicates the algorithm finds all the matches with fewer backtrackings, usually less elapsed time as well. 
Figure~\ref{fig:match_density} presents the averaged 
match density 
of
TurboFlux, SymBi, and CaLiG.
It is clear that 
{both TurboFlux and SymBi} have lower match density and one backtracking could only find about 0.2$\sim$10 matches on average, meaning most backtrackings are invalid and leaves much room to improve.

\subsection{Our Approach and Contributions}

{
As discussed above,
most previous methods are dedicated to candidate maintenance, rather
than directly reducing the cost of match generation.
Different from them, we propose two techniques, namely a novel index structure
CaLiG (to obtain tighter candidates) and a powerful incremental matching paradigm (to avoid unnecessary backtrackings),
aiming to minimize the overall cost of CSM.}

First, we propose a cost-effective structure, called \textit{candidate lighting graph (shorted as CaLiG)}, a directed graph where each node represents a matching pair of vertices $(u,v)$. CaLiG provides a tighter candidate space than previous methods, reducing the search space and total backtrackings. 
Cost-effective, means (that) it is worthwhile
to build up an index for tighter candidate maintenance because of the significant speed-up in the subsequent incremental match generation, even if a little bit more time may be required. 

In addition, we develop an efficient
incremental matching paradigm,  \textit{kernel-and-shell search (shorted as KSS)}.
The underlying principle is that some vertices are not necessary to be extended one by one following repeated backtrackings.
KSS decomposes the query into conditional kernel vertices and shell vertices. The partial matches for the kernel vertices are computed by backtracking first, then the complete incremental matches can be produced immediately by joining the candidates of shell vertices without any backtrackings.
We prove that finding the minimum conditional kernel set is an NP-hard problem, and thus propose an efficient greedy method.
To evaluate the efficiency of the proposed method, we conducted extensive experimental studies over real graphs. The results confirm the significant superiority of our method. 

In summary, we make the following contributions.
\begin{itemize}
    \item  We propose a  cost-effective index CaLiG that
    yields tighter candidate maintenance than the existing methods, which in turn reduces the total backtrackings.
    
    \item We design a novel subgraph matching paradigm, called kernel-and-shell search, which can produce incremental matches by simply joining candidates of the shell vertices, reducing the unpromising backtrackings.
    
    \item We formalize the problem of minimum conditional kernel set and prove its NP-hardness. To enhance continuous subgraph matching, KSS dynamically selects the best conditional kernel and shell sets for different update edges.
    
    \item Extensive experiments over real graphs have demonstrated that our proposed method outperforms the state-of-the-art algorithm orders of magnitude. 
\end{itemize}


\section{Problem Definition and Overview}
{
In this paper, we focus on undirected vertex-labeled graphs, though our algorithm can be extended to directed and edge-labeled graphs, where each vertex/edge has one label. 
For the edge-labeled graph, we can build a vertex-labeled graph by taking each edge as a vertex that connects two endpoints, where the label of the newly added vertex is that of the original edge. 

}

A graph is denoted as a tuple $G=\{V_G, E_G, L\}$, where $V_G$ and $E_G$ are the set of vertices and edges, respectively. $L$ is the function mapping a label, i.e., $L(v)$, to a vertex $v \in V_G$. Table~\ref{tab:notations} lists the frequently used notations in the paper.

\subsection{Problem Definition}

\begin{myDef}[Subgraph Isomorphism]\label{def:isomorphism}
Given a query graph $Q=\{V_Q, E_Q, L\}$ and a data graph $G=\{V_G, E_G, L\}$, $Q$ is subgraph isomorphic to $G$ if there exists an injective function \(f\):
\(V_Q \rightarrow V_G\), such that
\begin{enumerate}
\item
   $\forall$ \(u \in V_Q\), we have \(L(u) = L(f(u))\) where $f(u) \in V_G$, and
\item
  $\forall$ 
  \(e(u_1, u_2) \in E_Q\), we have
  \(e(f(u_1), f(u_2)) \in E_G\).
\end{enumerate}
\end{myDef}

\begin{table}[t]
  \caption{Notations}
  \label{tab:notations}
  \begin{tabular}{ll}
    \toprule
    Notations & Descriptions \\
    \midrule
\(Q\) and \(G\) & query graph and data graph \\
 \(u \in V_Q\) and \(v \in V_G\) &  query vertex and data vertex, respectively \\
\(e(v_i,v_j)\in E_G\) & edge between \(v_i\) and \(v_j\) \\
\(N_Q(u)\) & neighbors of \(u\) in graph \(Q\) \\
\(N_G(v)\) & neighbors of \(v\) in graph \(G\) \\
$d_Q$ and $d_G$ & {maximum degree of $Q$ and $G$}, respectively \\
($u,v$)-MP & a matching pair consists of $u$ and $v$\\
($u,v$)-MP.state & the lighting state of ($u,v$)-MP\\
$In_{CaLiG}(u,v)$ & in-neighbors of ($u,v$)-MP in CaLiG\\
$Out_{CaLiG}(u,v)$ & out-neighbors of ($u,v$)-MP in CaLiG\\
$BI(u,v)$ & the bigraph for ($u,v$)-MP\\
  \bottomrule
\end{tabular}
\end{table}
Subgraph matching returns all subgraphs of $G$ that are isomorphic to $Q$. Each match can be expressed as a set of one-to-one matching pairs $\{(u\leftrightarrow f(u))\}$. 
All the matches are denoted by $M$.

\begin{myDef}[Streaming Graph]
A graph $G$ is called a streaming graph if it changes dynamically following a sequence of update operations $\Delta G$ including four kinds of updates, i.e., vertex addition/deletion and edge addition/deletion.
\end{myDef}

Since the query graph in our task is connected, the newly added vertex, clearly an isolated vertex in the data graph, cannot be included in any subgraph matching result. The vertex deletion only happens to isolated vertices, and the deletion of a non-isolated vertex can be taken as deleting all its connected edges. Hence, 
we just consider edge addition and deletion by convention~\cite{symbi}.

\begin{Problem} (Continuous Subgraph Matching, shorted as CSM).
Given a query graph $Q$, a data graph $G$, and a graph update stream $\Delta G$, the continuous subgraph matching task is to find the incremental matches (i.e., newly added or decreased subgraph matches) of $Q$ for each update operation in $\Delta G$. 
\end{Problem}

\begin{example}
Let us consider the data graph in Figure~\autoref{fig:g}. Assume that there are two update operations, i.e., edge deletion $e(v_4, v_6)$ followed by edge addition $e(v_2, v_6)$.
The decreased match is
the left in Figure~\autoref{fig:um},
owing to deleting the edge $e(v_4, v_6)$. 
After adding the edge $e(v_2, v_6)$, a new match (right in Figure~\autoref{fig:um})
emerges.
\end{example}

\subsection{Overview of Our Approach}
\begin{figure*}[t]
    \centering
    \includegraphics[width=0.95\linewidth]{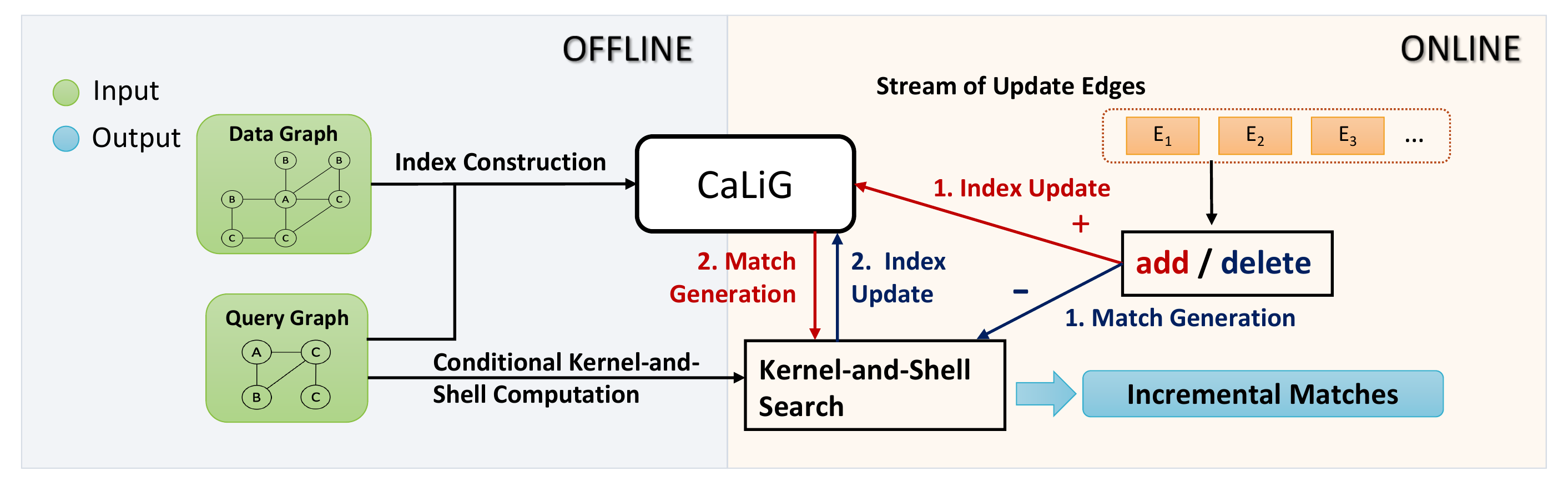}
        \vspace{-0.3cm}
    \caption{Overview of our approach.
    }
    \label{fig:process}
\end{figure*}

\autoref{fig:process} presents the framework of the proposed approach, consisting of the offline indexing phase and online querying phase.

In the offline phase, we develop a novel index \textbf{Ca}ndidate \textbf{Li}ghting \textbf{G}raph, shorted as CaLiG. 
The index structure CaLiG is constructed according to the input query graph $Q$ and data graph $G$. 

{
\begin{myDef}[Matching Pair]
A pair of vertices $u$ and $v$ form a matching pair, shorted as $(u,v)$-MP, if $L(u)$ = $L(v)$.
Each $(u,v)$-MP has a lighting state ``ON'' or ``OFF'', indicating whether $v$ can match $u$ or not, respectively. Let $(u,v)$-MP.state denote the lighting state.
\end{myDef}
}

{
CaLiG organizes candidate matching pairs in 
a vertex-pair graph, where each node
represents a pair ($u,v$) of query vertex $u$ and data vertex $v$. CaLiG imports the lighting state 
(``ON'' or ``OFF'') to indicate whether $v\in V_G$ matches $u \in V_Q$ or not, i.e., whether $v$ is a candidate of $u$.
}
Benefiting from capturing all the connecting relations in both the query graph and data graph, it is easy to maintain tighter candidates and support incremental match generation.
In addition, we compute a particular conditional kernel set and a shell set for each edge in the query graph. 

{
In the online phase, the system responds differently according to the received update operation.
}
For the edge addition, the index CaLiG is updated first to obtain new candidates, and then the subgraph matching is performed
(as marked using red lines in \autoref{fig:process}); For edge deletion, the subgraph matching is conducted first, and then CaLiG is updated
(as marked using deep blue lines in \autoref{fig:process}).

{
\textit{CaLiG update}. To support incremental subgraph matching for subsequent updates, the index CaLiG needs to be maintained in real-time. 
Either adding or deleting an edge
may lead to candidate updates and CaLiG changes including both the structure and the lighting states.
}
Moreover, the changed state of one pair could propagate to other matching pairs, starting recursive propagations over CaLiG.

\textit{Incremental match}. In order to complete the incremental match over the streaming graph efficiently, we design a new matching paradigm, kernel-and-shell search (KSS) in the subgraph matching phase. KSS divides the vertices of the query graph into conditional kernel vertices and shell vertices. It first
takes candidates of the update edge
 as initially partial matches and then expands the matches to cover all the kernel vertices. Finally, the incremental matches can be produced easily by
 joining candidates of the shell vertices without any backtrackings.


\section{CaLIG: Candidate Lighting Graph}

To reduce the intermediate results (i.e., failures and backtracking in the search), we resort to a carefully designed index
CaLiG.

\subsection{CaLiG Structure}\label{subsec:CaLiG}

{
According to the definition of subgraph isomorphism, the vertex $v \in V_G$ matching $u \in V_Q$ or not 
depends on whether neighbors of $v$ matches neighbors of $u$.
Therefore, CaLiG organizes the matching pairs $(u,v)$-MP as a graph over which the lighting states can be easily maintained.
}

\begin{myDef}[CaLiG Index]\label{def:CaLiG}
The CaLiG index for
$Q$ and $G$ is a directed graph where each node represents an MP. There is a directed edge from ($u_i,v_j$)-MP to ($u_k, v_l$)-MP if it holds that 
\begin{enumerate}[1)]
    \item $e(u_i, u_k) \in E_Q$ and  $e(v_j, v_l) \in E_G$; and,
    \item ($u_i,v_j$)-MP is ``ON'' or ($u_i,v_j$)-MP is turned ``OFF'' after ($u_k, v_l$)-MP. 
\end{enumerate}
\end{myDef}

In the CaLiG, we use the term ``node'' to distinguish vertices in $Q$ or $G$.
Note that the proposed index is designed for continuous subgraph matching and the lighting state of each matching pair may be updated according to the graph updates. Initially, the lighting states of all nodes in CaLiG are ``ON'',
 as all matching pairs are not pruned at first.

\begin{figure}

    \centering
    \includegraphics[width=0.6\linewidth]{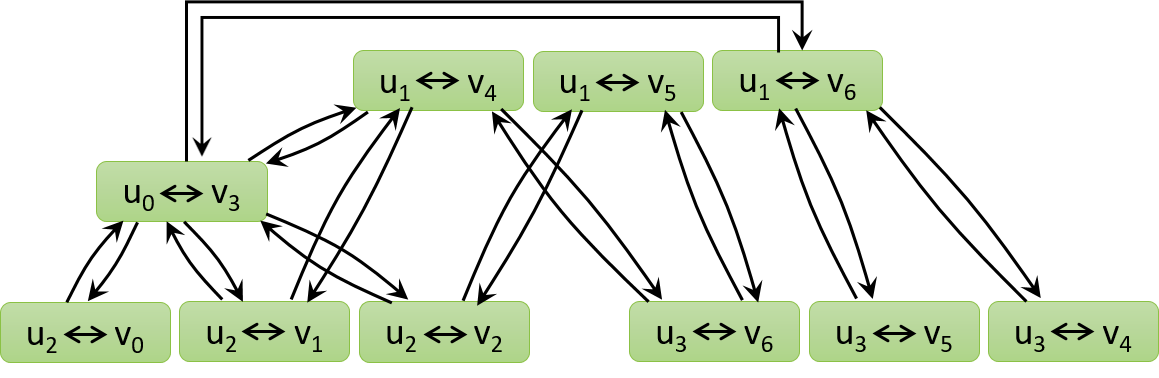}
    \caption{CaLiG for the query graph and data graph in Figure~\ref{fig:sm}, where the lighting states of all MPs are ``ON'' (as filled with green color).}
    \label{fig:calig}
\end{figure}

\begin{example}
Figure~\ref{fig:calig} shows the CaLiG Index constructed based on the query graph and data graph in Figure~\ref{fig:sm}. The query graph has 1 A vertex, 1 B vertex, and 2 C vertices, while the data graph has 1 A vertex, 3 B vertices, and 3 C vertices, so CaLiG contains $1\times 1 + 1\times 3 + 2\times 3 = 10$ nodes. Since $e(u_1,u_2) \in E_Q$, $e(v_1,v_4) \in E_G$, and $(u_1,v_4)$-MP is ``ON'', there is an edge from  $(u_1,v_4)$-MP to $(u_2,v_1)$-MP.
\end{example}

\begin{algorithm} [b] 
  \caption{ConstructCaLiG($G$, $Q$)}  
  \label{alg:build}  
  \KwIn{A data graph $G$, a query graph $Q$}  
  \KwOut{A candidate lighting graph $CaLiG$}  
  $CaLiG \gets$ GenerateEmptyCaLiG()\;
  \For{each $(u, v) \in (V_Q, V_G)$}
  {
      \If{$L(v)=L(u)$}{
        $(u,v)$-MP.state $\gets ON$\;
        add a node $(u,v)$-MP into CaLiG\;
      }
  }
  \For{each $(u,v)$-MP $\in CaLiG$}
  {
    \For{each $(u', v') \in (N_Q(u), N_G(v))$}{
      \If{$(u',v')$-MP $\in CaLiG$}{
        add an edge from $(u',v')$-MP to $(u,v)$-MP\;
      }
    }
  }
  IndexInitialization($CaLiG$)\;
  \Return{$CaLiG$}\;
\end{algorithm} 

Algorithm~\ref{alg:build} illustrates the process of constructing CaLiG and refining the lighting states of matching pairs (turning OFF the nodes that do meet the matching requirements).
First, we generate the matching pairs (nodes) of CaLiG (lines 2-5), where the initial lighting state of each matching pair is ON (line 4).
Next, we generate the edges in CaLiG (lines 6-9). For each node $(u,v)$-MP in CaLiG, we traverse all neighbors of $u$ in $Q$ and all neighbors of $v$ in $G$. If the pair of neighbors $(u', v')$ forms a matching pair in CaLiG, an edge from $(u',v')$-MP to $(u,v)$-MP is added as all the matching pairs are ``ON'' initially. 
Last, we need to initialize the structure of CaLiG by updating the lighting states (line 10), which will be introduced in detail in Section~\ref{subsec:candidate generation}.
Note that after CaLiG is initialized, the incremental matches if any must be embedded in all ON-state nodes.

\subsection{Lighting State Computation}\label{subsec:lighting state}

Based on the definition of subgraph isomorphism (Definition~\ref{def:isomorphism}), we know that $v$ matches $u$ only if $v$'s neighbors match $u$'s neighbors as well. 
{
The existing filter rules examine the existence
of candidates for the neighbors of each query vertex, but ignore the 
injective requirement necessary for subgraph isomorphism.
Although each neighbor of $u$ can find a candidate in $v$'s neighbors, $v$'s neighbors may still fail to match $u$'s neighbors when one neighbor of $v$ is taken as candidates of multiple neighbors of $u$, violating the injective requirement.
Thus, to compute the lighting state of nodes in CaLiG, we introduce a bipartite graph
for each matching pair.

\begin{myDef}[Bigraph for $(u,v)$-MP]\label{def:bigraph}
A bigraph for $(u,v)$-MP, denoted by BI$(u,v)$, is a bipartite graph with two 
disjoint sets of
vertices $N_Q(u)$ and $N_G(v)$, where there is an edge between $u_i \in N_Q(u)$ and $v_j \in N_G(v)$ if $(u_i,v_j)$-MP is an in-neighbor of $(u,v)$-MP in the CaLiG.
\end{myDef}
}

For a matching pair $(u,v)$, to determine whether $v$ matches $u$, we can just consider the in-neighbors of $(u,v)$-MP in CaLiG, rather than examine all the vertex pairs \{$(u_i,v_j)| u_i \in N_Q(u) \wedge v_j \in N_G(v) \wedge  L(u_i) = L(v_j)$\}.

\begin{figure} 
    \subfigure[No injective matching in the bigraph $BI(u_1,v_6)$ of $(u_1,v_6)$-MP.]{
         \centering
         \includegraphics[width=0.3\linewidth]{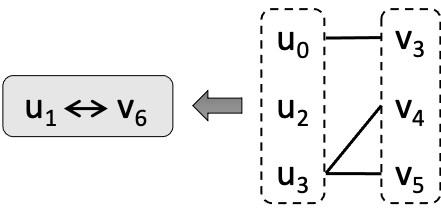}
         \label{fig:pm1}
         }
         \hspace{1.5cm}
     \subfigure[Having an injective matching in the bigragh $BI(u_1,v_4)$ of $(u_1,v_4)$-MP.]{
         \centering
         \includegraphics[width=0.3\linewidth]{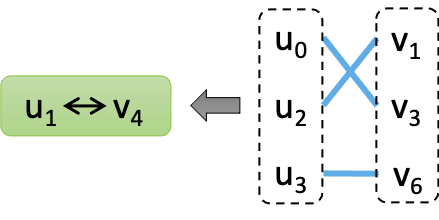}
         \label{fig:pm2}
         }
    \caption{Bigraphs for $(u_1,v_6)$-MP and $(u_1,v_4)$-MP respectively, where blue edges indicate an injective matching.}
    \label{fig:pm}
\end{figure}

\begin{example}
Let us consider $(u_1,v_6)$-MP and $(u_1,v_4)$-MP in Figure~\ref{fig:sm}. Their bigraphs are presented in Figure~\ref{fig:pm}.
The neighbor $u_2$ of $u_1$ does not have any candidates, which means when we take $(v_6 \leftrightarrow u_1)$ as a partial match, no candidates will be available for $u_2$, and this partial matching will fail.
That is, $v_6$ should not be a candidate of $u_1$.
Thus, the lighting state of the ($u_1,v_6$)-MP should be ``OFF''. Clearly, precise lighting states of MPs benefit tight candidates, improving the time efficiency of finding incremental matches.
\end{example}
{
Based on subgraph isomorphism, $v$'s neighbors can match $u$'s neighbors only if there is an injective matching for $BI(u,v)$.
}
{%
\begin{myDef}[Injective Matching]\label{injective matching}
Given a bigraph with two 
disjoint sets of
vertices {$X$ and $Y$}, there is an injective matching if each vertex $u$ in $X$is matched against a vertex in $Y$ via an edge and all the edges in the matching are independent of each other, i.e., all edges do not share vertices.
\end{myDef}
}

For a ($u,v$)-MP, if its bigraph does not have an {injective matching} for $N_Q(u)$, at least one vertex $u_i \in N_Q(u)$ cannot be matched. We can set the lighting state ``OFF'' safely.

\begin{lemma}\label{lemma:lighting}
 The bigraph as defined in Definition~\ref{def:bigraph} is sufficient to determine the lighting state of the matching pair ($u,v$)-MP. Given a ($u,v$)-MP, its lighting state is ``OFF'' only if there is no {injective matching} for $N_Q(u)$ in the bigraph $BI(u,v)$; Otherwise, the state is ``ON''.
\end{lemma}
\vspace{-0.25cm}
\begin{proof}
The edges of the bigraph represent in-neighbors of ($u,v$)-MP in CaLiG. According to Definition~\ref{def:CaLiG}, no matter what the state of $(u,v)$-MP is, the non-in-neighbors of $(u,v)$-MP are definitely OFF. An OFF-state node means it does not belong to any match, so we do not need to take the OFF-state node into account when we determine the state of $(u,v)$-MP. 
At least one vertex $u_i\in N_Q(u)$ cannot be matched if $BI(u,v)$ does not have any {injective matching} for $N_Q(u)$, leading to the OFF state of ($u,v$)-MP.
\end{proof}

\begin{example}
 \vspace{-0.1cm}
As shown in Figure~\ref{fig:pm1}, $BI(u_1,v_6)$ does not have any {injective matching}, so we change the state of $(u_1,v_6)$-MP to OFF. In Figure~\ref{fig:pm2}, there is an {injective matching} as marked in blue lines, so the state of $(u_1,v_4)$-MP remains ON.
\end{example}

Computing the {injective matching} for a bigraph (also called bipartite graph) could employ Hopcroft–Karp algorithm~\cite{Hopcroft} with the time complexity of $O(\sqrt{|V|}|E|)$, where $|V|$ is the number of vertices and $|E|$ is the number of edges in the bigraph.
In practice, it just matters whether such an {injective matching} exists or not rather than what the {injective matching} is. There are two straightforward cases in which no injective matching exists.

-Case 1: One vertex $u_i$ in $N_Q(u)$ has no incident edge in the bipartite graph, meaning has no candidate.

-Case 2: The vertices in $N_G(v)$ having at least one incident edge in the bipartite graph is less than $|N_Q(u)|$.

In the second case, we constrain the vertices having at least one incident edge since each edge represents a relation of candidate and the bipartite graph may change with the stream of edge updates. 

{
Computing the injective matching
is closely related to Hall's marriage theorem~\cite{hall}.
\begin{theorem} (Hall's Marriage Theorem~\cite{hall} )
Given a bigraph with two
disjoint sets of
vertices $X$ and $Y$, for a subset $W$ of $X$, let $N(W)$ denote the neighbors of $W$, i.e.,
a subset of $Y$ in which vertices are connected to some vertices of $W$.
There is an injective matching if and only if for every subset $W$ of $X$ we have $|W| < |N(W)|$. 
In other words, every subset $W$ of $X$ has sufficiently many neighbors in $Y$.
\end{theorem}

The two cases above 
are indeed special cases of Hall's marriage theorem, where  $W = \{u_i\}$ in Case 1 and  $W = X$ in Case 2.
}

\begin{lemma}
The vertex pairs connected by the edges in $BI(u,v)$ form a subset of \{$(u_i,v_j)| u_i \in N_Q(u) \wedge v_j \in N_G(v) \wedge  L(u_i) = L(v_j)$\}.
\end{lemma}
\vspace{-0.35cm}
\begin{proof}
The proof is straightforward according to definitions of CaLiG (Definition~\ref{def:CaLiG}) and bigraph (Definition~\ref{def:bigraph}).
\end{proof}

{Generally, }
the time cost of computing an injective matching can be reduced if the bigraph of $(u,v)$-MP contains fewer edges. It is ideal to just keep the necessary edges corresponding to in-neighbors of $(u,v)$-MP in CaLiG. 
The in-neighbors of $(u,v)$-MP could be further reduced through a propagation mechanism next.

\subsection{CaLiG Initialization}\label{subsec:candidate generation}

{
In CaLiG,
the lighting state of a node is
determined by the states of its neighbors.
When the lighting state of a node $(u,v)$-MP is turned ``OFF'',
i.e., $v$ is pruned from the candidate set of $u$, the neighbors of $(u,v)$-MP would be affected. If one of the neighbors is turned ``OFF'',
its neighbors
would be affected recursively.
Such a process is called OFF-state propagation which refines the candidates in CaLiG initialization.
}

\noindent \textbf{OFF-State Propagation}. 
By using Lemma~\ref{lemma:lighting}, the lighting state of each Matching Pair can be determined initially (i.e., the first round checking). The OFF-state matching pairs
cannot be included in any subgraph matches. Therefore, the edges corresponding to these
OFF-state matching pairs in the bigraphs of other matching pairs can be deleted safely, 
which may further turn off a set of matching pairs. The procedure proceeds iteratively until no matching pairs can be newly turned off. Finally, the ON-state matching pairs straightforwardly produce the candidates. The index CaLiG finishes initialization, ready for handling graph updates. 

\begin{algorithm} [b]
  \caption{IndexInitialization($CaLiG$)}  
  \label{alg:init}  
  \KwIn{A candidate lighting graph $CaLiG$}  
  \For{each $(u,v)$-MP $\in CaLiG$}{
    build a bigraph $BI(u,v)$ for $(u,v)$-MP\;
  }
  \For{each $(u,v)$-MP $\in CaLiG$}{
    \If{$(u,v)$-MP.state $= ON$ and $BI(u,v)$ has no injective matching}{
      $(u,v)$-MP.state $\gets OFF$\;
      OFF-Propagation($CaLiG$, $(u,v)$-MP)\;
    }
  }
\end{algorithm} 

Algorithm~\ref{alg:init} outlines the initialization process, where the procedure in line~6 (Algorithm~\ref{alg:off}) can be viewed as propagation on CaLiG. 
Let $Out_{CaLiG}(u,v)$ denote out-neighbors of ($u,v$)-MP in CaLiG.
For each node $(u',v')$-MP whose lighting state is ON (line 2), we remove the edge from $(u,v)$-MP to $(u',v')$-MP (line 3) as $(u,v)$-MP will not contribute to the state of $(u',v')$-MP.
Then we check the updated bigraph $BI(u',v')$. If $(u',v')$-MP is turned off, the procedure OFF-Propagation will be invoked recursively (lines~5-7).

\begin{algorithm}[b]
  \caption{OFF-Propagation($CaLiG$, $(u,v)$-MP)}
  \label{alg:off}
  \KwIn{$CaLiG$ and a matching pair $(u,v)$-MP that was turned off in the previous round} 
  \For{each $(u',v')$-MP $\in Out_{CaLiG}(u,v)$}{
      \If{$(u',v')$-MP.state $= ON$}{
        delete the edge from $(u,v)$-MP to $(u',v')$-MP\;
        update $BI(u',v')$\;
        \If{$BI(u',v')$ has no injective matching}{
          $(u',v')$-MP.state $\gets OFF$\;
          OFF-Propagation($CaLiG$, $(u',v')$-MP)\;
        }
      }
    
  }
\end{algorithm}
\begin{figure} [t]
    \subfigure[Step 1. Turn off all nodes that do not have any injective matchings.]{
         \centering
         \includegraphics[width=0.48\linewidth]{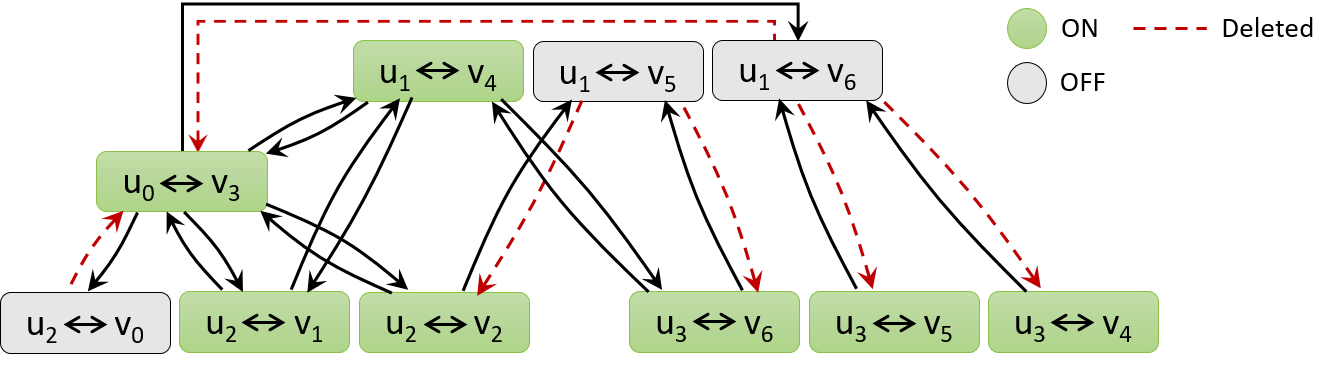}
         \label{fig:ft1}
         }
    \subfigure[Step 2. OFF-Propagation.]{
         \centering
         \includegraphics[width=0.46\linewidth]{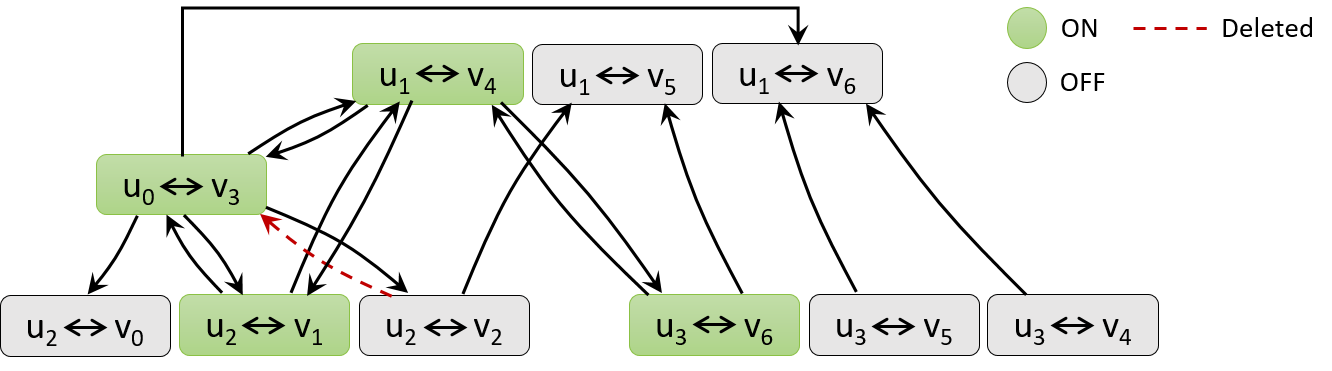}
         \label{fig:ft2}
         }
           \vspace{-0.25cm}
    \caption{CaLiG Initialization.}
    \label{fig:filter}
\end{figure}

\begin{example}
Let us consider the CaLiG in Figure~\ref{fig:calig}.
Since the bigraphs of $(u_2,v_0)$-MP, $(u_1,v_5)$-MP, and $(u_1,v_6)$-MP do not have any injective matching, they will be turned off in the first round.
At the same time, we remove the out-going edges (red dotted edges) as shown in Figure~\ref{fig:ft1}. In the second round, as shown in Figure \autoref{fig:ft2}, $(u_2,v_2)$-MP, $(u_3,v_5)$-MP and $(u_3,v_4)$-MP are turned off. In this round, only the edge from $(u_2,v_2)$-MP to $(u_0,v_3)$-MP is deleted as $(u_0,v_3)$-MP is the only ON-state out-neighbor.
Finally, no node can be further turned off and only nodes $(u_0,v_3)$-MP, $(u_2,v_1)$-MP, $(u_3,v_6)$-MP, and $(u_1,v_4)$-MP are ON. In fact, these four matching pairs can just form a subgraph match of the query graph.

\end{example}

\begin{figure*}[t] 
    \subfigure[Step 1. Delete all update edges in CaLiG.]{
         \centering
         \includegraphics[width=0.5\linewidth]{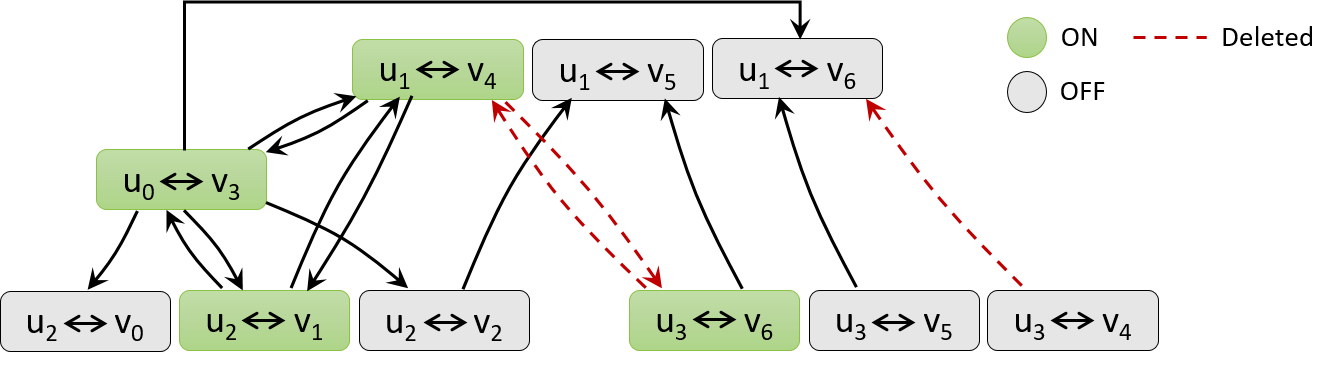}
         \label{fig:del1}
         }
    \subfigure[Step 2. Two nodes are turned off.]{
         \centering
         \includegraphics[width=0.45\linewidth]{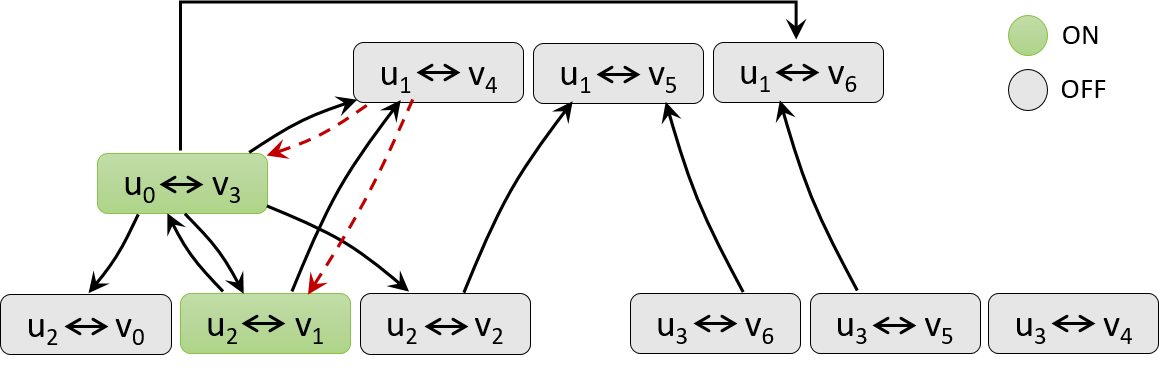}
         \label{fig:del2}
         }
         \vspace{-0.05cm}
    \subfigure[Step 3. OFF-Propagation.]{
         \centering
         \includegraphics[width=0.45\linewidth]{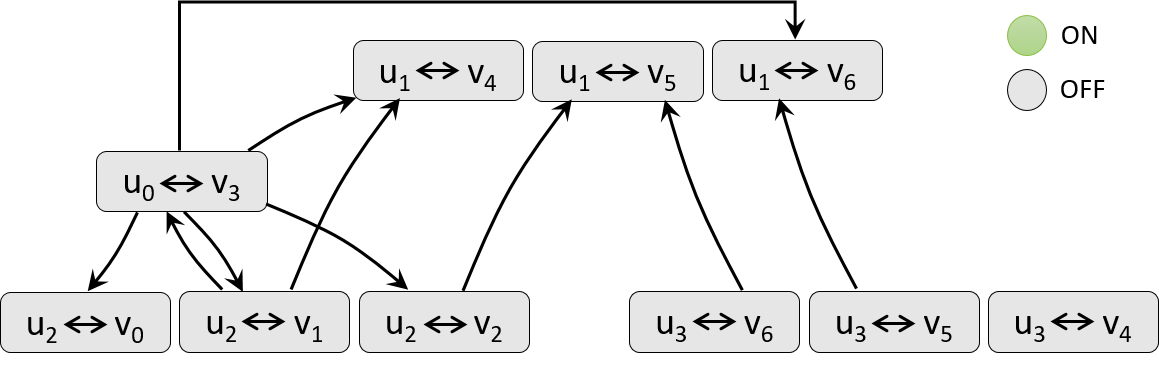}
         \label{fig:del3}
         }
         \vspace{-0.25cm}
    \caption{CaLiG update for deleting the edge $(v_4,v_6)$.}
    \label{fig:del}
\end{figure*}

The lighting process, i.e., CaLiG initialization, only needs to propagate to the ``ON''-state nodes.
For each ON node $(u,v)$-MP, its OFF in-neighbors have been deleted when the in-neighbors are turned off in the previous round, so the remaining matching pairs are all ON nodes. Then the state of $(u,v)$-MP will be re-checked by finding the injective matching. 
In short, the process of OFF propagation will only propagate among ON-state nodes, and the nodes that have been previously turned off will not be checked repeatedly.

\noindent \textbf{Compared with SymBi}. SymBi builds a query DAG and refines the candidates based on weak embeddings. It focuses on the existence (i.e., mapping from $N_Q(u)$ to $N_Q(v)$) such that the reused candidates could not be excluded. In contrast, CaLiG computes the injective matching from $N_Q(u)$ to $N_Q(v)$) and propagates along the edges in CaLiG.

Hence, \textit{the candidates computed by CaLiG are equal to or tighter than SymBi.}

\begin{lemma}
The candidates computed by CaLiG are equal to or tighter than SymBi.
\end{lemma}
\vspace{-0.25cm}
\begin{proof}
SymBi proposes an index $D_i[u,v]$ to maintain the dynamic candidate space. The recurrence process is described as:

\begin{itemize}
    \item $D_1[u,v] = 1$ iff $\exists v_p$ adjacent to $v$ such that $D_1[u_p,v_p] = 1$ for every parent $u_p$ of $u$ in DAG $ \hat q$;
    \item $D_2[u,v] = 1$ iff $D_1[u,v] = 1$ and $\exists v_c$ adjacent to $v$ such that $D_1[u_c,v_c] = 1$ for every parent $u_p$ of $u$ in DAG $\hat q$.
\end{itemize}
SymBi considers $\{v_i | D_2[u,v] = 1 \}$ as the candidates of query vertex $u$. A tighter filter rule could be obtained by extending the number of filtering iterations from only 2 to until the candidate space  converges and extending the vertex needed to check from parent/children to all the neighbors, i.e. 
\begin{itemize}
    \item The vertex $v$ in $Cand(u)$ iff
    $v_n$ adjacent to $v$ such that $v_n$  in $Cand(u)$ for every neighbor $u_p$ of $u$ in $q$.
\end{itemize}
which is the third constraint of CaLiG candidate space. Then The candidates computed by CaLiG are at least equal to SymBi.
\end{proof}

\noindent \textbf{Complexity Analysis}.
{
The number of nodes in CaLiG is $O(|V_G|\times |V_Q|)$ and the number of edges is $O(|E_G|\times |E_Q|)$. 
Correspondingly, the bigraph of node $(u,v)$-MP in CaLiG is a complete bipartite graph in the worst case, and the storage cost is $O(|N_Q(u)| \times |N_G(v)|)$, where $|N_Q(u)|$ and $|N_G(v)|$ are degrees of $u \in Q$ and $v \in G$, respectively. Thus, the overall space cost is $O(|E_G|\times |E_Q| + \sum_{(u,v)\in CaLiG}|N_Q(u)| \times |N_G(v)|) 
= O(|E_Q|\times |E_G|)
$.
In the CaLiG initialization, each edge in CaLiG is visited at most once and the injective matching is computed for the incident nodes 
with the time cost $O(d_Q^{2.5})$, where $d_Q$ is the maximum vertex degree of $Q$.
Thus, the overall time complexity is  
$O(|E_Q|\times |E_G| + |E_Q|\times |E_G| \times d_Q^{2.5}
) 
= O(|E_Q|\times |E_G|  \times d_Q^{2.5})
$.
}

\section{Dynamic update of CaLiG}

Generally, it is expected to produce tight candidates, taking as little time as possible. Benefiting from CaLiG, it is easy to {
update candidates by simply exploring CaLiG from the
incident nodes of newly added or deleted edges.
The core idea is that edge addition or deletion may cause the state alteration of some nodes, which would further propagate over CaLiG. To resolve the updates, 
we present how to address edge deletion (in Section~\ref{subsec:deletion}) and edge addition (in Section~\ref{subsec:addition}), respectively.
}

\subsection{Edge Deletion}\label{subsec:deletion}
Deleting an edge from the data graph $G$ may make some candidates fail to match the query vertices (turning off some matching pairs in CaLiG), decreasing the subgraph matches. 
{
When one edge is deleted from $G$, we first delete all the related edges from CaLiG and adopt the OFF-Propagation to refine the candidates by updating the lighting states.
}

\begin{algorithm}[b]
  \caption{UpdateCaLiGForDel($CaLiG$, $e(v_1,v_2)$)}
  \label{alg:del}
  \KwIn{$CaLiG$ and an updated edge $e(v_1,v_2)$ to delete}  
  \For{each $e(u_1,u_2) \in E_Q$}{
    \If{$L(u_1) = L(v_1)$ and $L(u_2) = L(v_2)$}{
      delete edges between $(u_1,v_1)$-MP and $(u_2,v_2)$-MP from CaLiG\;
      \If{$(u_1,v_1)$-MP.state $= ON$ and $BI(u_1,v_1)$ has no injective matching}{
        $(u_1,v_1)$-MP.state $\gets OFF$\;
        OFF-Propagation($CaLiG$, $(u_1,v_1)$-MP)\;
      }
      \If{$(u_2,v_2)$-MP.state $= ON$ and $BI(u_2,v_2)$ has no injective matching}{
        $(u_2,v_2)$-MP.state $\gets OFF$\;
        OFF-Propagation($CaLiG$, $(u_2,v_2)$-MP)\;
      }
    }
  }
\end{algorithm}

To be specific, let $e(v_1, v_2)$ denote the data edge (the edge in data graph $G$) to be deleted.
Since $e(v_1, v_2)$ may be a candidate of any query edge (the edge in query graph $Q$) $e(u_1, u_2)$  with the same label, it will result in edge deletions in CaLiG. The process is similar to CaLiG initialization and resorts to the OFF-Propagation (Algorithm~\ref{alg:off}). The details of handling edge deletions are outlined in Algorithm~\ref{alg:del}. First, it deletes the edges between $(u_1,v_1)$-MP and $(u_2,v_2)$-MP from CaLiG (lines 2-3). Then we check the two affected nodes $(u_1,v_1)$-MP and $(u_2,v_2)$-MP. If they are turned off due to the deleted edge $e(v_1, v_2)$, the OFF propagation will be invoked to further refine other nodes.

\begin{example}

Take Figure~\ref{fig:sm} as an example, where the edge $(v_4, v_6)$ is deleted.
As shown in Figure~\ref{fig:del1}, we delete (1) the edges between $(u_1,v_4)$-MP and $(u_3,v_6)$-MP; and (2) the edge from $(u_3,v_4)$-MP to $(u_1,v_6)$-MP. 
The nodes $(u_1,v_4)$-MP and $(u_3,v_6)$-MP are originally ON, so we perform bigraph checking separately. As shown in Figure \autoref{fig:del2}, neither of the nodes has an injective matching in the bigraph due to edge deletion, so they will be turned off and trigger the OFF-Propagation. Finally, all nodes will eventually be turned off, as shown in Figure \autoref{fig:del3}, indicating that there is no match for the query.
\end{example}

\noindent \textbf{Complexity Analysis}. Let $D$ represent the set of nodes turned off because of updating. For each node
in $D$,
{
we compute the injective matching for the bigraph $BI(u,v)$
and propagate the updating processing to its neighbors.
} Therefore, the time complexity of updating CaLiG for deleting an edge is $O(\sum_{(u,v)\in D}(|N_{CaLiG}(u,v)| +  |N_Q(u)|^{2.5}))$, where $N_{CaLiG}(u,v)$ is the set of neighbors of ($u,v$)-MP in CaLiG.
It seems that the time cost is a little bit high, but it is sensible considering
the benefit of reducing cumbersome backtrackings, 
that is, updating CaLiG is cost-effective.

\subsection{Edge Addition}
\label{subsec:addition}

\begin{figure*} [t]
    \subfigure[Step 1. Add an edge from $(u_1,v_6)$-MP to $(u_2,v_2)$-MP and compute the state of $(u_2,v_2)$-MP.]{
         \centering
         \includegraphics[width=0.45\linewidth]{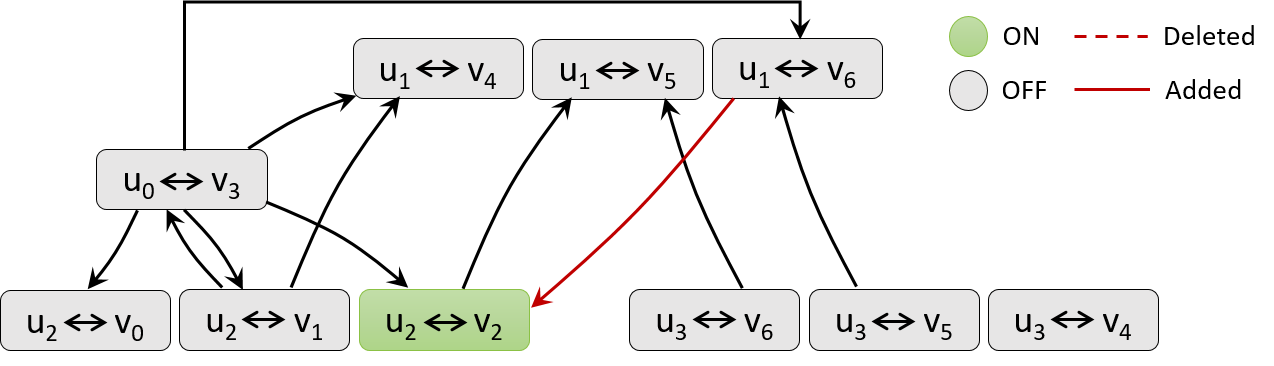}
         \label{fig:add1}
         }
    \subfigure[Step 2. Perform ON-Propagation.]{
         \centering
         \includegraphics[width=0.45\linewidth]{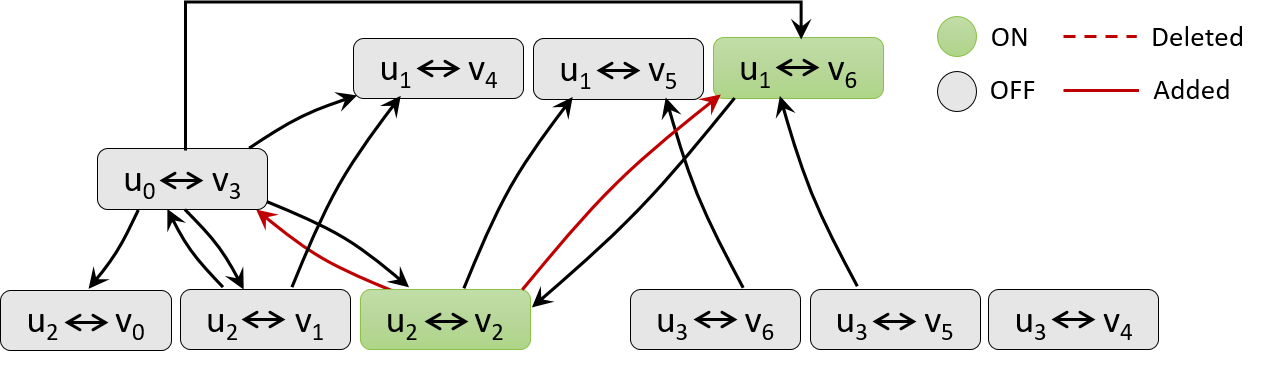}
         \label{fig:add2}
         }
    \subfigure[Step 3. Proceed ON-Propagation.]{
         \centering
         \includegraphics[width=0.45\linewidth]{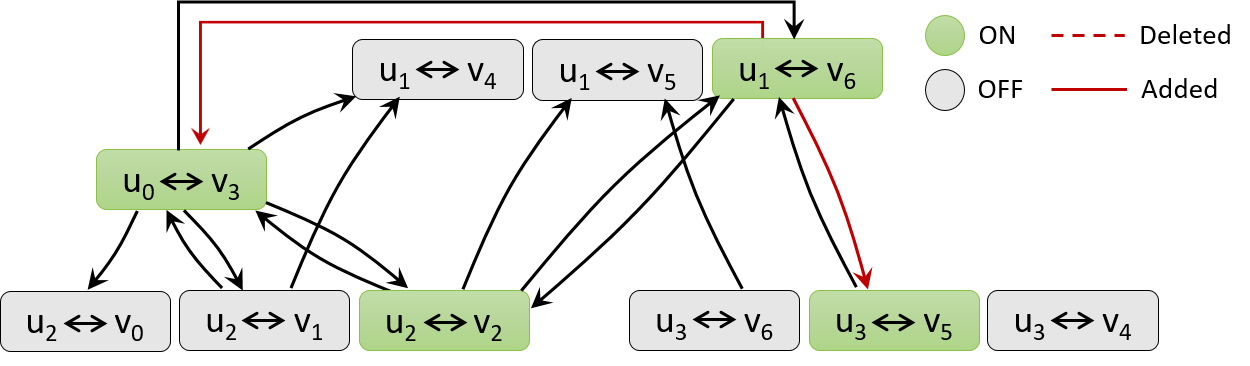}
         \label{fig:add3}
         }
    \subfigure[Step 4. ON-Propagation stops at $(u_2,v_1)$-MP.]{
        \centering
        \includegraphics[width=0.45\linewidth]{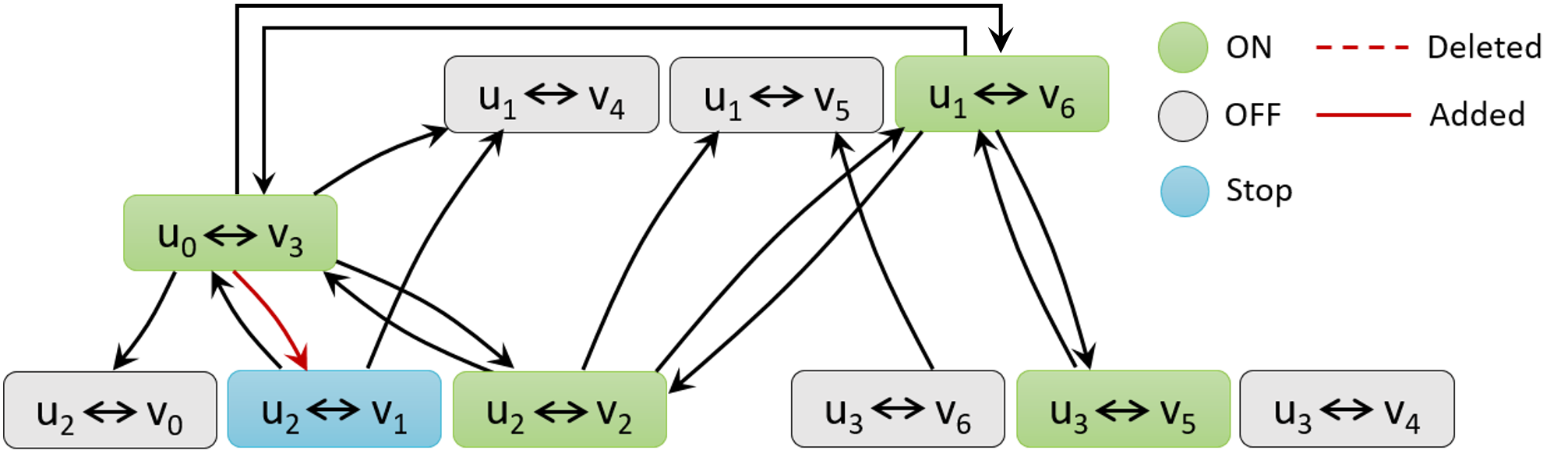}
        \label{fig:add4}
        }
        \vspace{-0.05cm}
    \subfigure[Step 5. Perform OFF-Propagation.]{
        \centering
        \includegraphics[width=0.45\linewidth]{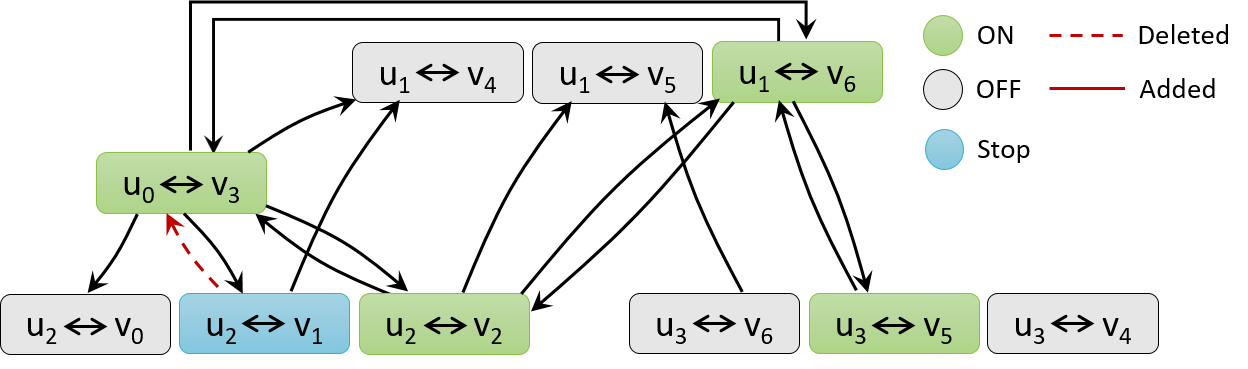}
        \label{fig:add5}
        }
        \vspace{-0.05cm}
    \subfigure[Step 6. Updated CaLiG after OFF-Propagation.]{
        \centering
        \includegraphics[width=0.45\linewidth]{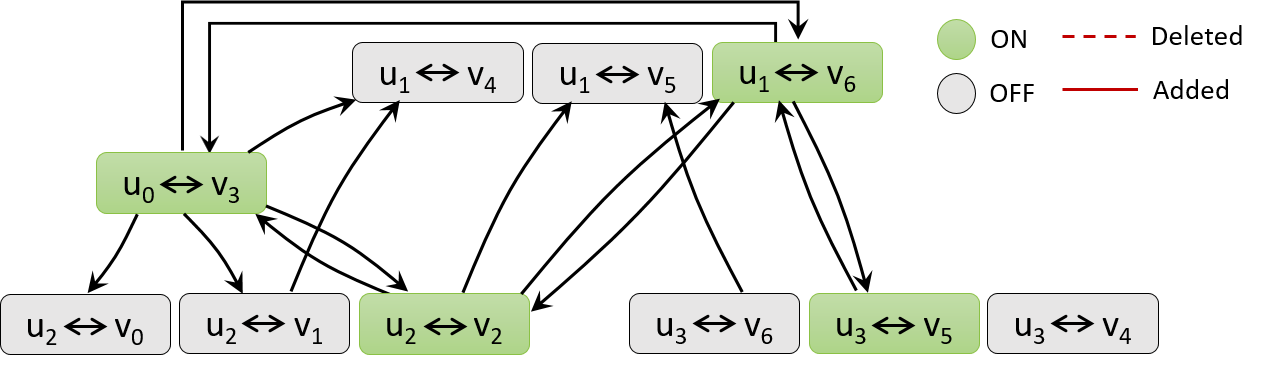}
        \label{fig:add6}
        }
    \caption{CaLiG updates for adding the edge $(v_2, v_6)$.}
    \label{fig:add}
\end{figure*}

Adding an edge to the data graph brings opportunities for turning on the OFF-state matching pairs. Based on CaLiG, it is
not difficult
to deliver the new candidates.
{
Intuitively, the updating process for edge addition
is similar to that for edge deletion.
We first add all the related edges to GaLiG,
and detect new candidates by ON-Propagation.
Before presenting the details of addressing edge addition, two proprieties regarding CaLiG are introduced as follows.
}

\begin{lemma} \label{lem:1}
For any node $(u,v)$-MP in CaLiG, all of its in-neighbors are either ON or turned OFF after $(u,v)$-MP.
\end{lemma}
\begin{proof}
\vspace{-0.15cm}
Assume ($u',v'$)-MP is an in-neighbor of $(u,v)$-MP and it is turned off before $(u,v)$-MP (i.e., $(u,v)$-MP is ON when $(u',v')$-MP has been turned off).
By the OFF-Propagation, the edge from $(u',v')$-MP to $(u,v)$-MP will be deleted once $(u',v')$-MP is turned off. Thus $(u',v')$-MP cannot
be the in-neighbor of $(u,v)$-MP.
\end{proof}

As we mentioned in Section~\ref{subsec:CaLiG}, the lighting states of all nodes are initialized to ON, and then gets updated by OFF-Propagation in the offline phase. However, this propagation process is irreversible, that is to say, if the turning off of $(u_1,v_1)$-MP leads to the turning off of $(u_2,v_2)$-MP, the turning on of $(u_2,v_2)$-MP because of edge addition will not turn $(u_1,v_1)$-MP on since $(u_1,v_1)$-MP has been turned off when $(u_2,v_2)$-MP is still ON.

\begin{lemma} \label{lem:2}
If there is an edge from the OFF-state $(u,v)$-MP to the OFF-state $(u',v')$-MP, the turning on of $(u,v)$-MP will not cause the turning on of $(u',v')$-MP.
\end{lemma}
\vspace{-0.3cm}
\begin{proof}
Assuming turning on $(u,v)$-MP causes  turning on $(u',v')$-MP, $(u',v')$-MP would not be turned off before $(u,v)$-MP
because $(u,v)$-MP was ON when computing its lighting state of $(u',v')$-MP, contradicting  that $(u',v')$-MP was turned off before $(u,v)$-MP.
\end{proof}

\begin{algorithm}[b]
  \caption{UpdateCaLiGForAdd($CaLiG$, $e(v_1,v_2)$)}
  \label{alg:add}
  \KwIn{$CaLiG$ and an updated edge $e(v_1,v_2)$ to add}  
  \For{each $e(u_1,u_2) \in E_Q$}{
    \If{$L(v_1) = L(u_1)$ and $L(v_2) = L(u_2)$}{ \label{alg:Add:if}
      add an edge from $(u_2,v_2)$-MP to $(u_1,v_1)$-MP\;
      \If{$(u_1,v_1)$-MP.state $= OFF$ and $BI(u_1,v_1)$ has an injective matching}{
        $(u_1,v_1)$-MP.$state$ $\gets ON$\;   \label{alg:Add:if end}
      }
      \If{$(u_1,v_1)$-MP.state $= ON$}{
        $StopSet$ \hspace{-0.045in} $\gets$ \hspace{-0.05in} ON-Propagation$(CaLiG, (u_1,v_1)$-MP)\;
        \While{$StopSet$ is not empty}{
          $(u,v)$-MP $\gets StopSet$.pop()\;
          OFF-Propagation($CaLiG$, $(u,v)$-MP)\;
        }
      }
    }
  }
\end{algorithm}

Let $e(v,v')$ denote the edge to add. According to Lemma~\ref{lem:1} and Lemma~\ref{lem:2}, to determine whether turning on matching pairs $(u,v)$-MP or $(u',v')$-MP, where $L(u)=L(v)$ and $L(u') = L(v')$, we only need to check its in-neighbors in CaLiG rather than considering all the vertex pairs \{$(u_i,v_j)| u_i \in N_Q(u) \wedge v_j \in N_G(v) \wedge  L(u_i) = L(v_j)$\} or \{$(u_i,v_j)| u_i \in N_Q(u') \wedge v_j \in N_G(v') \wedge  L(u_i) = L(v_j)$\}. The nodes that are not in-neighbors of $(u,v)$-MP or $(u',v')$-MP in CaLiG were turned off before $(u,v)$-MP or $(u',v')$-MP. Likewise, they will not be turned on
even if $(u,v)$-MP or $(u',v')$-MP is turned on.

\begin{algorithm}[htb]
  \caption{ON-Propagation($CaLiG$, $(u,v)$-MP)}
  \label{alg:on}
  \KwIn{$CaLiG$ and 
  a matching pair $(u,v)$-MP that was turned on in the previous round}  
  \KwOut{A set of stopping nodes $S$}
  $S \gets \emptyset$\;
  \For{each $(u',v')$-MP $\in In_{CaLiG}(u,v)$}{
      add an edge from $(u,v)$-MP to $(u',v')$-MP\;
      \If{$(u',v')$-MP.state $= OFF$}{
        \If{$BI(u',v')$ has an injective matching}{
          $(u',v')$-MP.state $\gets ON$\;
         $S \gets S$ $\cup$ ON-Propagation$(CaLiG$, $(u',v')$-MP)\;
        }
        \Else{
         $S \gets$  $S$ $\cup$  $(u',v')$-MP\;
          
        }
      }
    
  }
  \Return{
            $S$\;
          }
  
\end{algorithm}

Algorithm~\ref{alg:add} depicts the process of handling edge additions based on CaLiG. It involves two procedures ON-Propagation (Algorithm~\ref{alg:on}) and OFF-Propagation. Different from handling edge deletions, it just needs to add an edge from $(u_2,v_2)$-MP to $(u_1,v_1)$-MP and computes the lighting state of $(u_1,v_1)$-MP (lines~\ref{alg:Add:if}-\ref{alg:Add:if end}), since its turning on will propagate to $(u_2,v_2)$-MP through the ON-Propagation. If the state of $(u_1,v_1)$-MP is OFF, the state of $(u_2,v_2)$-MP will not change no matter whether it is ON or OFF.
For each ON-state $(u_1,v_1)$-MP, the procedure ON-Propagation is invoked to try to turn on more nodes. Note that before ON-Propagation we add an edge from $(u_2,v_2)$-MP to $(u_1,v_1)$-MP (line 3 in Algorithm~\ref{alg:add}), which indicates that we progressively take $(u_2,v_2)$-MP as an ON-state node. Hence, some nodes may be falsely turned on due to the propagation.
When a node is not turned on by ON-Propagation, it is recorded as a stopping node. All the stopping nodes constitute the set, called $StopSet$.
The OFF-Propagation is performed for each node in $StopSet$ to ensure that all the nodes that are newly turned on conforming to subgraph isomorphism. 

Algorithm~\ref{alg:on} presents the details of ON propagation. Based on Lemma~\ref{lem:2}, 
it only needs to propagate to its in-neighbor $(u',v')$-MP. If $(u',v')$-MP is OFF and its corresponding bigraph has an injective matching, we can turn $(u',v')$-MP off and invoke the ON-Propagation recursively. Otherwise, the node will be added into $S$, the set of stopping nodes.

\begin{example}\vspace{-1mm}
Take \autoref{fig:add} as an example, where the edge $(v_2, v_6)$ is added to the data graph, as shown by the red line in
Figure~\ref{fig:um}. 
 Firstly, as shown in Figure~\ref{fig:add1}, an edge is added from $(u_1,v_6)$-MP to $(u_2,v_2)$-MP in CaLiG. Then $(u_2,v_2)$-MP is turned on.
Secondly, we will perform ON-Propagation towards $(u_2,v_2)$-MP's in-neighbors as shown in Figure~\ref{fig:add2}, where the nodes $(u_1,v_6)$-MP and $(u_0,v_3)$-MP are originally OFF. Only $(u_2,v_2)$-MP is turned on as it has an injective matching. Then the propagation continues to try to turn on its OFF-state in-neighbors. Correspondingly, $(u_0,v_3)$-MP and $(u_3,v_5)$-MP are turned on as shown in Figure~\ref{fig:add3}. 
When propagating to $(u_2,v_1)$-MP, it is not turned on and marked as a stopping node as shown in Figure~\ref{fig:add4}. Finally, the OFF-Propagation is invoked starting from the stopping node $(u_2,v_1)$-MP, but $(u_0,v_3)$-MP remains ON-state after the OFF-Propagation (Figure~\ref{fig:add5}). The final updated CaLiG index is depicted in Figure~\ref{fig:add6}.

\end{example}

\noindent \textbf{Complexity Analysis}.
Let $D$ and $A$ represent the sets of nodes turned off and turned on because of edge addition, respectively.
The time complexity for edge addition is almost the same as that for edge deletion. The difference is we need to take the procedure of turning nodes in $A$ into consideration. The overall time complexity of updating CaLiG by adding an edge is $O(\sum_{(u,v)\in (D \cup A)}|N_{CaLiG}(u,v)|+ |N_Q(u)|^{2.5})$, where $N_{CaLiG}(u,v)$ is the set of neighbors of ($u,v$)-MP.


\section{KSS-based Subgraph Matching}

{
For continuous subgraph matching, we need to find all the new matches due to edge addition or the decreased matches due to edge deletion.
It is clear that the changed matches must contain the updated edges.
In detail, for each update edge $e(v_i, v_j)$, we need to find two connected ON-state nodes $(u_k,v_i)$-MP and $(u_l, v_j)$-MP in CaLiG that contain $v_i$ and $v_j$, respectively. These two nodes will be taken as a partial match
which can be extended to a complete match.
Thus, the widely used backtracking search could be utilized.
Such a backtracking search enumerates candidates of each query vertex one by one and examines constraints for the subgraph isomorphism at each step.
It performs intensive backtracks over the candidates to construct matches by trying to integrate the candidates, which incurs expensive time costs.

Most existing algorithms accelerate the backtracking by adjusting the matching order. In contrast, we seek a more powerful backtracking framework in this section. 
It is noticeable that the matches for the degree-one query vertices are independent, 
where the ``independent'' means that computing matches for one query vertex would not affect the matches of another.
Therefore, the matches of such vertices could be acquired together instead of 
exploring them one by one.

Thus, we develop a novel backtracking search framework, called Kernel-and-Shell Search (KSS),
for continuous subgraph matching.
KSS decomposes the query vertices into kernel and shell vertices, making it straightforward to deliver the incremental matches by joining the partial matches for kernel vertices and the candidate of shell vertices.
}

\subsection{Kernel and Shell Vertex}
{
To identify the independent vertices as discussed above, KSS defines the kernel set as the connected vertex cover, and the shell set is the complementary set. 
}

\begin{myDef}[Kernel Set and Shell Set]
\label{def:ks}
Given a query graph $Q$, its kernel set, also known as connected vertex cover, is a set of connected vertices s.t. each edge in $Q$ has at least one vertex in the set. Each vertex in the kernel set is called a kernel vertex.
The shell set is the complementary set of the kernel set, where vertices are independent of each other
and each of them is called a shell vertex.
\end{myDef}

{
The shell vertices are naturally
independent by definition.
Since the query graph is fixed in CSM, 
the corresponding kernel and shell vertices could be computed and stored in advance.
}

\begin{figure} 
    \subfigure[Query graph.]{
         \centering
         \includegraphics[scale=0.4]{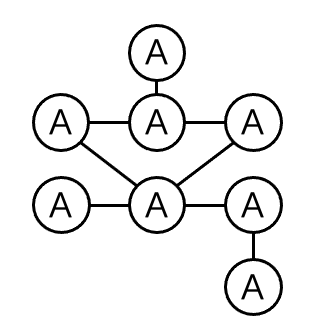}
         \label{fig:divide1}
         \vspace{-0.35cm}
         }
    \subfigure[Three kernel sets and shell sets.]{
         \centering
         \includegraphics[scale=0.4]{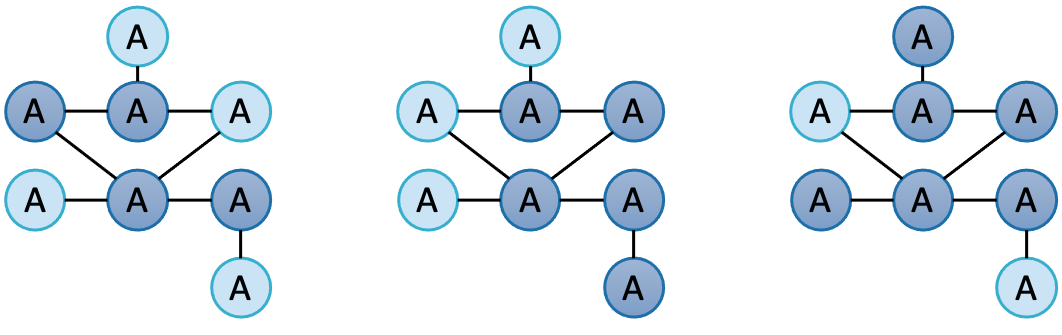}
         \label{fig:divide2}
         }
    \caption{An example of kernel vertices and shell vertices.
    }
    \label{fig:divide}
\end{figure}

\begin{example}
Figure~\autoref{fig:divide2} lists three different kernel sets and shell sets for the query graph in \autoref{fig:divide2}, where vertices in dark blue form the kernel set and the other vertices form the shell sets.

\end{example}

Given a query graph $Q$ and its kernel set and shell set, once the match of kernel set is determined, the candidates for each shell vertex are obtained by checking its adjacent matched vertex. As there is no edge dependency among shell vertices, the matches can be produced straightforwardly by joining the candidates of shell vertices, without any failing backtracks. Therefore, a good decomposition should have as many shell vertices as possible, indicating as few kernel vertices as possible. As discussed above, the desired incremental matches must contain the update edge $e(v_i, v_j)$ and the search starts from $e(v_i, v_j)$. Hence, the subgraph induced by the kernel set should contain the edge $e(u_k, u_l)$ such that the matching pairs $(u_k,v_i)$-MP and $(u_l,v_j)$-MP are ON in CaLiG.

\begin{myDef}[Conditional Kernel Set ]
\label{def:mcks}
Given a query graph $Q$ and an edge $e(u_k, u_l) \in E_Q$, the conditional kernel set, denoted by CKS, is the kernel set that contains vertices $u_k$ and $u_l$.
\end{myDef}

\begin{lemma}
\label{lem:mini kernet set}
Given a query graph $Q$, we get a new query graph $Q'$ by adding a virtual vertex $u_0$ and connecting $u_0$ with all vertices in $Q$. If $S \cup \{ u_0 \}$ is a minimum kernel set of $Q'$, $S$ is a minimum kernel set of $Q$, where $S \subseteq V_Q$. 
\end{lemma}
\vspace{-0.35cm}
\begin{proof}
Suppose $S$ is not a minimum kernel set of $V_Q$ and there is another kernel set $S'$ of $Q$ that is smaller than $S$. $S' \cup \{ u_0 \}$ will be a smaller kernel set of $Q'$, which leads to a contradiction.
\end{proof}

\begin{theorem}
Computing the minimum conditional kernel set (MCKS) for a query graph $Q$ and an edge $e(u_k, u_l) \in E_Q$ is NP-hard.
\end{theorem}\label{theorem:mcks}
\vspace{-0.35cm}
\begin{proof} 
The proof is achieved by reducing the NP-hard minimum connected vertex cover (MCVC) problem~\cite{NPC,CVC}. 

Let $Q'$ denote the graph obtained by adding a virtual vertex $u_0$ and connecting $u_0$ with all vertices in $Q$.
For each edge in $Q'$, we compute its MCKS. Let $S'$ represent the minimum MCKS among the $|E_{Q'}|$ MCKSs.  $S'$ is an MCVC (i.e., a minimum kernel set) of $Q'$, which can be proved by contradiction as follows.

Suppose $S'$ is not the MCVC of $Q'$. There must be another MCVC of $Q'$, denoted by $S$, such that $|S'| > |S|$. 
(1) When $S$ does not contain $u_0$, all the nodes $V_Q$ constitute the MCVC $S$ of $Q'$, i.e., $S = V_Q$. If $u_0$ is not contained in $S'$, $|S'| \le |V_Q|$, which contradicts $|S'| > |S|$. If $u_0$ is contained in $S'$, $S'=S^* \cup \{u_0\}$ such that $S^* \ge S$. Thus, we have $S'$=$V_Q \cup \{u_0\}$. As $S = V_Q$, $S$ is the MCKS for $Q'$ and an edge $e(u_x, u_y)$, where $e(u_x, u_y) \in E_{Q'}$. It contradicts that $S'$ is the smaller MCKS for $Q'$.
(2) When $S$ contains $u_0$, at least one neighbor $u_i$ of $u_0$ will be contained in $S$ as the kernel set is connected. $S$ is the MCKS for $Q'$ and the edge $e(u_0,u_i)$, which contradicts that $S'$ is the minimum one among all the possible edges.

If $u_0 \in S'$, $S'\setminus u_0$ is a minimum kernel set of $Q$ according to Lemma~\ref{lem:mini kernet set}; Otherwise, $S'$ is a minimum kernel set of $Q$. It is clear that the procedure of constructing the solution to MCVC is in polynomial time. Thus, if MCKS can be solved in polynomial time, the MCVC problem can be solved in polynomial time, which contradicts the NP-hardness of the MCVC problem.
\vspace{-0.3cm}
\end{proof}

\begin{algorithm}[t]
  \caption{FindMatches($CaLiG$, $e(v_1,v_2)$, $Kernel$, $Shell$)}
  \label{alg:m}
  \KwIn{$CaLiG$, an updated edge $e(v_1,v_2)$, the conditional kernel set $K$, and shell set $S$ for $e(v_1,v_2)$}
  \KwOut{Incremental matches due to $e(v_1,v_2)$}
  \For{each $u_1 \in Q$}{
    \If{$L(v_1) = L(u_1)$ and $(u_1,v_1)$-MP.state $= ON$}{
      $m[u_1] \gets v_1$\;
      \For{each $u_2 \in N_Q(u_1)$}{
        \If{$(u_2,v_2)$-MP $\in In_{CaLiG}(u_1,v_1)$}{
          $m[u_2] \gets v_2$\;
         \Return{KSS($m$, $K$, $S$).}
        }
      }
    }
  }
\end{algorithm}

Given the MCKS is NP-hard to compute,
a greedy algorithm is proposed in practice.
It adds  $u_k$ and $u_l$ into CKS and removes $u_k$, $u_l$, and their neighbors from $Q$. Then it enumerates a maximal collection of disjoint odd cycles in the remaining $Q'$. 
Let $t$ denote the number of odd cycles, $C_i$ denote the vertices of $i$-th cycle, and $C=\bigcup_i^t C_i$.
It computes the optimal vertex cover of the remaining bipartite graph $Q' \setminus C$, denoted as $W$.
If $C \cup W \cup \{e(u_k,u_l)\}$ is connected, return $C \cup W \cup \{e(u_k,u_l)\}$; otherwise, we apply approximation algorithms for Steiner tree to obtain one connected result.

\subsection{Kernel-and-Shell Search}
{
Powered by kernel and shell vertices, we develop a novel 
backtracking search framework.
When an edge is added or deleted, 
our method would first initialize partial matches based on the updated edge (Algorithm~\ref{alg:m}), and then invoke the kernel and shell search (Algorithm~\ref{alg:kernel}).
For kernel vertices, 
KSS computes the partial matches through the existing backtracking.
The incremental matches can be reported immediately by joining the partial matches and the candidates of shell vertices, without any unnecessary backtrackings.

\setlength{\textfloatsep}{14pt}%
\begin{algorithm}[t]
  \caption{\emph{KSS}($m$, $K$, $S$)}
  \label{alg:kernel}
  \KwIn{The partial match $m$, conditional kernel set $K$, and shell set $S$}
   \KwOut{Incremental matches due to $e(v_1,v_2)$}
  \If{$m.size < |K|$}{
    $th \gets m.size$\;
  $u \gets K[th]$\;
  $Cand(u) \gets$ generate $u$'s candidates\;
  \For{each $v \in Cand(u)$}{
    $m' \gets m$\;
    $m'[u] \gets v$\;
    \emph{KSS}$(m', K, S)$\;
    }
  }
  \Else{
    \For{each $u \in S$}{
        $Cand(u) \gets$ generate $u$'s candidates\;
    }
    \Return{$m \bowtie_{u\in S} Cand(u)$.}
  }
\end{algorithm}

Algorithm~\ref{alg:m} presents the process of finding incremental matches based on kernel-and-shell search (KSS). Given the update edge $(v_1,v_2)$, 
as a matching pair $(u_1,v_1)$-MP must be ON (lines 2-3).
The matching pair $(u_2,v_2)$-MP is then determined from the in-neighbors of $(u_1,v_1)$-MP (lines 4-6). Thus, we have found partial match $\{(u_1 \leftrightarrow  v_1, u_2 \leftrightarrow v_2)\}$, denoted as $m$. 
Next, we match the remaining query vertices in the order of kernel vertex first and then shell vertex by invoking the procedure \emph{KSS}.

Algorithm~\ref{alg:kernel} describes how to search matches for conditional kernel and shell vertices. 
First, it determines the query vertex to match according to the kernel set (lines 2-3), and then generates candidates for the vertex. Finally, it selects a candidate as the partial match to drive the matching process (lines 5-8).
After all the kernel vertices have been matched, we generate candidates for all shell vertices and join the candidates with the partial match to report the incremental matches (lines 10-12). 
The process is very efficient as we produce the incremental matches by a simple join operation without checking any constraints or backtracks. 

\noindent \textbf{Candidate generation.} 
Let $Cand(u)$ denote the candidates of query vertex $u$. The vertex $v$ in $Cand(u)$
meets the  constraints:
\begin{enumerate}[1)]\vspace{-0.015in}
    \item $v$ has not been used in the partial match.
    \item The node $(u,v)$-MP in CaLiG is ON.
    \item $v$ is a neighbor of $v'$ if $v' \in Cand(u')$ and $u' \in N_Q(u)$.
    \vspace{-0.01in}
\end{enumerate}
To generate $Cand(u)$ for  vertex 
$u$, two steps are conducted.

\textbf{Step 1.} For each $u_i \in N_Q(u)$ that been matched to $v_i$, we build a set $C_i = \{v'| (u', v')$-MP $\in In_{CaLiG}(u_i, v_i)$ and $(u', v')$-MP is ON$\}$. Then we intersect all of these sets $C_i$ to get a candidate set for $u$.

\textbf{Step 2.} We remove the data vertices that have already been used in the partial matching from the candidate set. After this step, we obtain the final candidate set of $u$.
}

\noindent \textbf{Pruning in advance.}
KSS computes the partial matches for kernel vertices first, which facilitates match generation while incurring new problems. For example, when we determine a partial match of all kernel vertices, the candidate set of a shell vertex may be empty, but we can only find this failed partial match when generating candidates for that shell vertex. Hence, we employ a pruning strategy that detects such failures as early as possible.
For each shell vertex $u$, once all its neighbors have been matched, we try to find a candidate for $u$ in advance. 
If no candidate vertex is found for $u$, it is safe to rule out the current partial match. Otherwise, it proceeds to find matches of the remaining kernel vertices.


 \noindent \textbf{Complexity Analysis}.
For KSS,  only the kernel part is matched by backtracking search, the worst-case complexity is induced to $O(|V_G|^{|K|})$, where $|K|$ is the size of kernel vertices, much smaller than traditional backtracking with the time complexity $O(|V_G|^{|V_Q|})$.

\section{Experimental Evaluation}

In this section, we evaluate our proposed method, denoted by CaLiG, and compare it with  {
two state-of-the-art algorithms  TurboFlux~\cite{turboflux} and SymBi~\cite{symbi}}.

\vspace{-0.25cm}
\subsection{Experimental Settings}

\textbf{Data graphs.}  
Table~\ref{tab:dataset} lists the graphs used in experiments that are downloaded from SNAP~\cite{snapnets}, except Netflow (downloaded from CAIDA~\cite{netflow}).
We randomly sample 5\% data edges as the streaming edges, with the ratio of deleted edges over added edges being 2:1.

\begin{table}[tbp]
  \centering
  \caption{Data graphs.}
  \setlength{\tabcolsep}{2.3mm}
  {
    \begin{tabular}{ccrrc}
    \toprule
    \textbf{Dataset} & \textbf{Abb.} & \multicolumn{1}{c}{\textbf{|V|}} & \multicolumn{1}{c}{\textbf{|E|}} & \textbf{Average degree} \\
    \midrule
    Lastfm & lfm   & 7,624  & 27,806  & 7.3  \\
     Facebook & fbm   & 22,470  & 170,823  & 15.2  \\
    Email & em    & 36,692  & 183,831  & 10.0  \\
    Github & gh    & 37,700  & 289,003  & 15.3  \\
    Deezer & dz    & 41,773  & 125,826  & 6.0  \\
    Twitch & tw    & 168,114  & 6,797,557  & 80.9  \\
    Skitter & sk    & 1,696,415  & 11,095,298  & 13.1  \\
    Netflow & nf    & 3,114,895  & 16,668,683  & 10.7  \\
    \bottomrule
    \end{tabular}
    }
  \label{tab:dataset}
\end{table}%
\noindent \textbf{Query graphs.} 
For each data graph, we sample 7 groups of subgraphs, denoted by $Q3$, $Q4$, $Q6$, $Q8$, $Q10$, $Q12$, $Q14$, and $Q16$, as query graphs by varying the number of vertices from 3 to 16. Each group $Qi$ contains 50 query graphs.

\noindent \textbf{Metrics.} 
The elapsed time is reported in milliseconds.
Due to the NP-hardness of subgraph matching, some queries may take an extremely long time. 
By convention, we set a timeout of 20 minutes for each query.
If one query cannot finish within the time limit, it is called \emph{uncompleted}. To evaluate an algorithm generally, we report the average elapsed time, as well as the completion rate, the average peak memory usage, and the 
match density.

All the algorithms are implemented in C++ and evaluated
 on a Linux Server equipped with Intel(R) Xeon(R) CPU E5-2640 @ 2.60GHz and 128G RAM.

\subsection{Experimental Results}

\noindent \textbf{Overall Performance.}
Table~\ref{tab:total} reports the average elapsed time of {400} query graphs on 8 data graphs respectively.
Table~\ref{tab:total}. As is observed, 
It shows that CaLiG runs much faster than   {
TurboFlux and SymBi in all graphs, achieving significant speedups over TurboFlux (from 22.10x to 5963.07x) and SymBi (from 18.05x to 978.69x) respectively.}
The last column $|\Delta m|$ represents the number of incremental matches. We can find that $|\Delta m|$ ranges from millions to more than 1 billion. The larger the number of results, the more time it takes.

\begin{table}[t]
  \centering
  \caption{Average elapsed time.}
    \begin{tabular}{crrrr}
    \toprule
          & \multicolumn{3}{c}{\textbf{Elapsed Time (ms)}} &  \\
         \cline{2-4}
    \textbf{Dataset} & \multicolumn{1}{c}{\textbf{TurboFlux}} & \multicolumn{1}{c}{\textbf{SymBi}} & \multicolumn{1}{c}{\textbf{CaLiG}} &
    \multicolumn{1}{c}{\textbf{$|\Delta m|$}} \\
    \midrule
    {Lastfm} & 49,341.9 & 13,848.3  & \textbf{641.6}    &7,626,438  \\
    {Facebook}   &   2,510,220.9  & 1,745,489.0  & \textbf{94,174.2}   & 443,817,336  \\
    {Email} &  2,836,668.9 & 1,873,555.0  & \textbf{82,417.3}   & 338,293,270  \\
    {Github} &  4,177,493.4  & 3,411,865.8  & \textbf{189,060.7}    & 1,679,993,941  \\
    {Deezer} & 368,280.5  & 99,535.6  & \textbf{4,669.0}    & 14,327,860  \\
    {Twitch} &  1,957,440.8  & 1,107,821.3  & \textbf{51,247.5}   & 464,933,035  \\
    {Skitter} &  1,460,160.6 & 321,649.7  & \textbf{9,447.0}    & 130,260,509  \\
    {Netflow} & 727,494.2 & 119,694.9  & \textbf{122.3}   & 48,801  \\
    \bottomrule
    \end{tabular}
  \label{tab:total}
\end{table}

\begin{figure}[t] 
    \subfigure[Lastfm.]{
         \centering
         \includegraphics[width=0.25\linewidth]{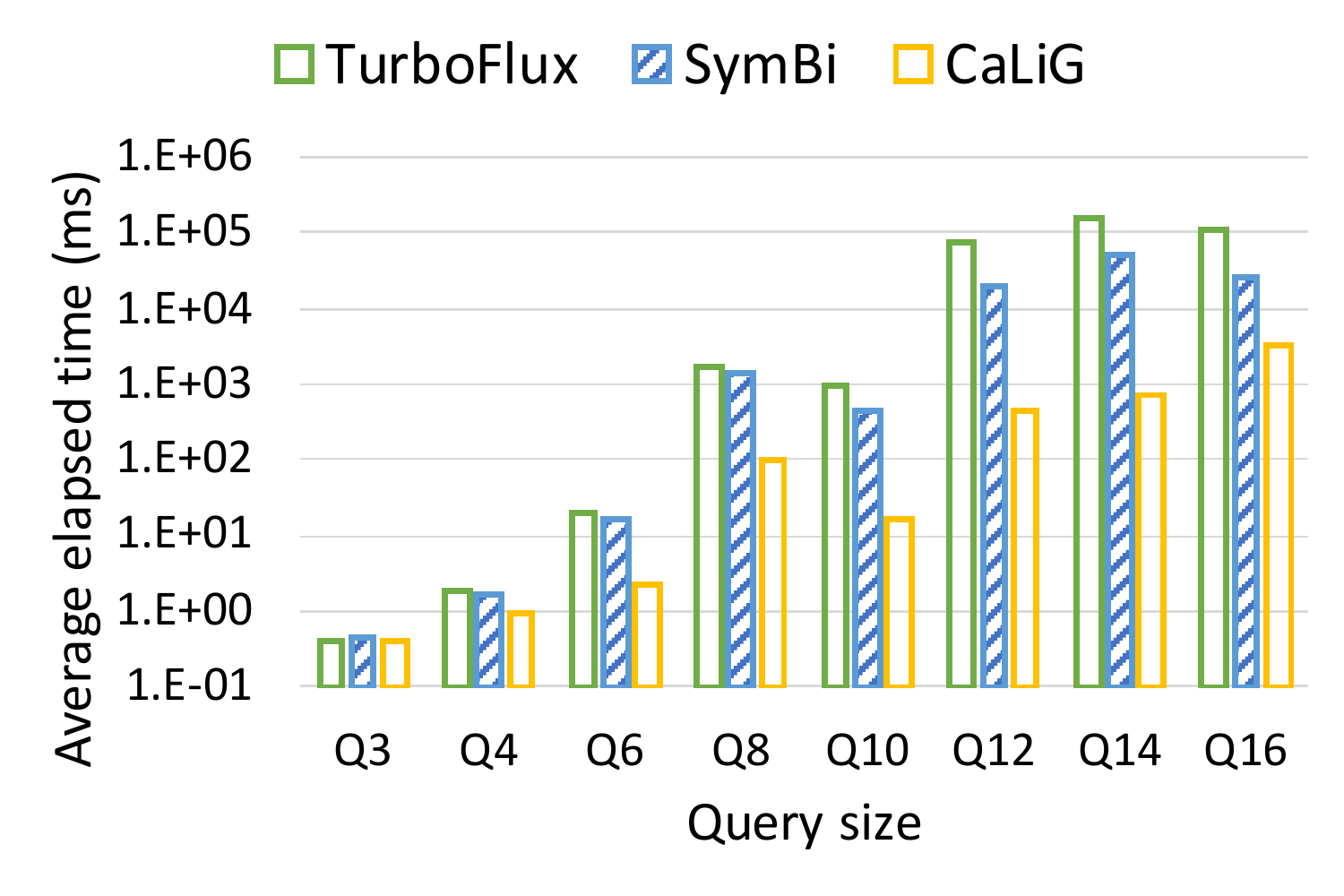}
         \label{fig:1_1}
         }
         \hspace{-0.5cm}
         \vspace{-0.05cm}
     \subfigure[Lastfm.]{
         \centering
         \includegraphics[width=0.25\linewidth]{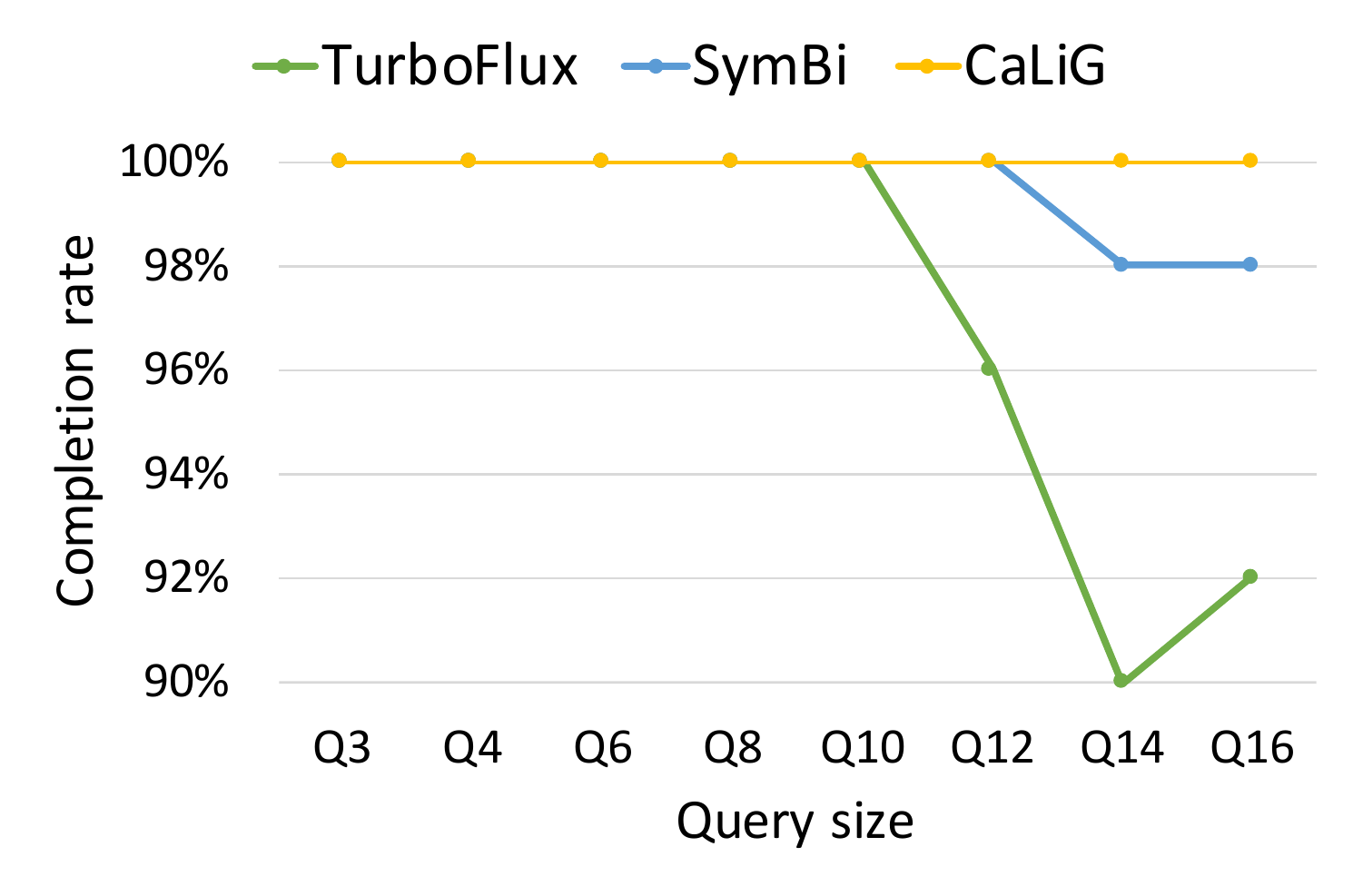}
         \label{fig:1_2}
         }
         \hspace{-0.5cm}
         \vspace{-0.05cm}
    \subfigure[Facebook.]{
         \centering
         \includegraphics[width=0.25\linewidth]{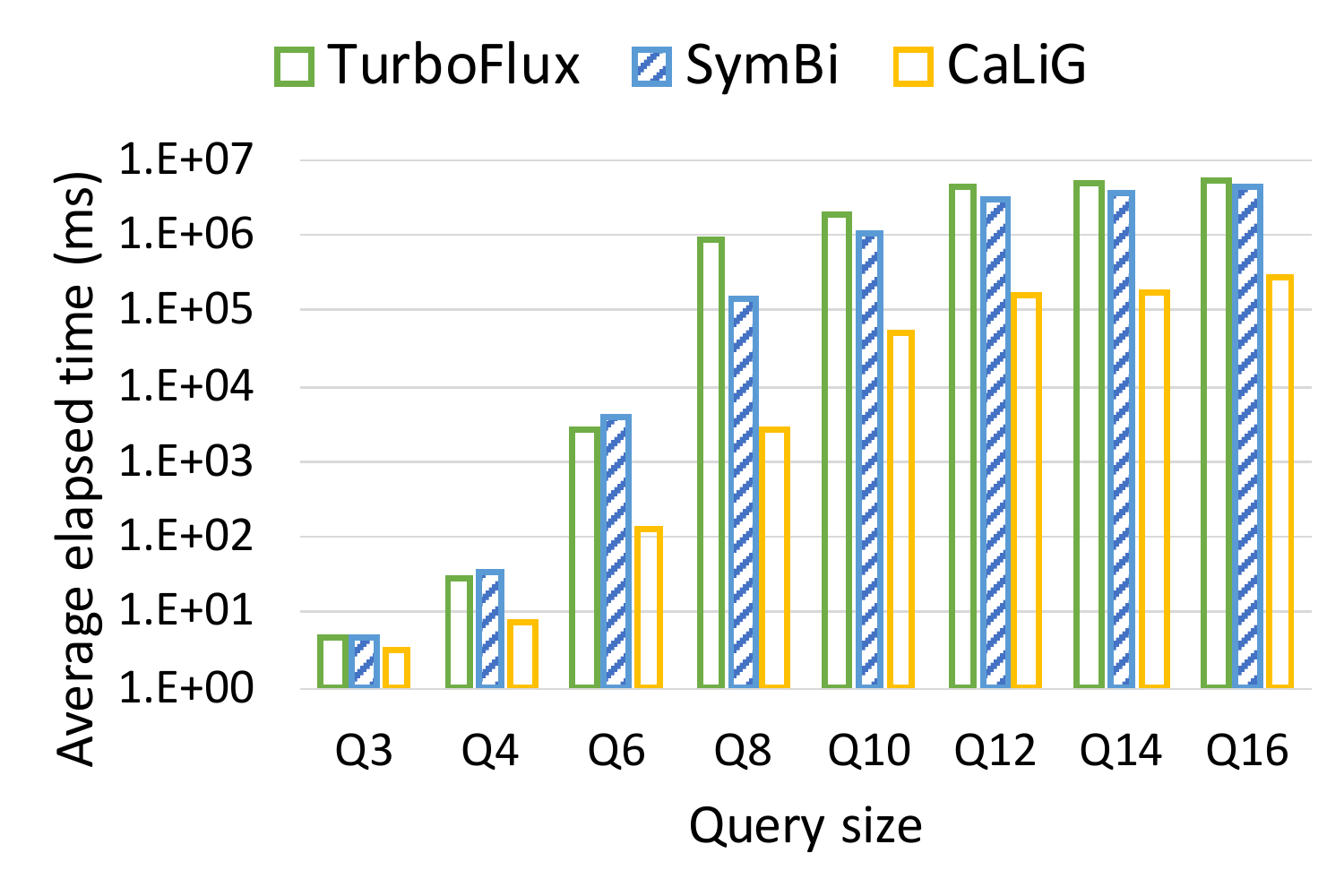}
         \label{fig:1_3}
         }
         \hspace{-0.5cm}
         \vspace{-0.05cm}
    \subfigure[Facebook.]{
         \centering
         \includegraphics[width=0.25\linewidth]{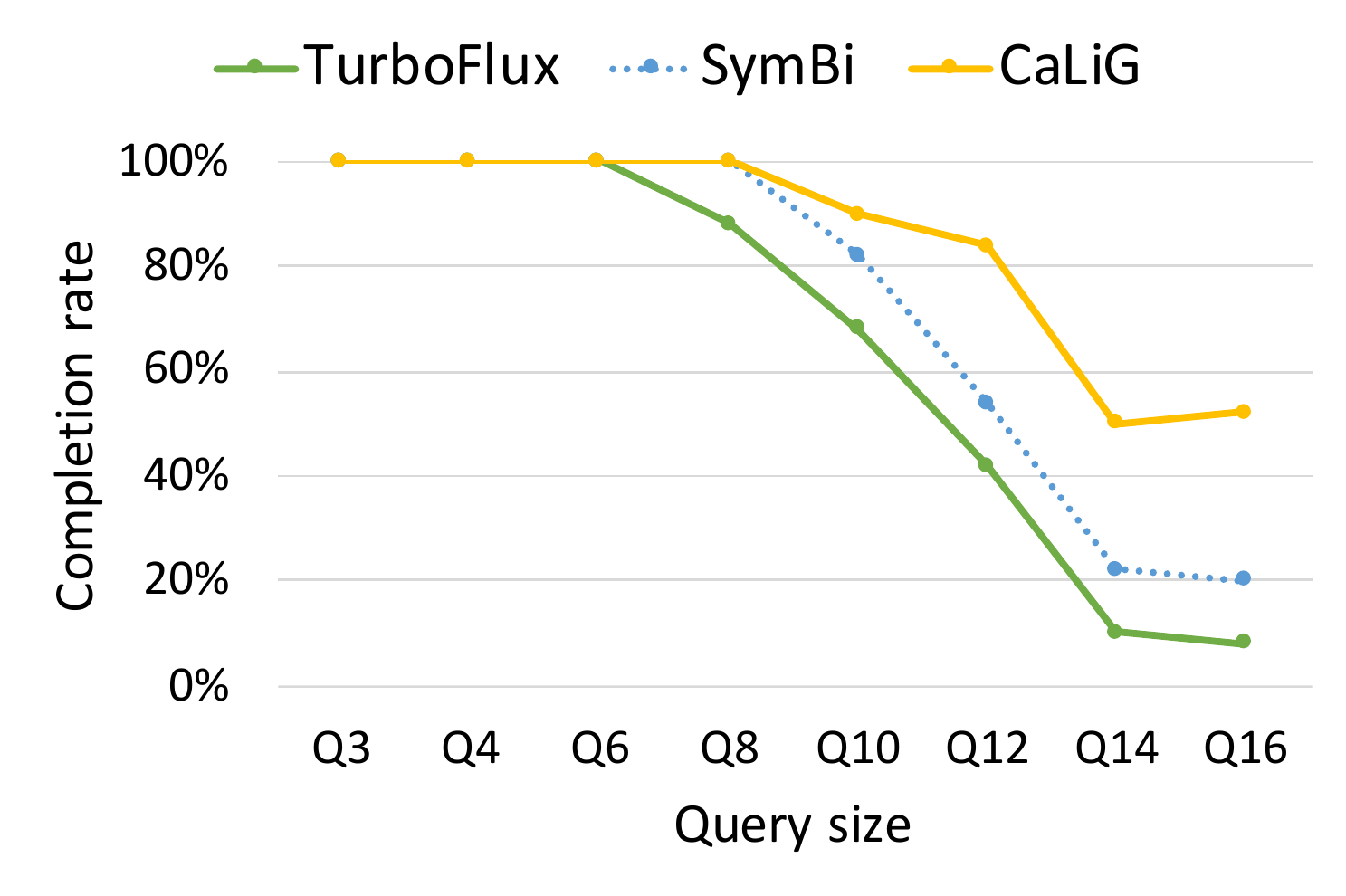}
         \label{fig:1_4}
         }
         \hspace{-0.5cm}
         \vspace{-0.05cm}
    \subfigure[Email.]{
         \centering
         \includegraphics[width=0.25\linewidth]{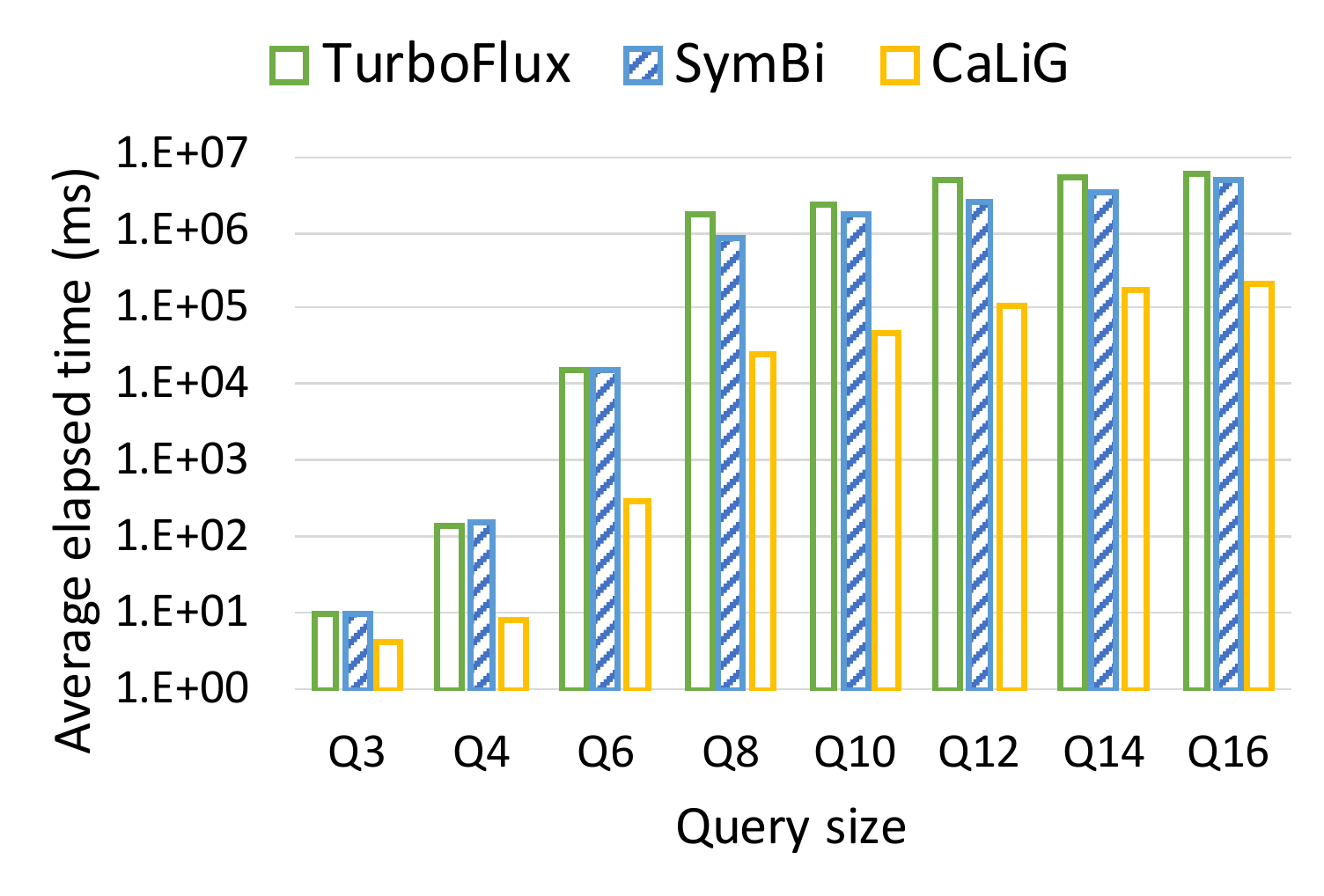}
         \label{fig:1_5}
         }
         \hspace{-0.5cm}
         \vspace{-0.05cm}
    \subfigure[Email.]{
         \centering
         \includegraphics[width=0.25\linewidth]{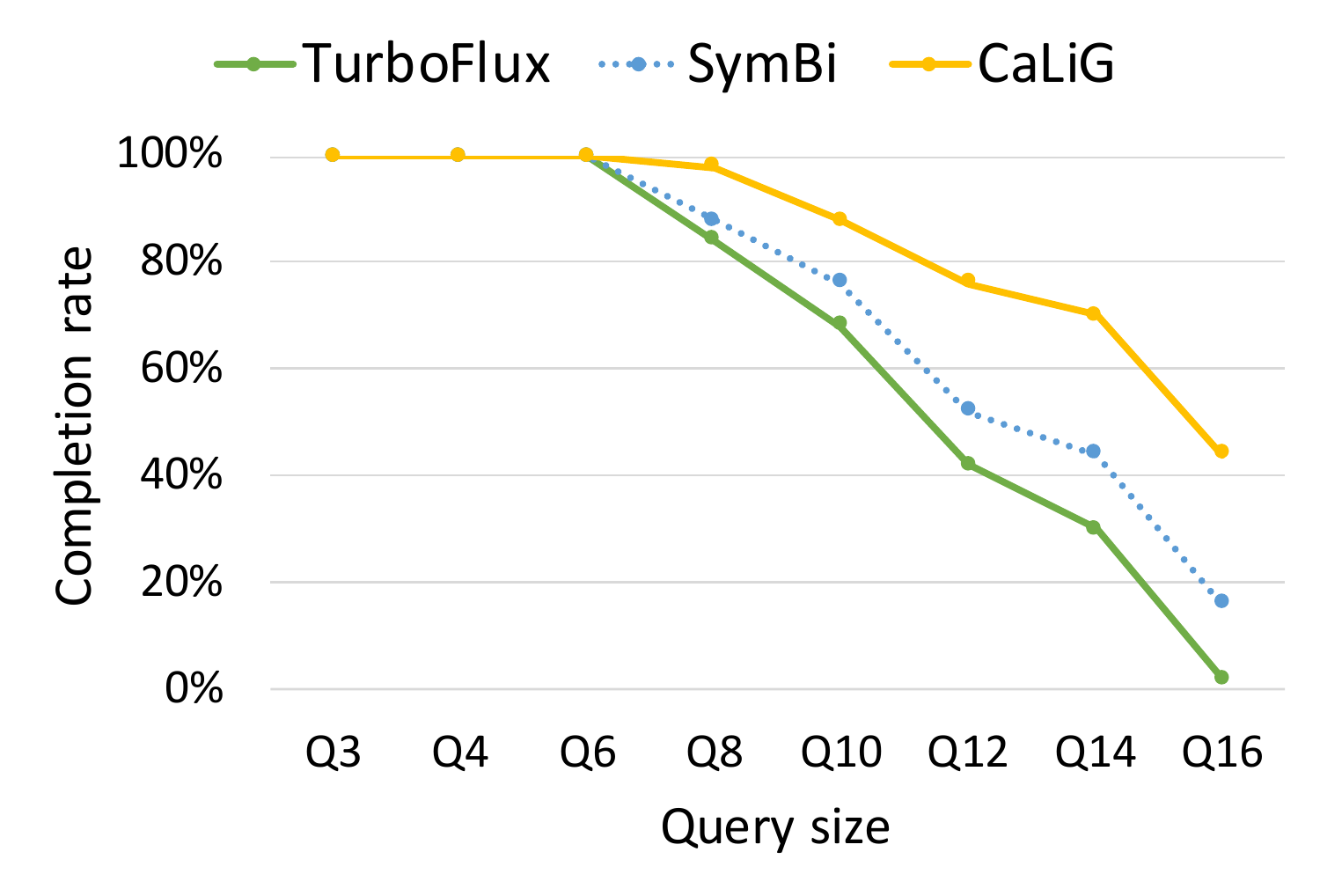}
         \label{fig:1_6}
         }
         \hspace{-0.5cm}
         \vspace{-0.05cm}
    \subfigure[Github.]{
         \centering
         \includegraphics[width=0.25\linewidth]{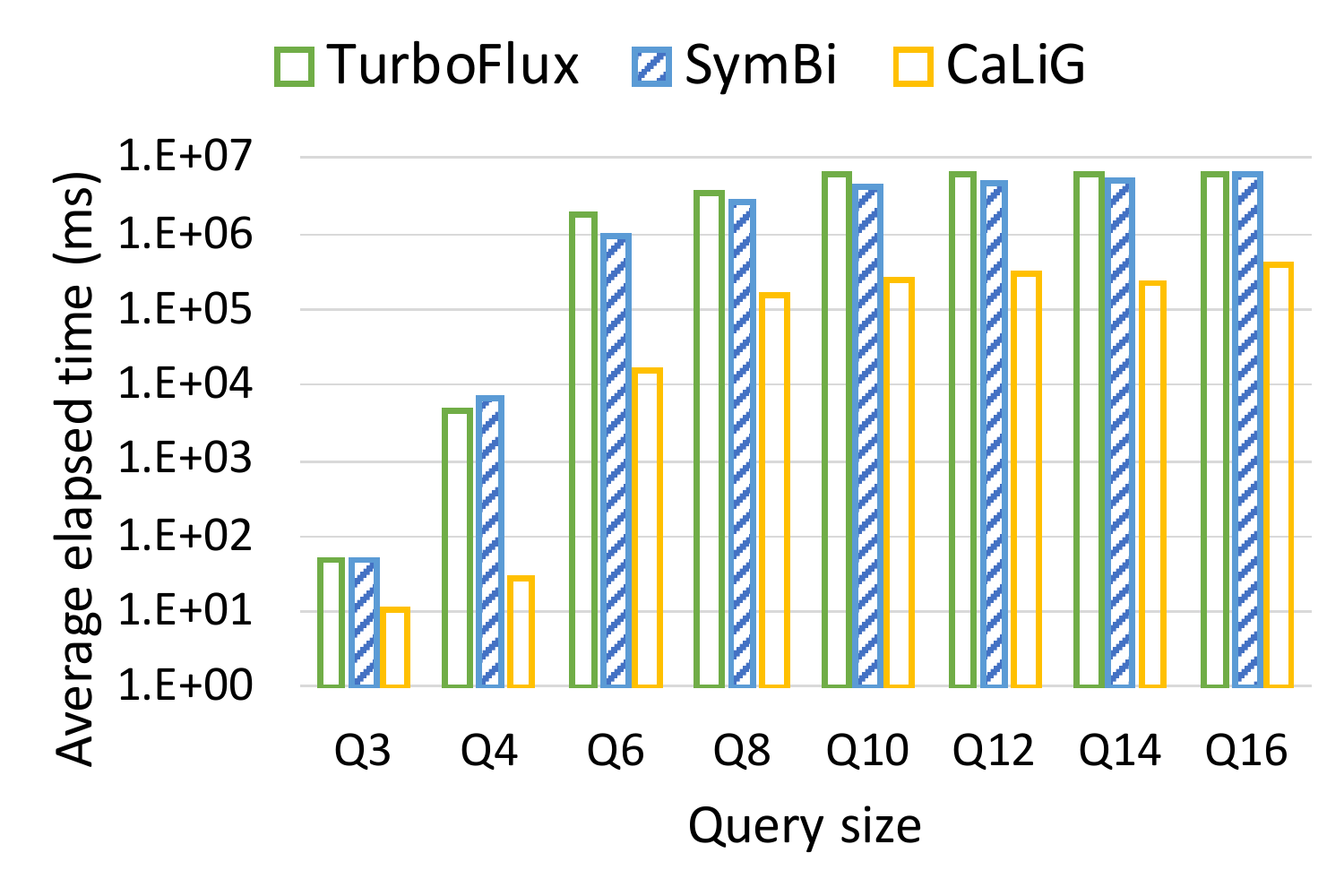}
         \label{fig:1_7}
         }
         \hspace{-0.5cm}
         \vspace{-0.05cm}
     \subfigure[Github.]{
         \centering
         \includegraphics[width=0.25\linewidth]{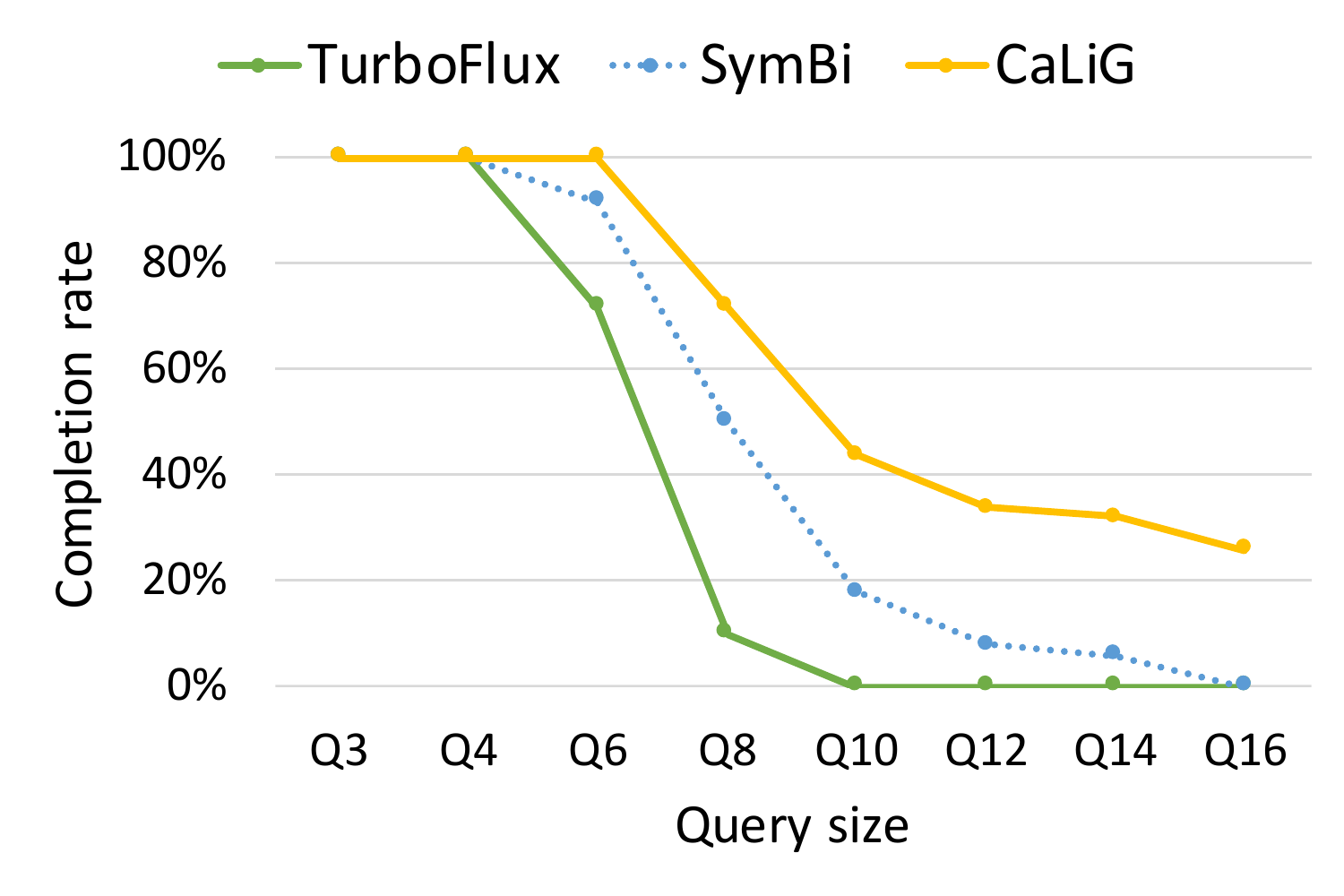}
         \label{fig:1_8}
         }
         \hspace{-0.5cm}
         \vspace{-0.05cm}
    \subfigure[Deezer.]{
         \centering
         \includegraphics[width=0.25\linewidth]{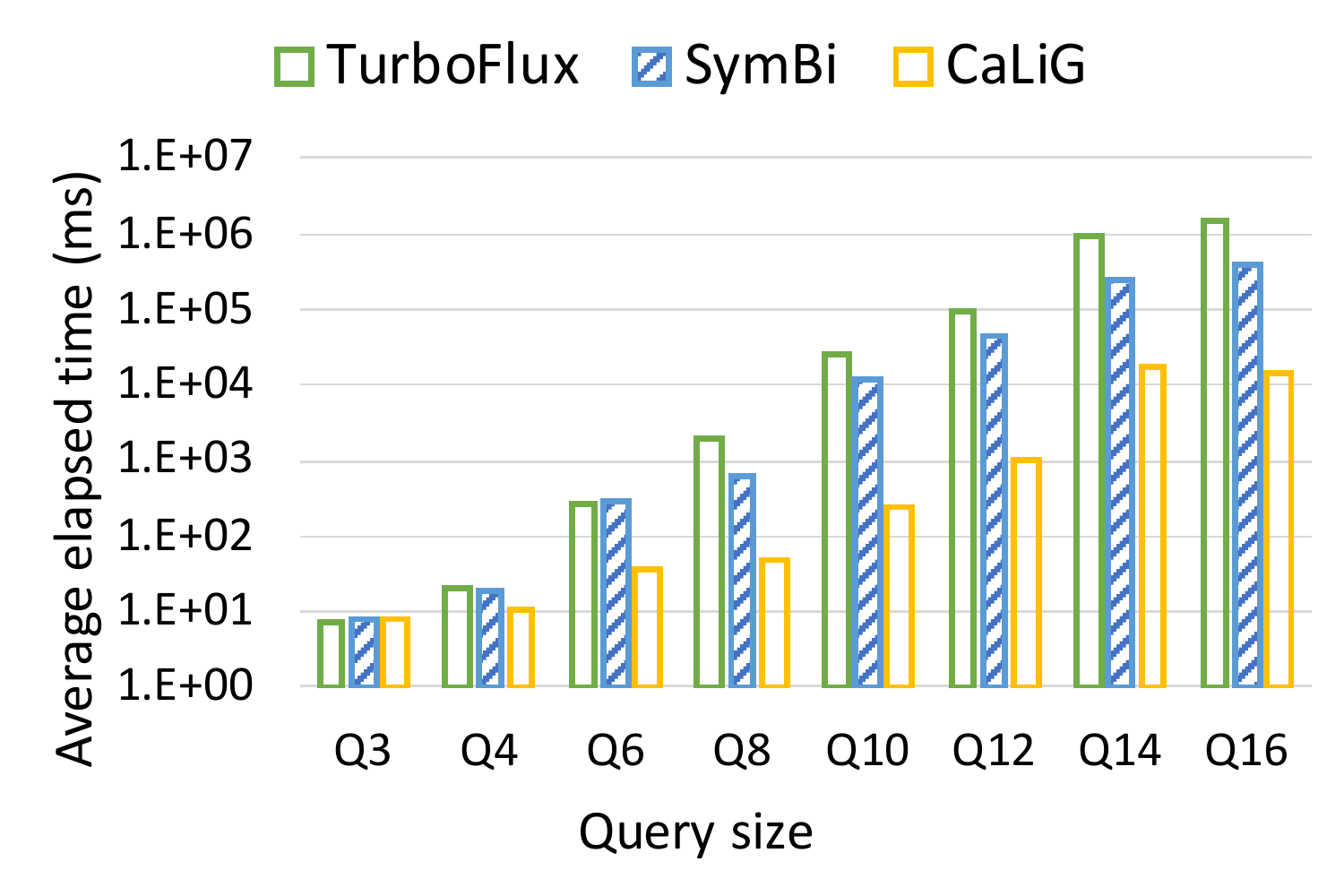}
         \label{fig:1_9}
         }
         \hspace{-0.5cm}
         \vspace{-0.05cm}
    \subfigure[Deezer.]{
         \centering
         \includegraphics[width=0.25\linewidth]{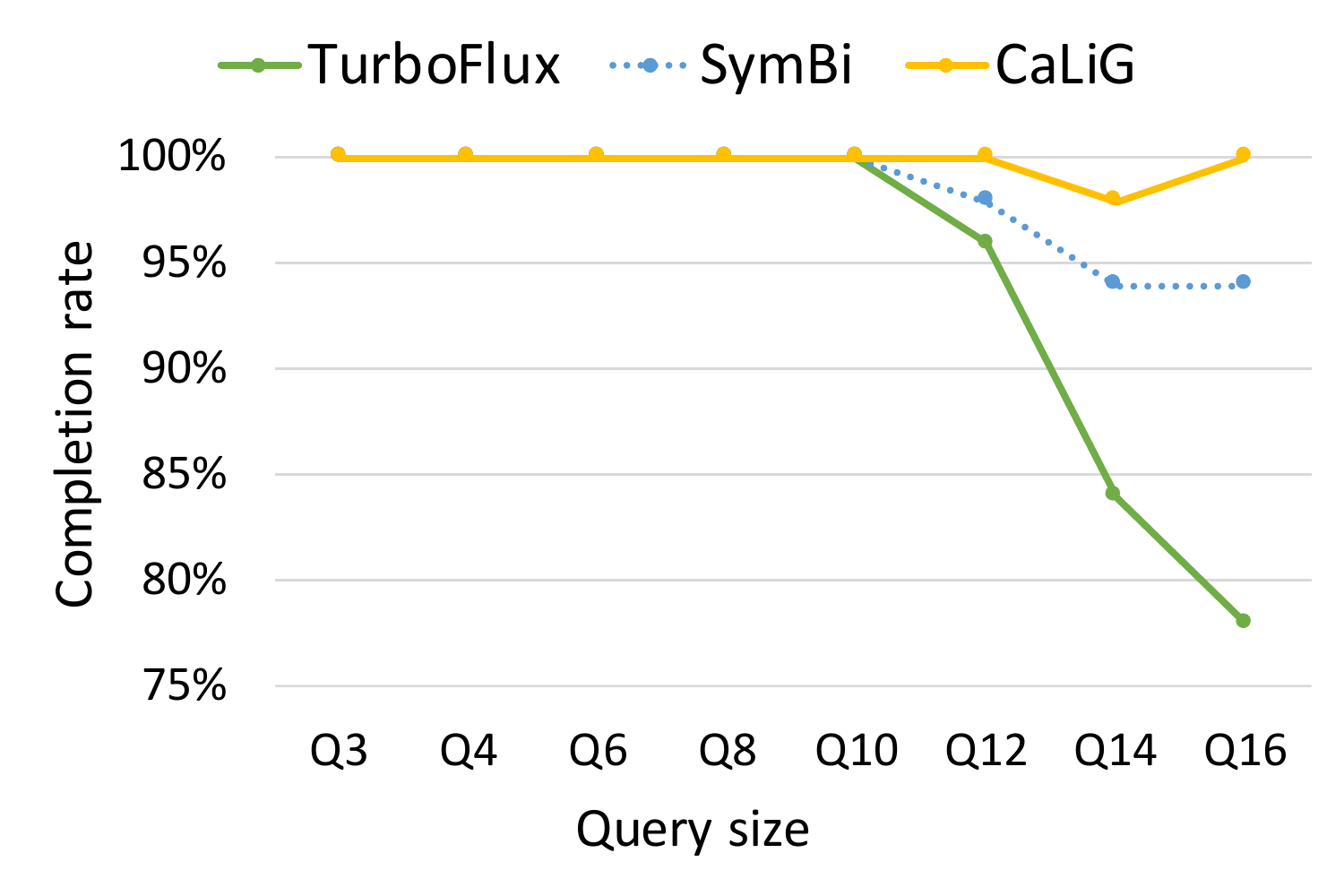}
         \label{fig:1_10}
         }
         \hspace{-0.5cm}
         \vspace{-0.05cm}
    \subfigure[Twitch.]{
         \centering
         \includegraphics[width=0.25\linewidth]{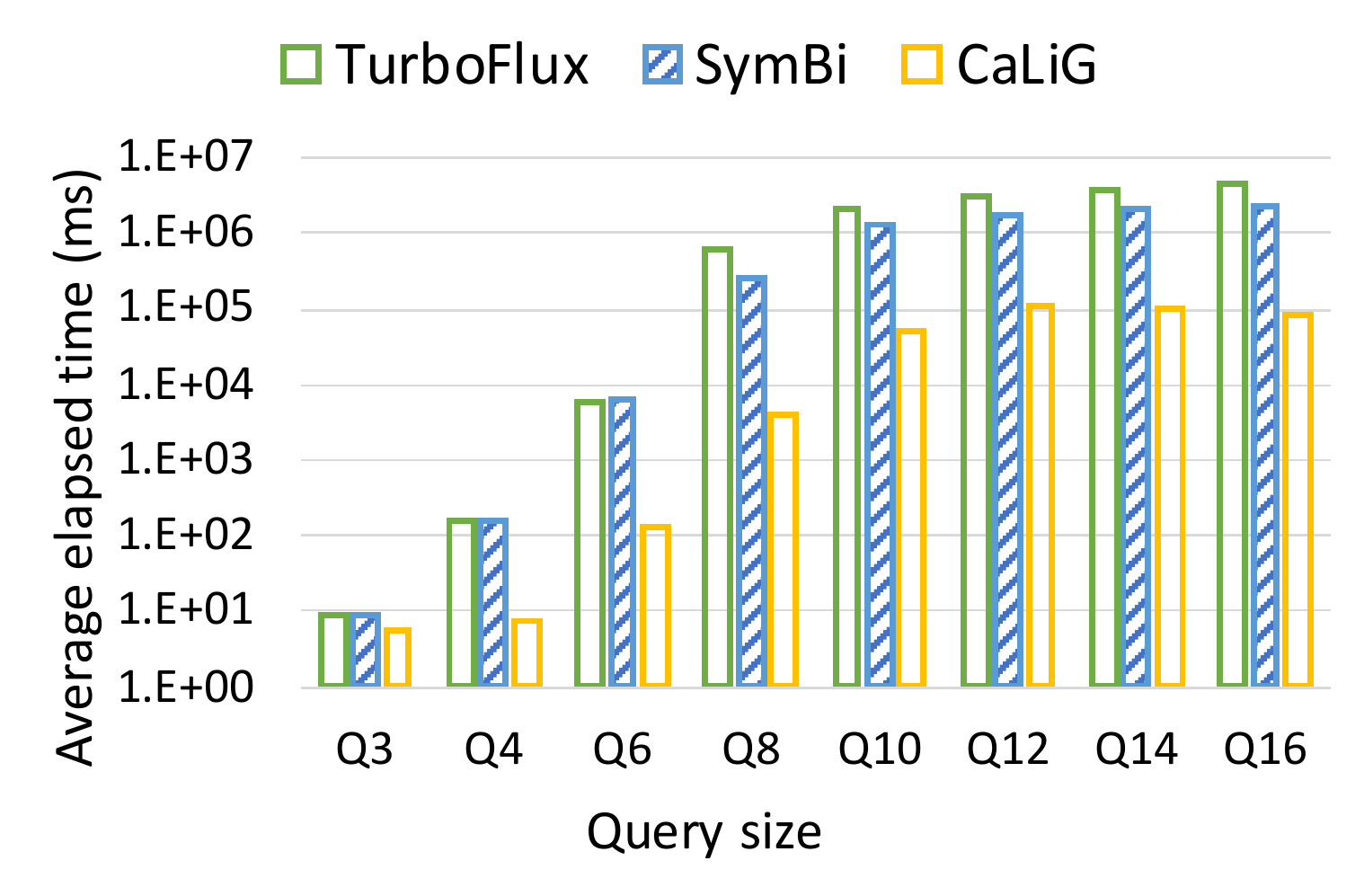}
         \label{fig:1_11}
         }
         \hspace{-0.5cm}
         \vspace{-0.05cm}
    \subfigure[Twitch.]{
         \centering
         \includegraphics[width=0.25\linewidth]{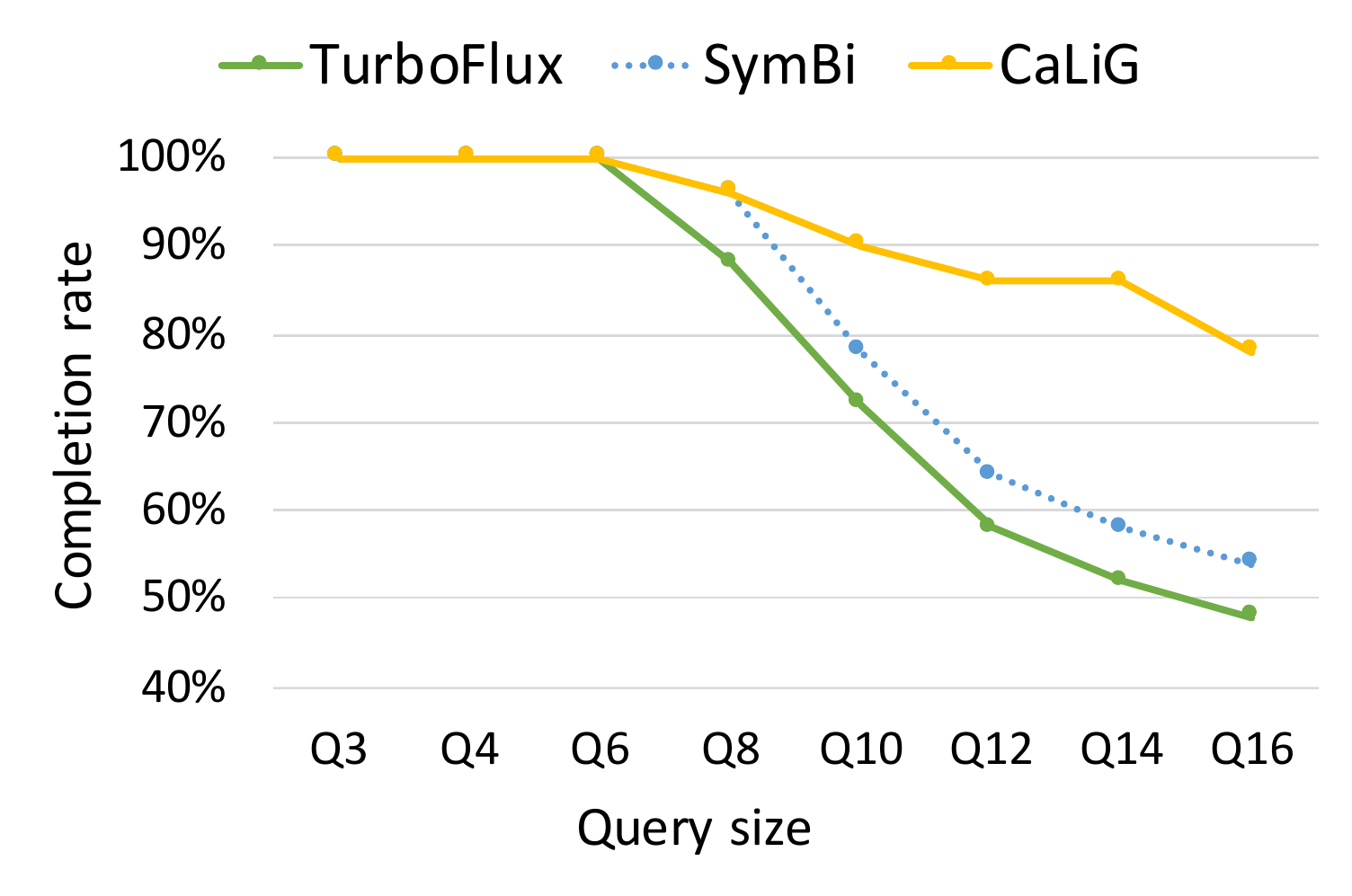}
         \label{fig:1_12}
         }
         \vspace{-0.5cm}
         \vspace{-0.05cm}
    \subfigure[Skitter.]{
         \centering
         \includegraphics[width=0.25\linewidth]{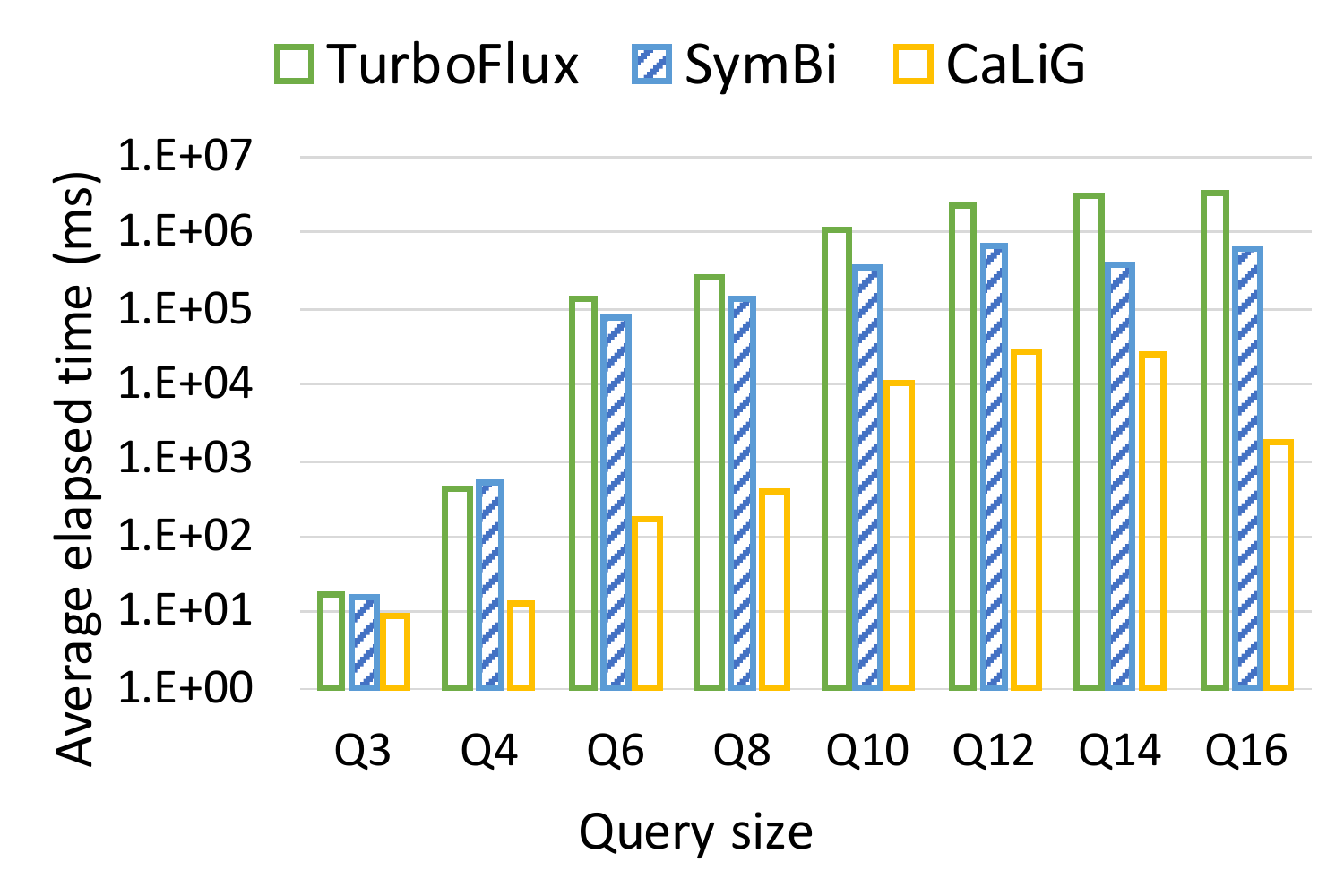}
         \label{fig:1_13}
         }
         \hspace{-0.5cm}
         \vspace{-0.1cm}
    \subfigure[Skitter.]{
         \centering
         \includegraphics[width=0.25\linewidth]{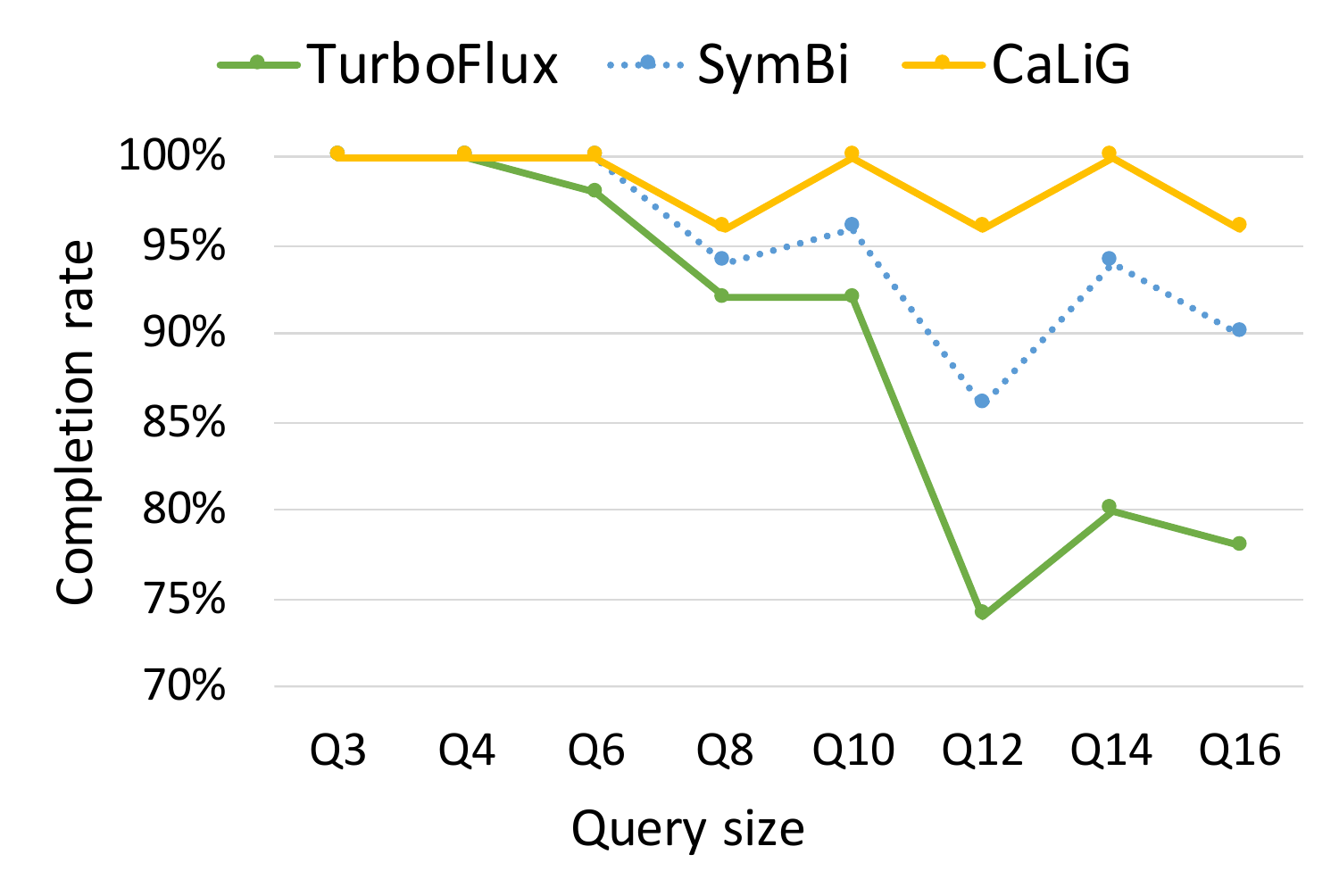}
         \label{fig:1_14}
         }
         \hspace{-0.5cm}
         \vspace{-0.1cm}
    \subfigure[Netflow.]{
         \centering
         \includegraphics[width=0.25\linewidth]{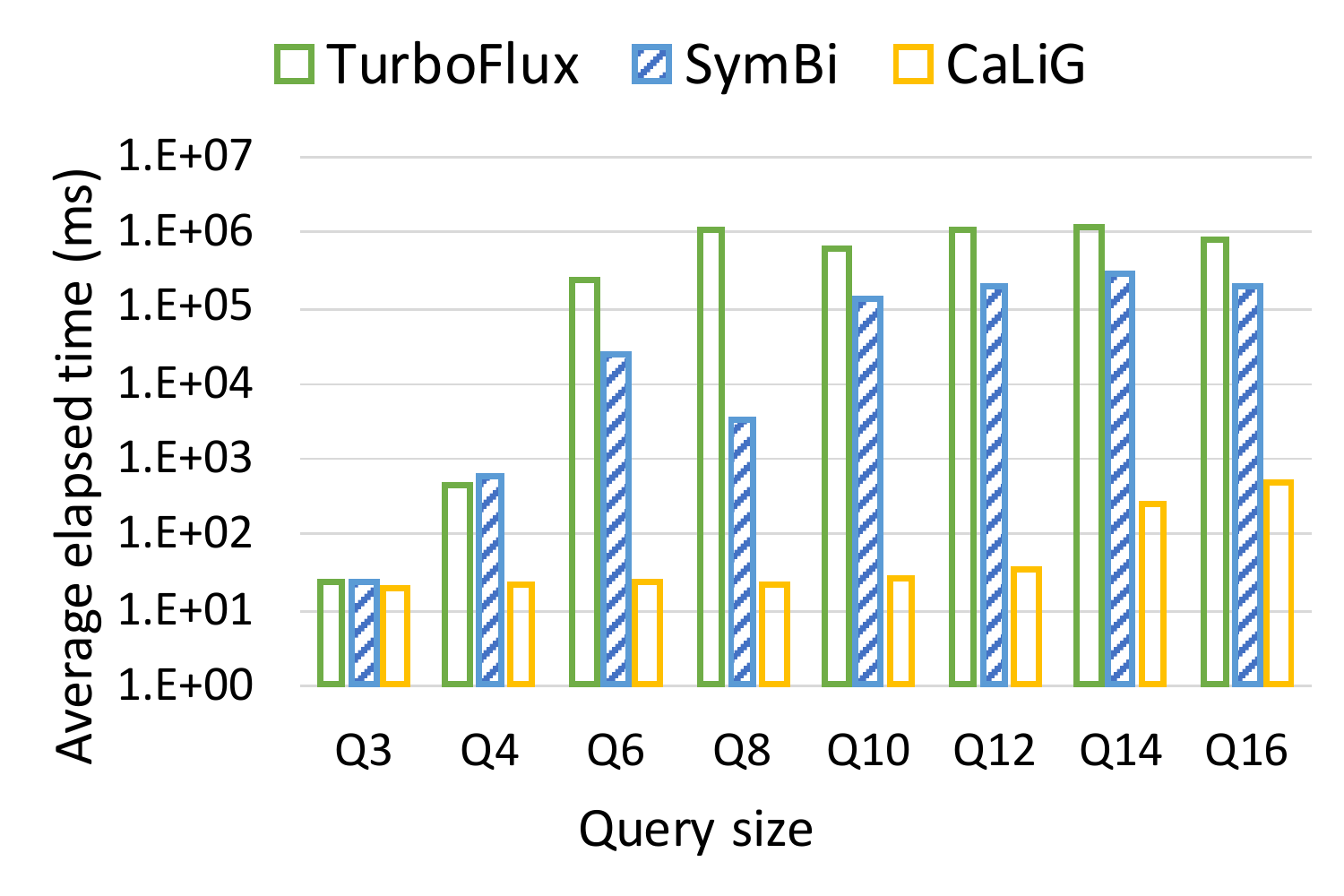}
         \label{fig:1_15}
         }
         \hspace{-0.5cm}
         \vspace{-0.1cm}
    \subfigure[Netflow.]{
         \centering
         \includegraphics[width=0.25\linewidth]{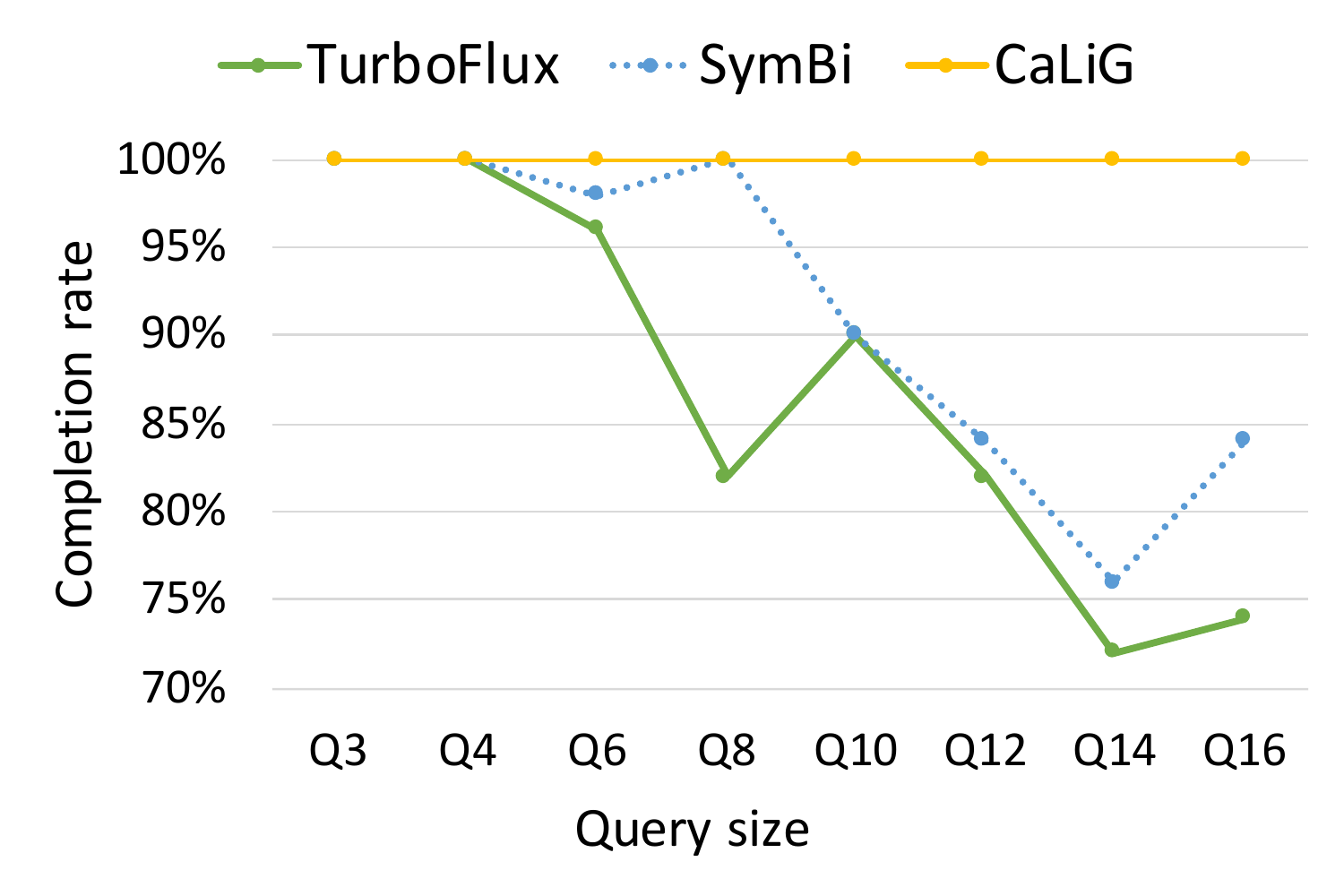}
         \label{fig:1_16}
         }
    \caption{Varying query graph in different data graphs.}
    \label{fig:vqg}
\end{figure}

\noindent \textbf{Varying the size of query graph.}
The effect of query size (i.e., the number of vertices) is evaluated by varying $|V_Q|$ from {3 to 16}.
Figure~\ref{fig:vqg} shows the detailed results, where each dataset has two sub-figures representing the
average elapsed time and the completion rate for each group of queries in the same size.
We can see that CaLiG exhibits significant advantages over the competitors. 

For all data graphs, CaLiG always approximately outperforms SymBi one to three orders of magnitude faster in terms of time efficiency, while achieving a higher completion rate. The average speedup on the group of queries $Q12$ reaches 5472x (on Netflow). {
Moreover, CaliG achieves better speedups over TurboFlux.
}
As the query size grows, the time cost increases as well. 
For those larger queries, the completion rate of CaLiG is significantly higher.

\begin{figure}[h]
    \subfigure[Varying the scale of data graph.]{
    \centering
    \includegraphics[width=0.45\linewidth]{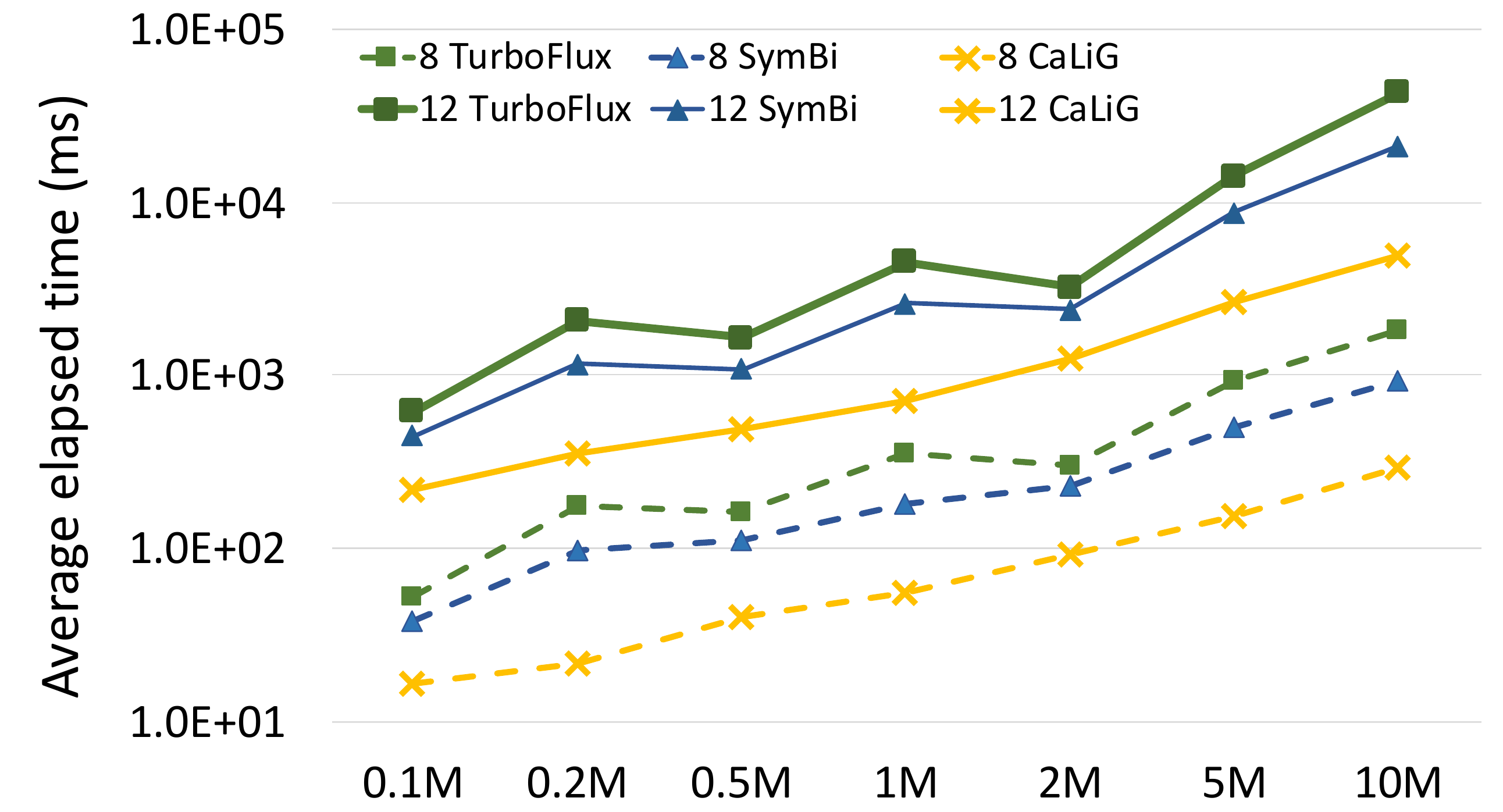}
    \label{fig:2_1}
    }
    \subfigure[Varying the number of vertex labels.]{
    \centering
    \includegraphics[width=0.45\linewidth]{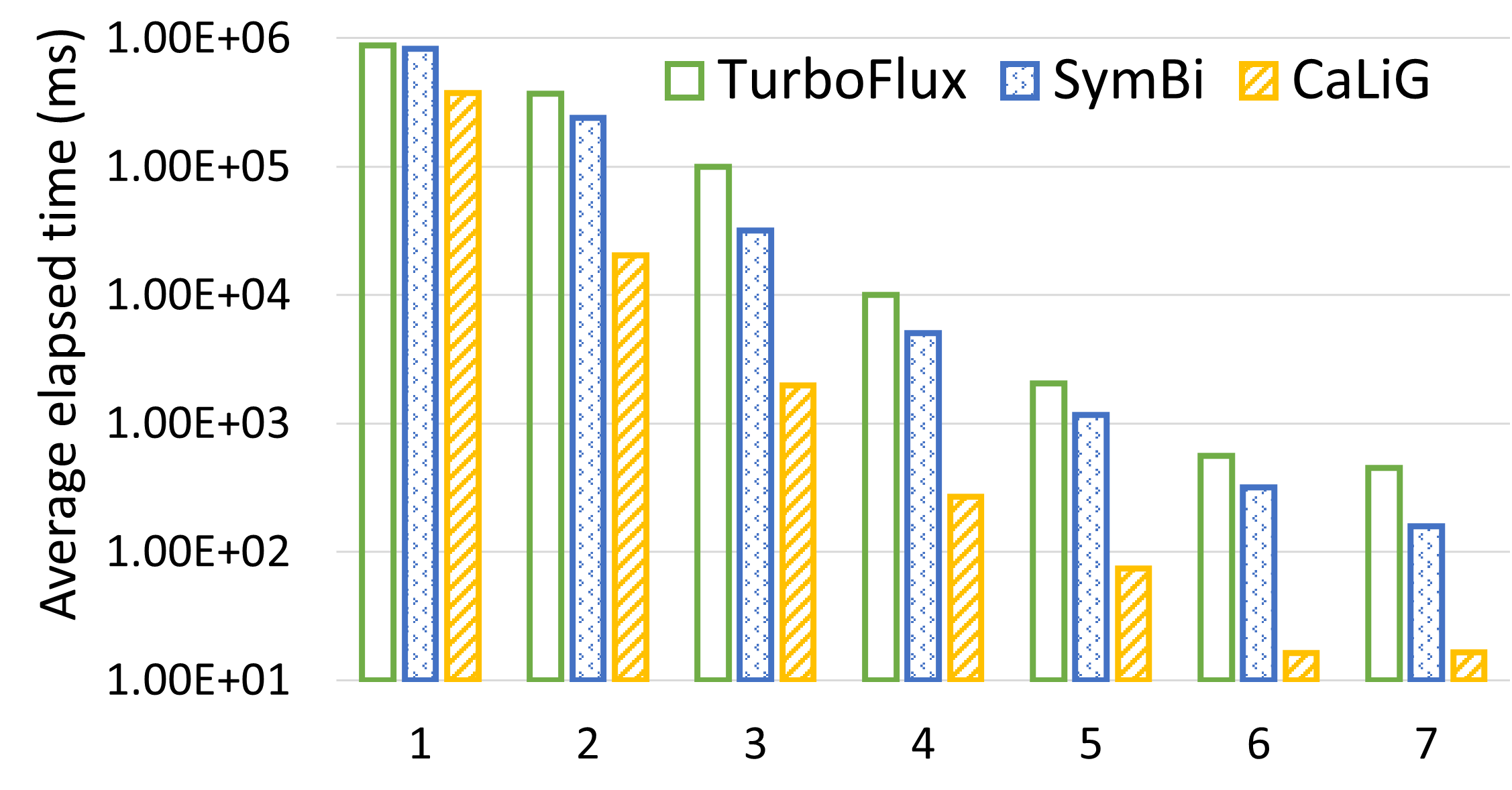}
    \label{fig:2_2}
    }
    \caption{Varying number of vertex labels and the scale of the data graph.}
\end{figure}

{
\noindent \textbf{Varying the size of data graph.} To evaluate the scalability, 
we randomly generate 7 data graphs of different sizes through LSBench~\cite{lsbench}: 0.1, 0.2, 0.5, 1, 2, 5, and 10 million vertices, while the other characters of the data graphs are the same.
Two groups of query graphs with 8 and 12 vertices are used in this experiment.
The results are shown in Figure~\ref{fig:2_1}. The dotted lines represent the results of the 8-vertex query graphs, and the solid lines represent the results of the 12-vertex query graphs.
Compared with TurboFlux and SymBi, CaLiG is consistently faster regardless of the size of the data graph.
In addition, the elapsed time of all algorithms tends to increase with the growth of the dataset size.
}

\noindent \textbf{Varying the number of vertex labels for data graph.} 
We evaluate the effect number of labels by randomly assigning labels to vertices. Totally 7 varied graphs, with the number of labels 1, 2, 3, 4, 5, 7, and 9, are generated for each original data graph. 
Figure~\ref{fig:2_2} presents the results on the data graph \emph{Deezer}.
We can see that the CaLiG algorithm is 2.2 to 19.7 times faster than SymBi, {and 2.4 to 50.5 times faster than TurboFlux.}
The speedup increases with the growth of the number of labels.
In addition, we can find that the number of labels affects the elapsed time. Basically, the time cost is inversely proportional to the number of labels, that is, the more labels, the less elapsed time for continuous subgraph matching.

\begin{figure}[h]
     \subfigure[Varying the scale of streaming data.]{
             \centering
    \includegraphics[width=0.5\linewidth]{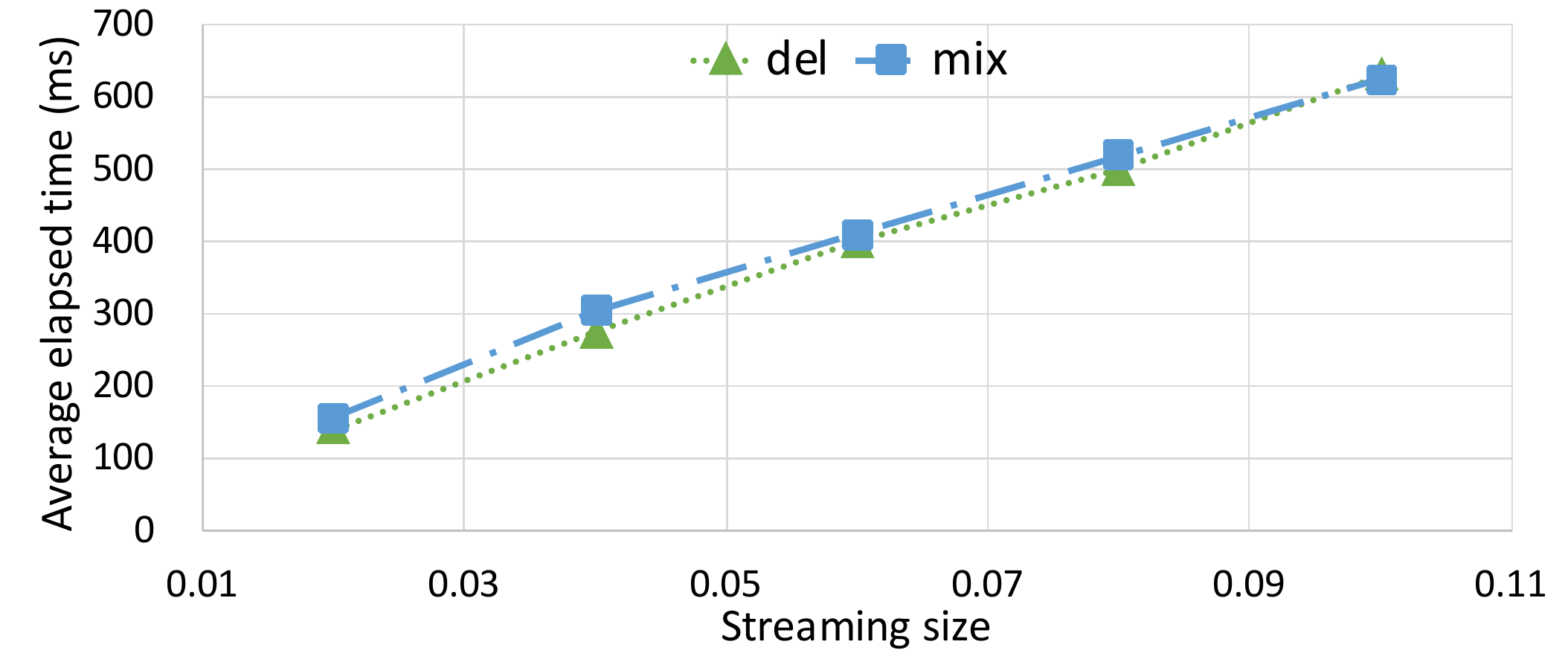}
    \label{fig:update}
    }
   \subfigure[Memory usage.]{
         \centering
    \includegraphics[width=0.43\linewidth]{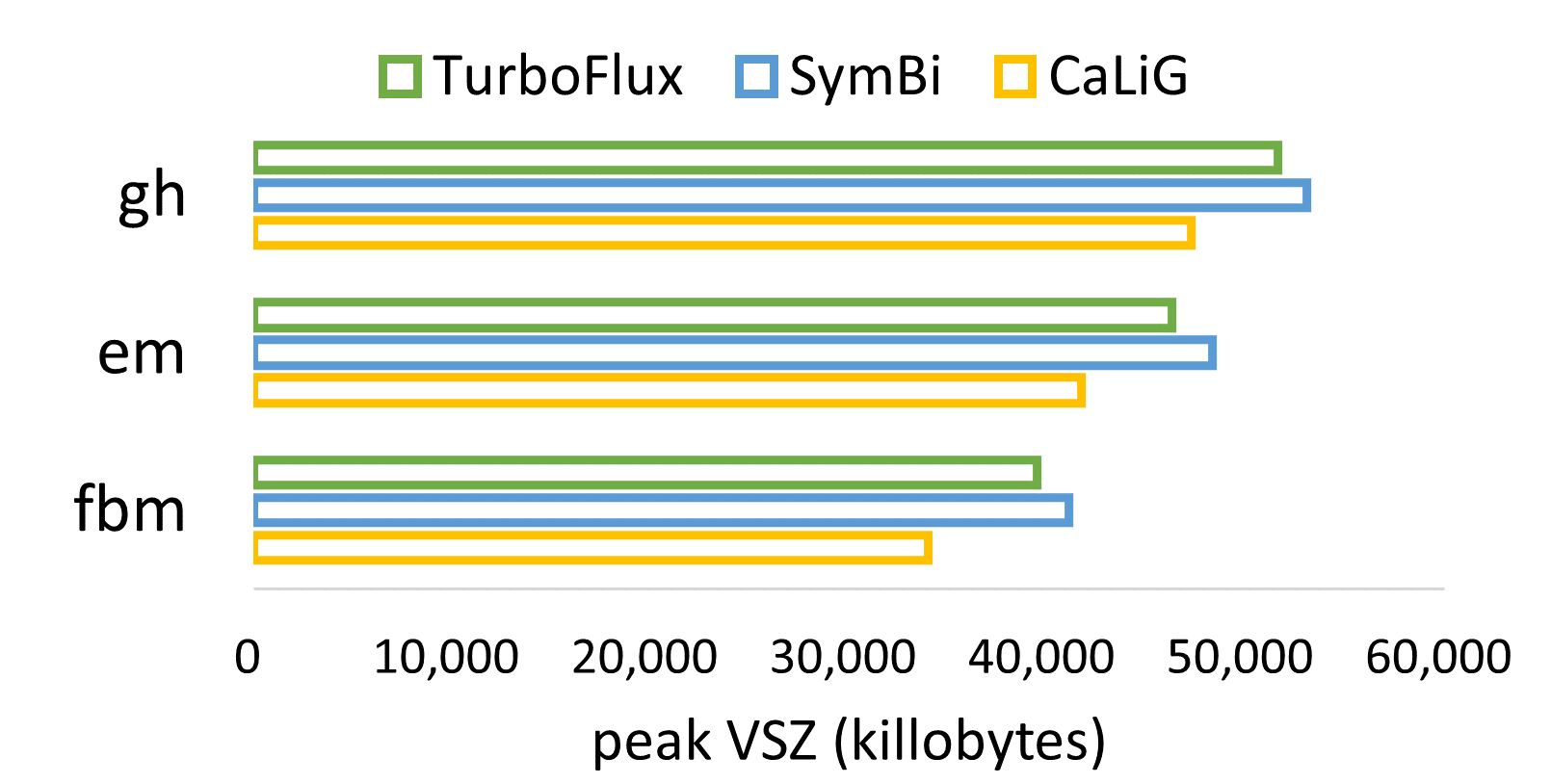}
    \label{fig:mem}
    }
    \caption{Varying the scale of streaming data and Memory usage.}
\end{figure}

\noindent \textbf{Varying the scale of streaming data.}
We explore how the stream size affects our algorithm next.
Two sets of experiments are set up respectively, one set to test the mixed update flow, that is, there are deleted edges and added edges in the update flow, and the ratio of them is 2:1. In the other set, only deleted edges are allowed in the stream. The two sets of update streams are randomly sampled from the original data edges, and their scales are set to 2\%, 4\%, 6\%, 8\%, and 10\% of the number of data edges, respectively.
Figure~\ref{fig:update} reports the results on the data graph \emph{Email}. It shows that the performance is linear to the scale of update streams, whether it is a mixed stream or an edge-deletion stream, indicating good scalability of CaLiG. 
In addition, we can also find that CaLiG exhibits a similar performance in processing edge additions and edge deletions.
\begin{table}[tbp]
  \centering
  \caption{Match Density under different query sizes}
    \begin{tabular}{cccccccccc}
    \toprule
    \textbf{Method} & \textbf{Dataset} & 4 & 6 & 8 & 10 & 12 & 14 \\
    \midrule
    \multirow{3}*{TurboFlux} & \emph{dz} &
0.530 &
0.240 &
0.040 &
0.048 &
0.021 &
0.0001 
\\
& \emph{lfm} &
1.858 &
1.190 &
0.152 &
0.005 &
0.002 &
0.002 
\\
& \emph{sk} &
4.562 &
1.514 & 
0.214 &
0.123 &
0.119 &
0.009 
\\
    \midrule
    \multirow{3}*{SymBi} & \emph{dz} &
0.557 &
0.392 &
0.252 &
0.110 &
0.024 &
0.0009 
\\
& \emph{lfm} &
2.012 &
1.265 &
0.509 &
0.371 &
0.104 &
0.270 
\\
& \emph{sk} &
\textbf{35.83} &
12.59 & 
1.243 &
0.567 &
0.183 &
0.070 
\\
\midrule
\multirow{3}*{CaLiG} & \emph{dz} &
\textbf{
2.139} &
\textbf{22.88} &
\textbf{102.8} &
\textbf{52.83} &
\textbf{10.79} &
\textbf{77.05} 
\\
& \emph{lfm} &
\textbf{9.834} &
\textbf{199.7} &
\textbf{65.36} &
\textbf{422.4} &
\textbf{156.8} &
\textbf{1028} 
\\
& \emph{sk} &
{30.05} &
\textbf{2617} &
\textbf{156588} &
\textbf{504142} &
\textbf{113778} &
\textbf{357529} \\
    \bottomrule
    \end{tabular}
  \label{tab:md}

\end{table}

\noindent \textbf{Match Density.} To discuss the effectiveness of the backtracking search, match density (MD) is defined as 
$$
\text{MD} = \dfrac{\text{\# of Incremental Matches}}{\text{\# of Backtrackings}}
$$

With higher MD, the same number of matches could be found in fewer backtrackings, and in less time. 
Table~\ref{tab:md} lists the results on \emph{dz}, \emph{lfm}, and \emph{sk}.
{
The match density of CaLiG is larger than that of TurboFlux and SymBi up to 7 orders of magnitude:
one backtracking in TurboFlux or SymBi fails to generate one match for most queries, }
while CaLiG could generate thousands of matches in one backtracking. Moreover, with the growth of query size, MDs of TurboFlux and SymBi decrease, but the MD of CaLiG increases at the same time, demonstrating CaLiG would have a more significant advance in solving large queries.

\noindent \textbf{Memory usage.}
We evaluate the memory usage by reporting the peak memory that is defined as the maximum of the virtual set size (VSZ) in the ``ps'' utility output. Figure~\ref{fig:mem} gives the results on graphs \emph{Facebook}, \emph{Email}, and \emph{Github} for query graphs with 8 vertices.
CaLiG has a smaller memory usage than {TurboFlux and} SymBi on all the graphs.

\noindent \textbf{Ablation test.}
We evaluate the effect of each individual technique proposed in this paper. 
The methods we used for comparison include
\begin{enumerate}
\item
  \textit{CaLiG}, the complete version of the proposed algorithm;
 {
\item
  \textit{+CaIM}, 
  caching the computed injective matching to avoid some unnecessary recomputation. 
    If the update edge is not involved in the cached injective matching, it is not required to recompute the injective matching;
  } 
  
\item
  \textit{-InjM},
  disabling the
  injective matching
  checking, but only considering whether each vertex in the $N_Q(u)$ has a connecting edge in the bigraph;
\item
  \textit{-NState}, 
  disabling the lighting state update and checking;
\item
  \textit{-KSS}, 
  replacing the KSS technique with the way of matching one expanded vertex at a time.
\end{enumerate}

\begin{figure} 
    \subfigure[Facebook for Q8.]{
         \centering
         \includegraphics[width=0.35\linewidth]{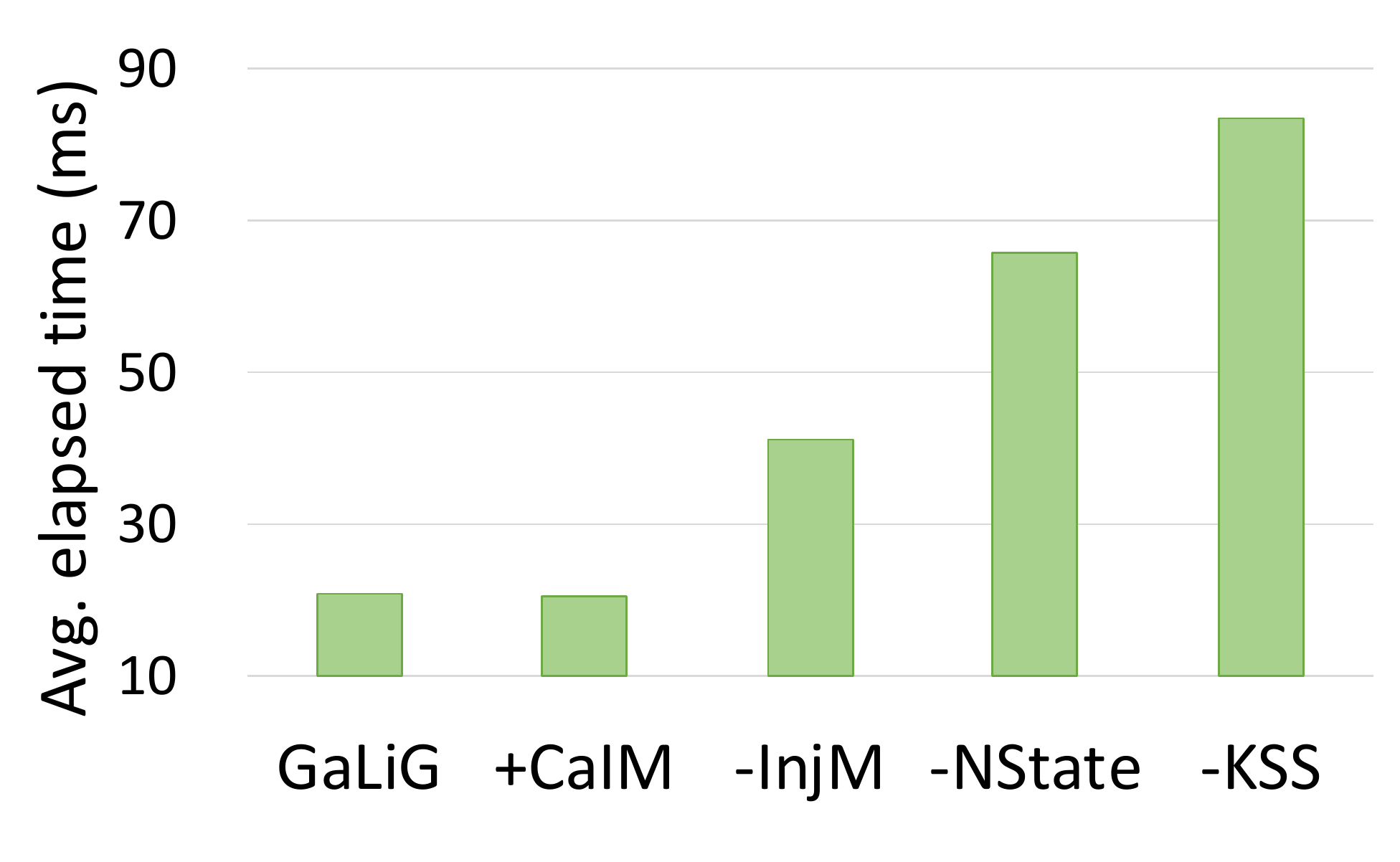}
         \label{fig:comp1}
         }
     \subfigure[Facebook for Q10.]{
         \centering
         \includegraphics[width=0.35\linewidth]{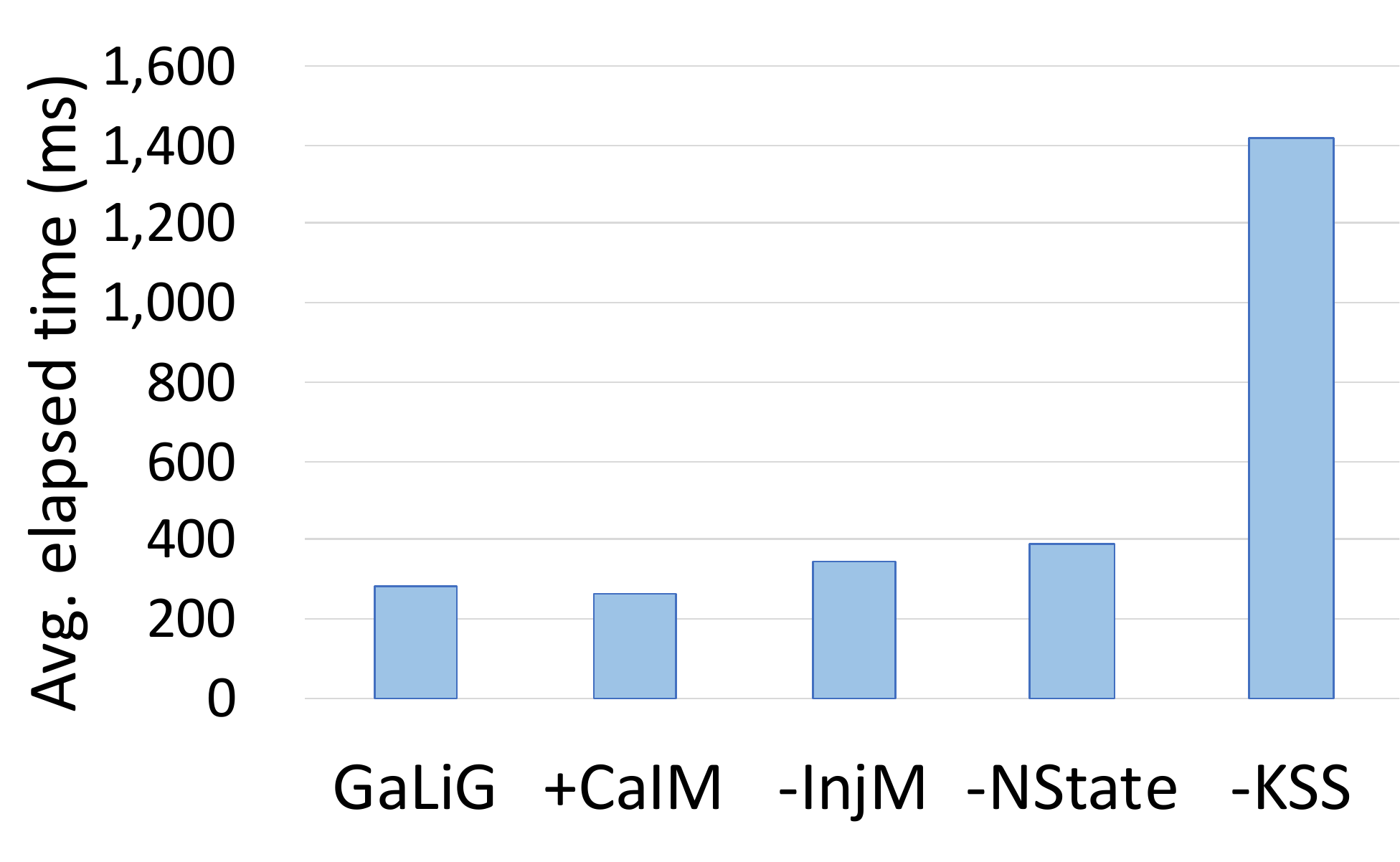}
         \label{fig:comp2}
         }
         \vspace{-0.05cm}
         
    \subfigure[Github for Q8.]{ \vspace{-0.25cm}
         \centering
         \includegraphics[width=0.35\linewidth]{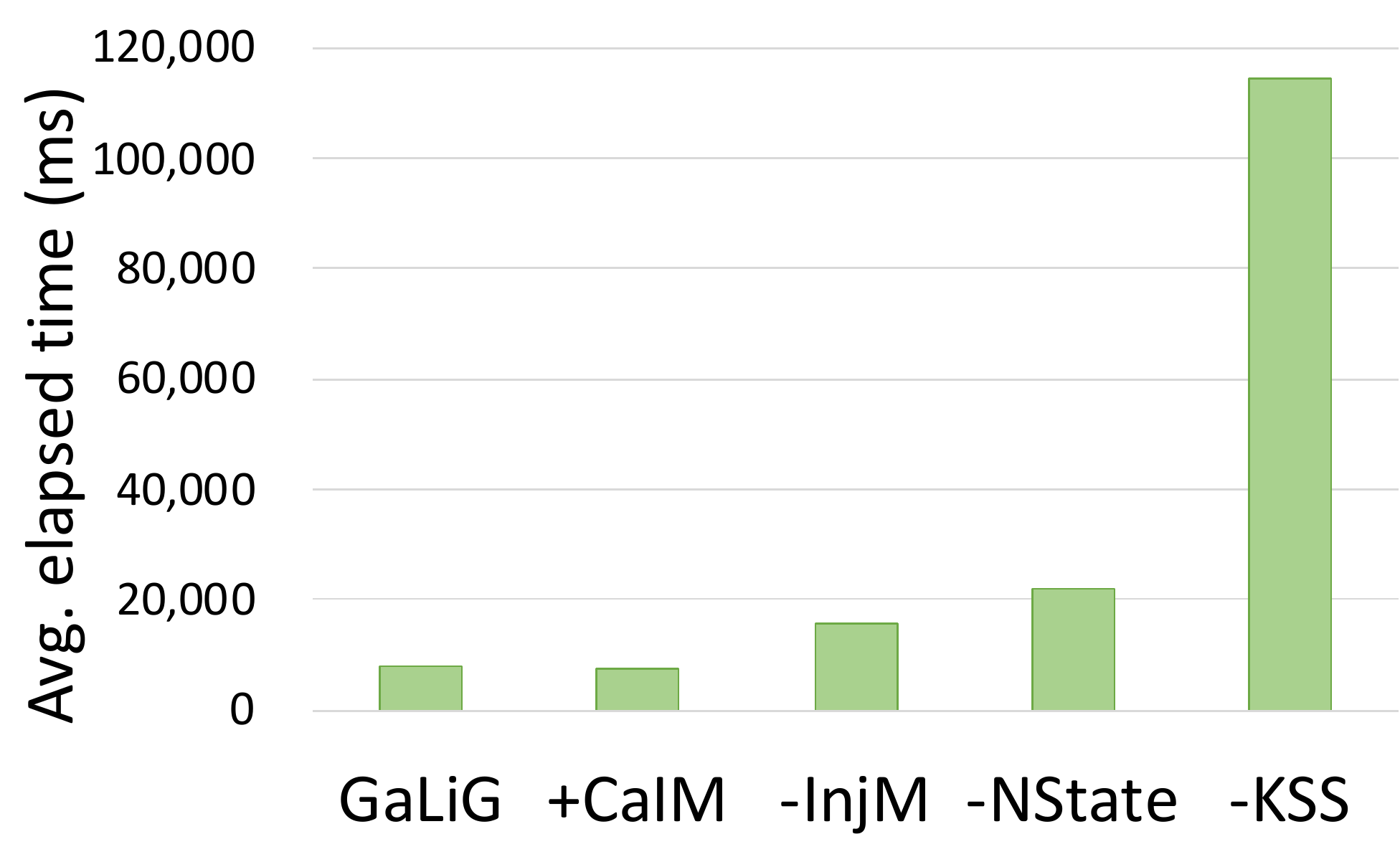}
         \label{fig:comp3}
         }
         \vspace{-0.05cm}
    \subfigure[Github for Q10.]{  \vspace{-0.25cm}
         \centering
         \includegraphics[width=0.35\linewidth]{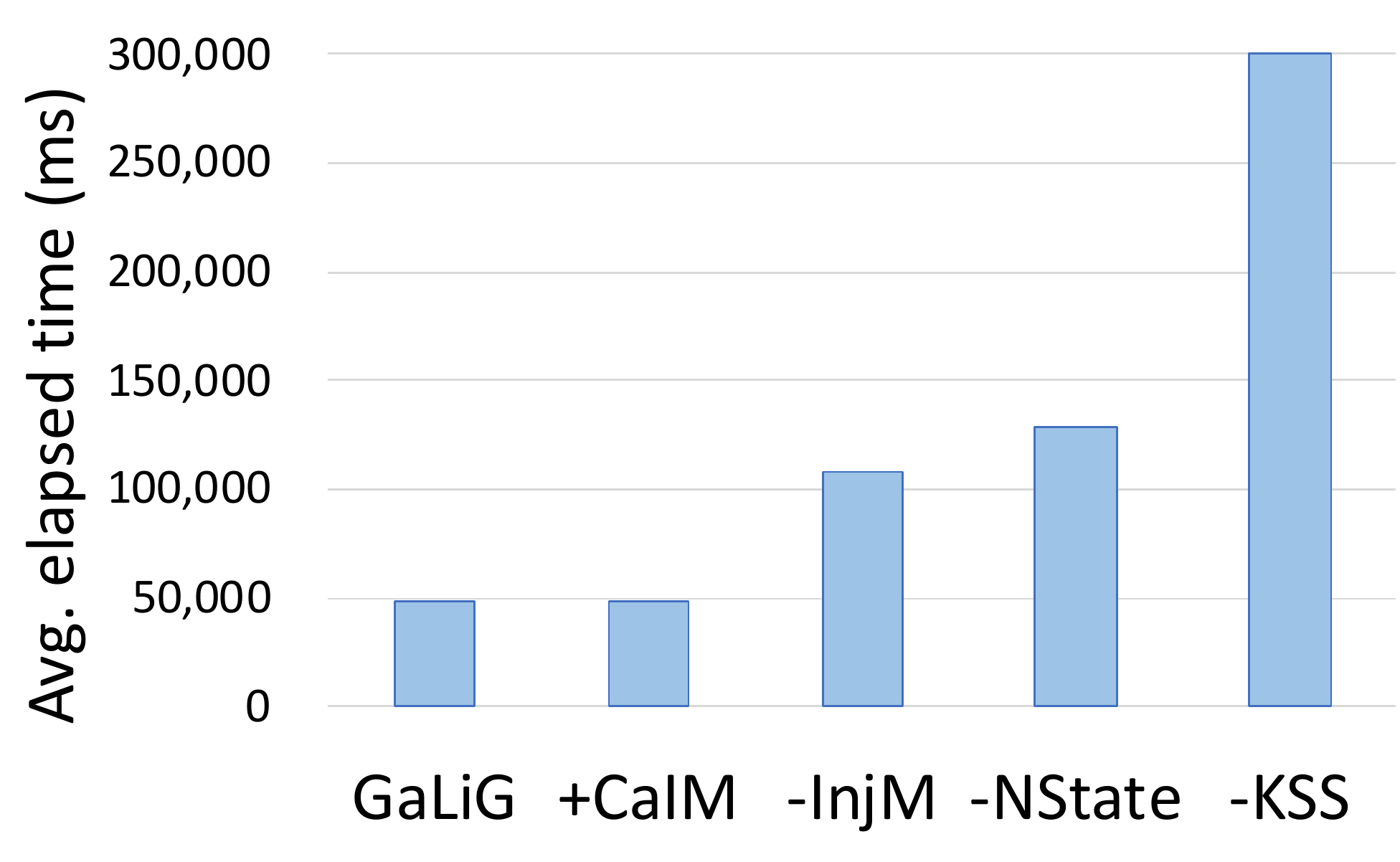}
         \label{fig:comp4}
         }
    \caption{Effect of different components of CaLiG.}
    \label{fig:comp}

\end{figure}

\autoref{fig:comp} presents the results on Facebook and Github, where $Q8$ and $Q10$ represent the query graph that has 8 and 10 vertices, respectively.
{We can observe that the improvement is marginal by caching the injective matching as the cost of computing the injective matching is much smaller than the overall cost. In contrast, the elapsed time will significantly increase when other techniques are disabled.
Among them,  KSS exhibits the most superiority, achieving 4.0X$\thicksim$14.7X speedups on average, since incremental match generation dominates the total cost.}


\section{Related Work}
\textbf{Streaming graph engines.} 
The recent research on streaming graph engines has provided vertex or edge-centric programming models for iterative incremental graph computation.
Kineograph~\cite{kineograph} and GraphInc~\cite{10.1145/2390021.2390023} are two systems that enable incremental computation for monotonic algorithms like Shortest Path.
Chronos~\cite{chronos} assumes streaming graph data is stored persistently and simultaneously computes general graph algorithms on several snapshots. Auxo~\cite{auxo} is complementary to Chronos and could be used as its data source.
GraphBolt~\cite{graphbolt} proposes a generalized incremental model to handle non-monotonic algorithms like Belief Propagation, but involves more overheads than KickStarter for monotonic algorithms.
GraphS~\cite{10.14778/3229863.3229874} designs a real-time streaming system called GraphS for cycle detection.
RisGraph~\cite{risgraph} targets per-update analysis to provide low latency and detailed information in comparison.
However, the systems above do not pay attention to the problem of continuous subgraph matching.

\noindent \textbf{Subgraph matching.}
Subgraph matching is a fundamental requirement for graph databases and has been studied extensively in recent decades. 
QuickSI~\cite{quicksi} follows the direct enumeration framework, which directly explores the data graph to enumerate all results. Most state-space based representation models (that is, each state represents an intermediate result) use this framework, e.g., Ri~\cite{ri}, VF2++~\cite{vf2}, Graphql~\cite{graphql}, CFL\cite{cfl}, CECI\cite{ceci}, and DP-ISO\cite{dpiso}.
Due to the effective filtering and sorting methods, they have achieved great improvements in overall performance.
However, these algorithms focus on the problem in a static data graph. For streaming graphs, it is unacceptable for the real-time requirement to perform subgraph matching over the graphs
prior to and after updating respectively.

\noindent \textbf{Continuous subgraph matching.}
IncIsoMat~\cite{10.1145/2489791} and Graphflow~\cite{graphflow} expand matches from the updated edges without any additional index structures, showing limited real-time performance.
SJ-Tree~\cite{osti_1183625} materializes all the partial matches of subgraphs, leading to a huge storage and index update cost.
 TurboFlux~\cite{turboflux} employs a concise structure DCG, a complete multigraph, to represent intermediate results, reducing the storage cost. Similarly, CEPDG~\cite{DBLP:journals/www/ZhangGZW21} employs TreeMat to store the partial matches of trees.
 It supports changes in both pattern graphs and data graphs.
However, TurboFlux and CEPDG 
exploit a spanning tree of the query graph to filter the candidates, 
leaving the non-tree edges to be maintained by DCG. 
SymBi~\cite{symbi} introduces an auxiliary data structure DCS to store weak embeddings of directed acyclic graphs as intermediate results.
Compared to 
 DCG used in
TurboFlux,
DCS considers both tree and non-tree edges in a DAG of the query graph.
Nevertheless, the index constructed by SymBi is not cost-effective enough, that is, the candidates could be tightened to archive less total cost.


\section{Conclusion}
In this paper, we 
propose a  cost-effective index CaLiG for continuous subgraph matching. 
 CaLiG yields tighter candidates than previous methods with low cost, contributing to
considerable speed up for match generation.
To further accelerate incremental generation, we develop a novel subgraph matching paradigm, called KSS. With the partial matches of kernel vertices, KSS can produce incremental matches by just joining the candidates of shell vertices without 
 any backtrackings.
The empirical experiments demonstrate the efficiency of our proposed method.

\begin{acks}
This work was supported by 
 the Research Grants Council of Hong Kong, China (No. 14203618, No. 14202919, and No. 14205520). Weiguo Zheng is the corresponding author.
\end{acks}



\balance
\bibliographystyle{ACM-Reference-Format}
\bibliography{sample-base}


\begin{thebibliography}{40}


\ifx \showCODEN    \undefined \def \showCODEN     #1{\unskip}     \fi
\ifx \showDOI      \undefined \def \showDOI       #1{#1}\fi
\ifx \showISBNx    \undefined \def \showISBNx     #1{\unskip}     \fi
\ifx \showISBNxiii \undefined \def \showISBNxiii  #1{\unskip}     \fi
\ifx \showISSN     \undefined \def \showISSN      #1{\unskip}     \fi
\ifx \showLCCN     \undefined \def \showLCCN      #1{\unskip}     \fi
\ifx \shownote     \undefined \def \shownote      #1{#1}          \fi
\ifx \showarticletitle \undefined \def \showarticletitle #1{#1}   \fi
\ifx \showURL      \undefined \def \showURL       {\relax}        \fi
\providecommand\bibfield[2]{#2}
\providecommand\bibinfo[2]{#2}
\providecommand\natexlab[1]{#1}
\providecommand\showeprint[2][]{arXiv:#2}

\bibitem[net({[n.\,d.]})]%
        {netflow}
 \bibinfo{year}{[n.\,d.]}\natexlab{}.
\newblock \bibinfo{title}{Anonymized Internet Traces 2013}.
\newblock
  \bibinfo{howpublished}{\url{https://catalog.caida.org/details/dataset/passive_2013_pcap}}.
\newblock


\bibitem[Abdelhamid et~al\mbox{.}(2018)]%
        {8509462}
\bibfield{author}{\bibinfo{person}{Ehab Abdelhamid}, \bibinfo{person}{Mustafa
  Canim}, \bibinfo{person}{Mohammad Sadoghi}, \bibinfo{person}{Bishwaranjan
  Bhattacharjee}, \bibinfo{person}{Yuan-Chi Chang}, {and}
  \bibinfo{person}{Panos Kalnis}.} \bibinfo{year}{2018}\natexlab{}.
\newblock \showarticletitle{Incremental Frequent Subgraph Mining on Large
  Evolving Graphs}. In \bibinfo{booktitle}{\emph{2018 IEEE 34th International
  Conference on Data Engineering (ICDE)}}. \bibinfo{pages}{1767--1768}.
\newblock


\bibitem[Bhattarai et~al\mbox{.}(2019)]%
        {ceci}
\bibfield{author}{\bibinfo{person}{Bibek Bhattarai}, \bibinfo{person}{Hang
  Liu}, {and} \bibinfo{person}{H.~Howie Huang}.}
  \bibinfo{year}{2019}\natexlab{}.
\newblock \showarticletitle{CECI: Compact Embedding Cluster Index for Scalable
  Subgraph Matching}. In \bibinfo{booktitle}{\emph{Proceedings of the 2019
  International Conference on Management of Data}} (Amsterdam, Netherlands)
  \emph{(\bibinfo{series}{SIGMOD '19})}. \bibinfo{publisher}{Association for
  Computing Machinery}, \bibinfo{address}{New York, NY, USA},
  \bibinfo{pages}{1447–1462}.
\newblock
\showISBNx{9781450356435}


\bibitem[Bi et~al\mbox{.}(2016)]%
        {cfl}
\bibfield{author}{\bibinfo{person}{Fei Bi}, \bibinfo{person}{Lijun Chang},
  \bibinfo{person}{Xuemin Lin}, \bibinfo{person}{Lu Qin}, {and}
  \bibinfo{person}{Wenjie Zhang}.} \bibinfo{year}{2016}\natexlab{}.
\newblock \showarticletitle{Efficient Subgraph Matching by Postponing Cartesian
  Products}. In \bibinfo{booktitle}{\emph{Proceedings of the 2016 International
  Conference on Management of Data}} (San Francisco, California, USA)
  \emph{(\bibinfo{series}{SIGMOD '16})}. \bibinfo{publisher}{Association for
  Computing Machinery}, \bibinfo{address}{New York, NY, USA},
  \bibinfo{pages}{1199–1214}.
\newblock
\showISBNx{9781450335317}
\urldef\tempurl%
\url{https://doi.org/10.1145/2882903.2915236}
\showDOI{\tempurl}


\bibitem[Bonnici et~al\mbox{.}(2013)]%
        {ri}
\bibfield{author}{\bibinfo{person}{Vincenzo Bonnici}, \bibinfo{person}{Rosalba
  Giugno}, \bibinfo{person}{Alfredo Pulvirenti}, \bibinfo{person}{Dennis
  Shasha}, {and} \bibinfo{person}{Alfredo Ferro}.}
  \bibinfo{year}{2013}\natexlab{}.
\newblock \showarticletitle{A subgraph isomorphism algorithm and its
  application to biochemical data}.
\newblock \bibinfo{journal}{\emph{BMC Bioinformatics}} \bibinfo{volume}{14},
  \bibinfo{number}{SUPPL7} (\bibinfo{date}{22 April} \bibinfo{year}{2013}).
\newblock
\showISSN{1471-2105}
\urldef\tempurl%
\url{https://doi.org/10.1186/1471-2105-14-S7-S13}
\showDOI{\tempurl}
\newblock
\shownote{Funding Information: This article is published as part of a
  supplement. The publication costs for this article were funded by PO grant -
  FESR 2007-2013 Linea di intervento 4.1.1.2, CUP G23F11000840004.}.


\bibitem[Boshmaf et~al\mbox{.}(2011)]%
        {10.1145/2076732.2076746}
\bibfield{author}{\bibinfo{person}{Yazan Boshmaf}, \bibinfo{person}{Ildar
  Muslukhov}, \bibinfo{person}{Konstantin Beznosov}, {and}
  \bibinfo{person}{Matei Ripeanu}.} \bibinfo{year}{2011}\natexlab{}.
\newblock \showarticletitle{The Socialbot Network: When Bots Socialize for Fame
  and Money}. In \bibinfo{booktitle}{\emph{Proceedings of the 27th Annual
  Computer Security Applications Conference}}. \bibinfo{pages}{93–102}.
\newblock


\bibitem[Cai et~al\mbox{.}(2012)]%
        {10.1145/2390021.2390023}
\bibfield{author}{\bibinfo{person}{Zhuhua Cai}, \bibinfo{person}{Dionysios
  Logothetis}, {and} \bibinfo{person}{Georgos Siganos}.}
  \bibinfo{year}{2012}\natexlab{}.
\newblock \showarticletitle{Facilitating Real-Time Graph Mining}. In
  \bibinfo{booktitle}{\emph{Proceedings of the Fourth International Workshop on
  Cloud Data Management}}. \bibinfo{publisher}{Association for Computing
  Machinery}, \bibinfo{pages}{1–8}.
\newblock
\showISBNx{9781450317085}


\bibitem[Cheng et~al\mbox{.}(2012)]%
        {kineograph}
\bibfield{author}{\bibinfo{person}{Raymond Cheng}, \bibinfo{person}{Ji Hong},
  \bibinfo{person}{Aapo Kyrola}, \bibinfo{person}{Youshan Miao},
  \bibinfo{person}{Xuetian Weng}, \bibinfo{person}{Ming Wu},
  \bibinfo{person}{Fan Yang}, \bibinfo{person}{Lidong Zhou},
  \bibinfo{person}{Feng Zhao}, {and} \bibinfo{person}{Enhong Chen}.}
  \bibinfo{year}{2012}\natexlab{}.
\newblock \showarticletitle{Kineograph: Taking the Pulse of a Fast-Changing and
  Connected World}. In \bibinfo{booktitle}{\emph{Proceedings of the 7th ACM
  European Conference on Computer Systems}}. \bibinfo{publisher}{Association
  for Computing Machinery}, \bibinfo{pages}{85–98}.
\newblock


\bibitem[Choudhury et~al\mbox{.}(2013)]%
        {streamworks}
\bibfield{author}{\bibinfo{person}{Sutanay Choudhury},
  \bibinfo{person}{Lawrence Holder}, \bibinfo{person}{George Chin},
  \bibinfo{person}{Abhik Ray}, \bibinfo{person}{Sherman Beus}, {and}
  \bibinfo{person}{John Feo}.} \bibinfo{year}{2013}\natexlab{}.
\newblock \showarticletitle{StreamWorks: A System for Dynamic Graph Search}. In
  \bibinfo{booktitle}{\emph{Proceedings of the 2013 ACM SIGMOD International
  Conference on Management of Data}}. \bibinfo{pages}{1101–1104}.
\newblock


\bibitem[Choudhury et~al\mbox{.}(2015)]%
        {osti_1183625}
\bibfield{author}{\bibinfo{person}{Sutanay Choudhury},
  \bibinfo{person}{Lawrence~B. Holder}, \bibinfo{person}{George~Chin Jr.},
  \bibinfo{person}{Khushbu Agarwal}, {and} \bibinfo{person}{John Feo}.}
  \bibinfo{year}{2015}\natexlab{}.
\newblock \showarticletitle{A Selectivity based approach to Continuous Pattern
  Detection in Streaming Graphs}. In \bibinfo{booktitle}{\emph{Proceedings of
  the 18th International Conference on Extending Database Technology, {EDBT}
  2015, Brussels, Belgium, March 23-27, 2015}}. \bibinfo{pages}{157--168}.
\newblock


\bibitem[Fan et~al\mbox{.}(2013)]%
        {10.1145/2489791}
\bibfield{author}{\bibinfo{person}{Wenfei Fan}, \bibinfo{person}{Xin Wang},
  {and} \bibinfo{person}{Yinghui Wu}.} \bibinfo{year}{2013}\natexlab{}.
\newblock \showarticletitle{Incremental Graph Pattern Matching}.
\newblock \bibinfo{journal}{\emph{ACM Trans. Database Syst.}}
  \bibinfo{volume}{38}, \bibinfo{number}{3}, Article \bibinfo{articleno}{18}
  (\bibinfo{date}{sep} \bibinfo{year}{2013}), \bibinfo{numpages}{47}~pages.
\newblock


\bibitem[Feng et~al\mbox{.}(2021)]%
        {risgraph}
\bibfield{author}{\bibinfo{person}{Guanyu Feng}, \bibinfo{person}{Zixuan Ma},
  \bibinfo{person}{Daixuan Li}, \bibinfo{person}{Shengqi Chen},
  \bibinfo{person}{Xiaowei Zhu}, \bibinfo{person}{Wentao Han}, {and}
  \bibinfo{person}{Wenguang Chen}.} \bibinfo{year}{2021}\natexlab{}.
\newblock \bibinfo{booktitle}{\emph{RisGraph: A Real-Time Streaming System for
  Evolving Graphs to Support Sub-Millisecond Per-Update Analysis at Millions
  Ops/s}}.
\newblock \bibinfo{pages}{513–527}.
\newblock


\bibitem[Garey and Johnson(1990)]%
        {NPC}
\bibfield{author}{\bibinfo{person}{Michael~R. Garey} {and}
  \bibinfo{person}{David~S. Johnson}.} \bibinfo{year}{1990}\natexlab{}.
\newblock \bibinfo{booktitle}{\emph{Computers and Intractability; A Guide to
  the Theory of NP-Completeness}}.
\newblock \bibinfo{publisher}{W. H. Freeman; Co.}, \bibinfo{address}{USA}.
\newblock
\showISBNx{0716710455}


\bibitem[Gupta et~al\mbox{.}(2014)]%
        {10.14778/2733004.2733010}
\bibfield{author}{\bibinfo{person}{Pankaj Gupta}, \bibinfo{person}{Venu
  Satuluri}, \bibinfo{person}{Ajeet Grewal}, \bibinfo{person}{Siva Gurumurthy},
  \bibinfo{person}{Volodymyr Zhabiuk}, \bibinfo{person}{Quannan Li}, {and}
  \bibinfo{person}{Jimmy Lin}.} \bibinfo{year}{2014}\natexlab{}.
\newblock \showarticletitle{Real-Time Twitter Recommendation: Online Motif
  Detection in Large Dynamic Graphs}.
\newblock \bibinfo{journal}{\emph{Proc. VLDB Endow.}} \bibinfo{volume}{7},
  \bibinfo{number}{13} (\bibinfo{date}{aug} \bibinfo{year}{2014}),
  \bibinfo{pages}{1379–1380}.
\newblock


\bibitem[Hall(1935)]%
        {hall}
\bibfield{author}{\bibinfo{person}{P. Hall}.} \bibinfo{year}{1935}\natexlab{}.
\newblock \showarticletitle{On Representatives of Subsets}.
\newblock \bibinfo{journal}{\emph{Journal of the London Mathematical Society}}
  \bibinfo{volume}{s1-10}, \bibinfo{number}{1} (\bibinfo{year}{1935}),
  \bibinfo{pages}{26--30}.
\newblock
\urldef\tempurl%
\url{https://doi.org/10.1112/jlms/s1-10.37.26}
\showDOI{\tempurl}
\showeprint{https://londmathsoc.onlinelibrary.wiley.com/doi/pdf/10.1112/jlms/s1-10.37.26}


\bibitem[Han et~al\mbox{.}(2019a)]%
        {dpiso}
\bibfield{author}{\bibinfo{person}{Myoungji Han}, \bibinfo{person}{Hyunjoon
  Kim}, \bibinfo{person}{Geonmo Gu}, \bibinfo{person}{Kunsoo Park}, {and}
  \bibinfo{person}{{Wook Shin} Han}.} \bibinfo{year}{2019}\natexlab{a}.
\newblock \showarticletitle{Efficient subgraph matching: Harmonizing dynamic
  programming, adaptive matching order, and failing set together}. In
  \bibinfo{booktitle}{\emph{SIGMOD 2019 - Proceedings of the 2019 International
  Conference on Management of Data}}. \bibinfo{pages}{1429--1446}.
\newblock


\bibitem[Han et~al\mbox{.}(2019b)]%
        {auxo}
\bibfield{author}{\bibinfo{person}{Wentao Han}, \bibinfo{person}{Kaiwei Li},
  \bibinfo{person}{Shimin Chen}, {and} \bibinfo{person}{Wenguang Chen}.}
  \bibinfo{year}{2019}\natexlab{b}.
\newblock \showarticletitle{Auxo: a temporal graph management system}.
\newblock \bibinfo{journal}{\emph{Big Data Mining and Analytics}}
  \bibinfo{volume}{2}, \bibinfo{number}{1} (\bibinfo{year}{2019}),
  \bibinfo{pages}{58--71}.
\newblock
\urldef\tempurl%
\url{https://doi.org/10.26599/BDMA.2018.9020030}
\showDOI{\tempurl}


\bibitem[Han et~al\mbox{.}(2014)]%
        {chronos}
\bibfield{author}{\bibinfo{person}{Wentao Han}, \bibinfo{person}{Youshan Miao},
  \bibinfo{person}{Kaiwei Li}, \bibinfo{person}{Ming Wu},
  \bibinfo{person}{Vijayan Prabhakaran}, \bibinfo{person}{Wenguang Chen},
  \bibinfo{person}{Enhong Chen}, \bibinfo{person}{Fan Yang}, {and}
  \bibinfo{person}{Lidong Zhou}.} \bibinfo{year}{2014}\natexlab{}.
\newblock \showarticletitle{Chronos: A Graph Engine for Temporal Graph
  Analysis}. In \bibinfo{booktitle}{\emph{EuroSys}}.
\newblock


\bibitem[He and Singh(2006)]%
        {closuretree}
\bibfield{author}{\bibinfo{person}{Huahai He} {and} \bibinfo{person}{A.K.
  Singh}.} \bibinfo{year}{2006}\natexlab{}.
\newblock \showarticletitle{Closure-Tree: An Index Structure for Graph
  Queries}. In \bibinfo{booktitle}{\emph{22nd International Conference on Data
  Engineering (ICDE'06)}}. \bibinfo{pages}{38--38}.
\newblock
\urldef\tempurl%
\url{https://doi.org/10.1109/ICDE.2006.37}
\showDOI{\tempurl}


\bibitem[He and Singh(2008)]%
        {graphql}
\bibfield{author}{\bibinfo{person}{Huahai He} {and} \bibinfo{person}{Ambuj~K.
  Singh}.} \bibinfo{year}{2008}\natexlab{}.
\newblock \showarticletitle{Graphs-at-a-Time: Query Language and Access Methods
  for Graph Databases}. In \bibinfo{booktitle}{\emph{Proceedings of the 2008
  ACM SIGMOD International Conference on Management of Data}} (Vancouver,
  Canada) \emph{(\bibinfo{series}{SIGMOD '08})}.
  \bibinfo{publisher}{Association for Computing Machinery},
  \bibinfo{address}{New York, NY, USA}, \bibinfo{pages}{405–418}.
\newblock
\showISBNx{9781605581026}
\urldef\tempurl%
\url{https://doi.org/10.1145/1376616.1376660}
\showDOI{\tempurl}


\bibitem[Hopcroft and Karp(1971)]%
        {Hopcroft}
\bibfield{author}{\bibinfo{person}{John~E. Hopcroft} {and}
  \bibinfo{person}{Richard~M. Karp}.} \bibinfo{year}{1971}\natexlab{}.
\newblock \showarticletitle{A N5/2 Algorithm for Maximum Matchings in
  Bipartite}. In \bibinfo{booktitle}{\emph{Proceedings of the 12th Annual
  Symposium on Switching and Automata Theory (Swat 1971)}}
  \emph{(\bibinfo{series}{SWAT '71})}. \bibinfo{publisher}{IEEE Computer
  Society}, \bibinfo{address}{USA}, \bibinfo{pages}{122–125}.
\newblock
\urldef\tempurl%
\url{https://doi.org/10.1109/SWAT.1971.1}
\showDOI{\tempurl}


\bibitem[Jüttner and Madarasi(2018)]%
        {vf2}
\bibfield{author}{\bibinfo{person}{Alpár Jüttner} {and}
  \bibinfo{person}{Péter Madarasi}.} \bibinfo{year}{2018}\natexlab{}.
\newblock \showarticletitle{VF2++—An improved subgraph isomorphism
  algorithm}.
\newblock \bibinfo{journal}{\emph{Discrete Applied Mathematics}}
  \bibinfo{volume}{242} (\bibinfo{year}{2018}), \bibinfo{pages}{69--81}.
\newblock
\showISSN{0166-218X}
\urldef\tempurl%
\url{https://doi.org/10.1016/j.dam.2018.02.018}
\showDOI{\tempurl}
\newblock
\shownote{Computational Advances in Combinatorial Optimization}.


\bibitem[Kankanamge et~al\mbox{.}(2017)]%
        {graphflow}
\bibfield{author}{\bibinfo{person}{Chathura Kankanamge},
  \bibinfo{person}{Siddhartha Sahu}, \bibinfo{person}{Amine Mhedbhi},
  \bibinfo{person}{Jeremy Chen}, {and} \bibinfo{person}{Semih Salihoglu}.}
  \bibinfo{year}{2017}\natexlab{}.
\newblock \showarticletitle{Graphflow: An Active Graph Database}. In
  \bibinfo{booktitle}{\emph{Proceedings of the 2017 ACM International
  Conference on Management of Data}}. \bibinfo{pages}{1695–1698}.
\newblock


\bibitem[Kim et~al\mbox{.}(2018)]%
        {turboflux}
\bibfield{author}{\bibinfo{person}{Kyoungmin Kim}, \bibinfo{person}{In Seo},
  \bibinfo{person}{Wook-Shin Han}, \bibinfo{person}{Jeong-Hoon Lee},
  \bibinfo{person}{Sungpack Hong}, \bibinfo{person}{Hassan Chafi},
  \bibinfo{person}{Hyungyu Shin}, {and} \bibinfo{person}{Geonhwa Jeong}.}
  \bibinfo{year}{2018}\natexlab{}.
\newblock \showarticletitle{TurboFlux: A Fast Continuous Subgraph Matching
  System for Streaming Graph Data}. In \bibinfo{booktitle}{\emph{Proceedings of
  the 2018 International Conference on Management of Data}}.
  \bibinfo{pages}{411–426}.
\newblock
\showISBNx{9781450347037}


\bibitem[Ko et~al\mbox{.}(2021)]%
        {iturbograph}
\bibfield{author}{\bibinfo{person}{Seongyun Ko}, \bibinfo{person}{Taesung Lee},
  \bibinfo{person}{Kijae Hong}, \bibinfo{person}{Wonseok Lee},
  \bibinfo{person}{In Seo}, \bibinfo{person}{Jiwon Seo}, {and}
  \bibinfo{person}{Wook-Shin Han}.} \bibinfo{year}{2021}\natexlab{}.
\newblock \bibinfo{booktitle}{\emph{ITurboGraph: Scaling and Automating
  Incremental Graph Analytics}}.
\newblock \bibinfo{publisher}{Association for Computing Machinery},
  \bibinfo{pages}{977–990}.
\newblock


\bibitem[Kumar and Huang(2019)]%
        {graphone}
\bibfield{author}{\bibinfo{person}{Pradeep Kumar} {and}
  \bibinfo{person}{H.~Howie Huang}.} \bibinfo{year}{2019}\natexlab{}.
\newblock \showarticletitle{{GraphOne}: A Data Store for Real-time Analytics on
  Evolving Graphs}. In \bibinfo{booktitle}{\emph{17th USENIX Conference on File
  and Storage Technologies (FAST 19)}}. \bibinfo{pages}{249--263}.
\newblock


\bibitem[Le-Phuoc et~al\mbox{.}(2012)]%
        {lsbench}
\bibfield{author}{\bibinfo{person}{Danh Le-Phuoc}, \bibinfo{person}{Minh
  Dao-Tran}, \bibinfo{person}{Minh-Duc Pham}, \bibinfo{person}{Peter Boncz},
  \bibinfo{person}{Thomas Eiter}, {and} \bibinfo{person}{Michael Fink}.}
  \bibinfo{year}{2012}\natexlab{}.
\newblock \showarticletitle{Linked Stream Data Processing Engines: Facts and
  Figures}. In \bibinfo{booktitle}{\emph{The Semantic Web -- ISWC 2012}},
  \bibfield{editor}{\bibinfo{person}{Philippe Cudr{\'e}-Mauroux},
  \bibinfo{person}{Jeff Heflin}, \bibinfo{person}{Evren Sirin},
  \bibinfo{person}{Tania Tudorache}, \bibinfo{person}{J{\'e}r{\^o}me Euzenat},
  \bibinfo{person}{Manfred Hauswirth}, \bibinfo{person}{Josiane~Xavier
  Parreira}, \bibinfo{person}{Jim Hendler}, \bibinfo{person}{Guus Schreiber},
  \bibinfo{person}{Abraham Bernstein}, {and} \bibinfo{person}{Eva Blomqvist}}
  (Eds.). \bibinfo{publisher}{Springer Berlin Heidelberg},
  \bibinfo{address}{Berlin, Heidelberg}, \bibinfo{pages}{300--312}.
\newblock
\showISBNx{978-3-642-35173-0}


\bibitem[Leskovec and Krevl(2014)]%
        {snapnets}
\bibfield{author}{\bibinfo{person}{Jure Leskovec} {and} \bibinfo{person}{Andrej
  Krevl}.} \bibinfo{year}{2014}\natexlab{}.
\newblock \bibinfo{title}{{SNAP Datasets}: {Stanford} Large Network Dataset
  Collection}.
\newblock \bibinfo{howpublished}{\url{http://snap.stanford.edu/data}}.
\newblock


\bibitem[Li et~al\mbox{.}(2019)]%
        {CVC}
\bibfield{author}{\bibinfo{person}{Yuchao Li}, \bibinfo{person}{Wei Wang},
  {and} \bibinfo{person}{Zishen Yang}.} \bibinfo{year}{2019}\natexlab{}.
\newblock \showarticletitle{The Connected Vertex Cover Problem in K-Regular
  Graphs}.
\newblock \bibinfo{journal}{\emph{J. Comb. Optim.}} \bibinfo{volume}{38},
  \bibinfo{number}{2} (\bibinfo{date}{aug} \bibinfo{year}{2019}),
  \bibinfo{pages}{635–645}.
\newblock
\showISSN{1382-6905}
\urldef\tempurl%
\url{https://doi.org/10.1007/s10878-019-00403-3}
\showDOI{\tempurl}


\bibitem[Mariappan and Vora(2019)]%
        {graphbolt}
\bibfield{author}{\bibinfo{person}{Mugilan Mariappan} {and}
  \bibinfo{person}{Keval Vora}.} \bibinfo{year}{2019}\natexlab{}.
\newblock \showarticletitle{GraphBolt: Dependency-Driven Synchronous Processing
  of Streaming Graphs}. In \bibinfo{booktitle}{\emph{Proceedings of the
  Fourteenth EuroSys Conference 2019}}. \bibinfo{publisher}{Association for
  Computing Machinery}, Article \bibinfo{articleno}{25}.
\newblock


\bibitem[McGregor(2014)]%
        {10.1145/2627692.2627694}
\bibfield{author}{\bibinfo{person}{Andrew McGregor}.}
  \bibinfo{year}{2014}\natexlab{}.
\newblock \showarticletitle{Graph Stream Algorithms: A Survey}.
\newblock \bibinfo{journal}{\emph{SIGMOD Rec.}} (\bibinfo{date}{may}
  \bibinfo{year}{2014}), \bibinfo{pages}{9–20}.
\newblock


\bibitem[Min et~al\mbox{.}(2021)]%
        {symbi}
\bibfield{author}{\bibinfo{person}{Seunghwan Min}, \bibinfo{person}{Sung~Gwan
  Park}, \bibinfo{person}{Kunsoo Park}, \bibinfo{person}{Dora Giammarresi},
  \bibinfo{person}{Giuseppe~F. Italiano}, {and} \bibinfo{person}{Wook{-}Shin
  Han}.} \bibinfo{year}{2021}\natexlab{}.
\newblock \showarticletitle{Symmetric Continuous Subgraph Matching with
  Bidirectional Dynamic Programming}.
\newblock \bibinfo{journal}{\emph{Proc. {VLDB} Endow.}} \bibinfo{volume}{14},
  \bibinfo{number}{8} (\bibinfo{year}{2021}), \bibinfo{pages}{1298--1310}.
\newblock


\bibitem[Namaki et~al\mbox{.}(2017)]%
        {beams}
\bibfield{author}{\bibinfo{person}{Mohammad~Hossein Namaki},
  \bibinfo{person}{Keyvan Sasani}, \bibinfo{person}{Yinghui Wu}, {and}
  \bibinfo{person}{Tingjian Ge}.} \bibinfo{year}{2017}\natexlab{}.
\newblock \showarticletitle{BEAMS: Bounded Event Detection in Graph Streams}.
  In \bibinfo{booktitle}{\emph{2017 IEEE 33rd International Conference on Data
  Engineering (ICDE)}}. \bibinfo{pages}{1387--1388}.
\newblock


\bibitem[Qiu et~al\mbox{.}(2018)]%
        {10.14778/3229863.3229874}
\bibfield{author}{\bibinfo{person}{Xiafei Qiu}, \bibinfo{person}{Wubin Cen},
  \bibinfo{person}{Zhengping Qian}, \bibinfo{person}{You Peng},
  \bibinfo{person}{Ying Zhang}, \bibinfo{person}{Xuemin Lin}, {and}
  \bibinfo{person}{Jingren Zhou}.} \bibinfo{year}{2018}\natexlab{}.
\newblock \showarticletitle{Real-Time Constrained Cycle Detection in Large
  Dynamic Graphs}.
\newblock \bibinfo{journal}{\emph{Proc. VLDB Endow.}} (\bibinfo{date}{aug}
  \bibinfo{year}{2018}), \bibinfo{pages}{1876–1888}.
\newblock


\bibitem[Sahu et~al\mbox{.}(2017)]%
        {quicksi}
\bibfield{author}{\bibinfo{person}{Siddhartha Sahu}, \bibinfo{person}{Amine
  Mhedhbi}, \bibinfo{person}{Semih Salihoglu}, \bibinfo{person}{Jimmy Lin},
  {and} \bibinfo{person}{M.~Tamer \"{O}zsu}.} \bibinfo{year}{2017}\natexlab{}.
\newblock \showarticletitle{The Ubiquity of Large Graphs and Surprising
  Challenges of Graph Processing}.
\newblock \bibinfo{journal}{\emph{Proc. VLDB Endow.}} \bibinfo{volume}{11},
  \bibinfo{number}{4} (\bibinfo{date}{dec} \bibinfo{year}{2017}),
  \bibinfo{pages}{420–431}.
\newblock
\showISSN{2150-8097}


\bibitem[Shasha et~al\mbox{.}(2002)]%
        {10.1145/543613.543620}
\bibfield{author}{\bibinfo{person}{Dennis Shasha}, \bibinfo{person}{Jason T.~L.
  Wang}, {and} \bibinfo{person}{Rosalba Giugno}.}
  \bibinfo{year}{2002}\natexlab{}.
\newblock \showarticletitle{Algorithmics and Applications of Tree and Graph
  Searching}. In \bibinfo{booktitle}{\emph{Proceedings of the Twenty-First ACM
  SIGMOD-SIGACT-SIGART Symposium on Principles of Database Systems}}.
  \bibinfo{pages}{39–52}.
\newblock


\bibitem[Song et~al\mbox{.}(2014)]%
        {10.14778/2735496.2735504}
\bibfield{author}{\bibinfo{person}{Chunyao Song}, \bibinfo{person}{Tingjian
  Ge}, \bibinfo{person}{Cindy Chen}, {and} \bibinfo{person}{Jie Wang}.}
  \bibinfo{year}{2014}\natexlab{}.
\newblock \showarticletitle{Event Pattern Matching over Graph Streams}.
\newblock \bibinfo{journal}{\emph{Proc. VLDB Endow.}} \bibinfo{volume}{8},
  \bibinfo{number}{4} (\bibinfo{date}{dec} \bibinfo{year}{2014}),
  \bibinfo{pages}{413–424}.
\newblock


\bibitem[Tang et~al\mbox{.}(2016)]%
        {40999f14773f48b7a3e910d0e899be67}
\bibfield{author}{\bibinfo{person}{Nan Tang}, \bibinfo{person}{Qing Chen},
  {and} \bibinfo{person}{Prasenjit Mitra}.} \bibinfo{year}{2016}\natexlab{}.
\newblock \showarticletitle{Graph stream summarization: From big bang to big
  crunch}. In \bibinfo{booktitle}{\emph{SIGMOD 2016 - Proceedings of the 2016
  International Conference on Management of Data}}.
  \bibinfo{pages}{1481--1496}.
\newblock


\bibitem[Wang(2010)]%
        {5741690}
\bibfield{author}{\bibinfo{person}{Alex~Hai Wang}.}
  \bibinfo{year}{2010}\natexlab{}.
\newblock \showarticletitle{Don't follow me: Spam detection in Twitter}. In
  \bibinfo{booktitle}{\emph{2010 International Conference on Security and
  Cryptography (SECRYPT)}}. \bibinfo{pages}{1--10}.
\newblock


\bibitem[Zhang et~al\mbox{.}(2021)]%
        {DBLP:journals/www/ZhangGZW21}
\bibfield{author}{\bibinfo{person}{Qianzhen Zhang}, \bibinfo{person}{Deke Guo},
  \bibinfo{person}{Xiang Zhao}, {and} \bibinfo{person}{Xi Wang}.}
  \bibinfo{year}{2021}\natexlab{}.
\newblock \showarticletitle{Continuous matching of evolving patterns over
  dynamic graph data}.
\newblock \bibinfo{journal}{\emph{World Wide Web}} \bibinfo{volume}{24},
  \bibinfo{number}{3} (\bibinfo{year}{2021}), \bibinfo{pages}{721--745}.
\newblock


\end{thebibliography}

\received{April 2022}
\received[revised]{July 2022}
\received[accepted]{August 2022}

\end{document}